\documentclass[12pt]{article}

\usepackage{verbatim}
\usepackage{multirow}
\usepackage{amsmath}
\usepackage{scicite}
\usepackage{times}
\usepackage{graphicx} 
\usepackage{hyperref}
\usepackage{braket}
\usepackage[usenames,dvipsnames]{color}
\usepackage{siunitx}

\usepackage{xr}
\usepackage[sort&compress]{cleveref}

\usepackage[dvipsnames]{xcolor}

\definecolor{cwcolor}{RGB}{0, 0, 255} 
\newcommand\txtblue[1]{{\color{black}#1}}

\definecolor{swcolor}{RGB}{255, 165, 255}

\graphicspath{{Figures/}}

\newenvironment{sciabstract}{%
\begin{quote} \bf}
{\end{quote}}

\title{Operating semiconductor quantum processors with hopping spins}

\author{Chien-An Wang,$^{1}$ Valentin John,$^{1}$ Hanifa Tidjani,$^{1}$ Cécile X. Yu,$^{1}$\\
Alexander S. Ivlev,$^{1}$ Corentin Déprez,$^{1}$ Floor van Riggelen-Doelman,$^{1}$\\
Benjamin D. Woods,$^{2}$ Nico W. Hendrickx,$^{1}$ William I. L. Lawrie,$^{1}$\\
Lucas E. A. Stehouwer,$^{1}$ Stefan D. Oosterhout,$^{3}$ Amir Sammak,$^{3}$\\
Mark Friesen,$^{2}$ Giordano Scappucci,$^{1}$ Sander L. de Snoo,$^{1}$\\ 
Maximilian Rimbach-Russ,$^{1}$ Francesco Borsoi,$^{1, \dagger}$ Menno Veldhorst$^{1, \dagger, \ast}$\\
\\
\normalsize{$^{1}$QuTech and Kavli Institute of Nanoscience, Delft University of Technology,}\\
\normalsize{P.O. Box 5046, 2600 GA Delft, The Netherlands}\\
\normalsize{$^{2}$Department of Physics, University of Wisconsin-Madison}\\ 
\normalsize{Madison, WI, 53706, USA}\\
\normalsize{$^{3}$QuTech and Netherlands Organisation for Applied Scientific Research (TNO), }\\ 
\normalsize{2628 CK Delft, The Netherlands}\\
\\
\normalsize{$^\dagger$ These authors jointly supervised this work}\\
\normalsize{$^\ast$To whom correspondence should be addressed; E-mail:  m.veldhorst@tudelft.nl}
}

\date{}

\begin{document} 


\baselineskip24pt


\maketitle 

\begin{sciabstract}
Qubits that can be efficiently controlled are essential for the development of scalable quantum hardware. While resonant control is used to execute high-fidelity quantum gates, the scalability is challenged by the integration of high-frequency oscillating signals, qubit crosstalk and heating. 
Here, we show that by engineering the hopping of spins between quantum dots with site-dependent spin quantization axis, quantum control can be established with discrete signals.
We demonstrate hopping-based quantum logic and obtain single-qubit gate fidelities of 99.97\%, coherent shuttling fidelities of 99.992\% per hop, and a two-qubit gate fidelity of 99.3\%, corresponding to error rates that have been predicted to allow for quantum error correction. We also show that hopping spins constitute a tuning method by statistically mapping the coherence of a 10-quantum dot system.  
Our results show that dense quantum dot arrays with sparse occupation could be developed for efficient and high-connectivity qubit registers. 
\end{sciabstract}

Loss and DiVincenzo proposed hopping of electrons between two quantum dots as an efficient method for coherent spin control~\cite{Loss1998QuantumDots}. By applying discrete pulses to the quantum dot gates, a single spin can be transferred between qubit sites with differently oriented spin quantization axes, thereby enabling two-axis control of the qubit. Universal quantum logic is then achieved through tunable exchange interaction between spins residing in different quantum dots. That work initiated the field of semiconductor spin qubits and inspired over two decades of extensive research, but a successful implementation of their initial proposal has remained elusive due to experimental challenges~\cite{Burkard2023}. 

Alternative methods for coherent single-spin control have emerged, including \txtblue{electron} spin resonance~\cite{Koppens2006, Veldhorst2014} and \txtblue{electric} dipole spin resonance using either micromagnets~\cite{Pioro2007, Yoneda2018} or spin-orbit interaction~\cite{Bulaev2007, Nadj-Perge2010, Wang2022_NatureComm, Hendrickx2019_2Qubits} to enable a coupling between the electric field and the spin degree of freedom. However, all these methods rely on resonant Rabi driving and require high-power, and high-frequency analog control signals that already limit qubit performance in small quantum processors~\cite{Noiri2022, Xue2022, Philips2022}. The development of local, efficient, and low-power control mechanisms of semiconductor spins is \txtblue{now a key driver}~\cite{Takeda2018,Undseth2023, Undseth2024}. To this end, qubits encoded in multiple spins and in multiple quantum dots, such as singlet-triplet, hybrid, and exchange-only qubits,  have been investigated as \txtblue{possible} platforms~\cite{Burkard2023}. While these qubit encodings enabled digital single-qubit control, they also come with new challenges in coherence, control and creation of quantum links. 
For example, the exchange-only qubits are susceptible to leakage outside their computational subspace, require four exchange pulses to execute an arbitrary single-qubit gate and over 12 exchange pulses for a single two-qubit gate~\cite{russThreeelectronSpinQubits2017, Andrews2019, Weinstein2023}. 

Here, we demonstrate that single-spin qubits can be operated using \txtblue{baseband} control signals, as envisaged in the original proposal for quantum computation with quantum dots~\cite{Loss1998QuantumDots}. 
We use hole spins in germanium quantum dots, where the strong spin-orbit interaction gives rise to an anisotropic \textit{g}-tensor that is strongly dependent on the electrostatic and strain environment~\cite{Scappucci2021}. 
We harness the resulting differences in the spin quantization axis between quantum dots~\cite{Vanriggelendoelman2023,Jadot2021} to achieve high-fidelity single-qubit control using discrete pulses by shuttling the spin between quantum dot sites. 
A key advantage in such hopping-based operation is that the spin rotation frequency is given by the Larmor precession. The latter remains sizeable even at small magnetic fields where quantum coherence is substantially improved~\cite{Lawrie2023, Hendrickx2024}. 
This enables us to perform universal quantum control with error rates exceeding thresholds predicted for practical quantum error correction\cite{Fowler2012}, while also operating with low-frequency \txtblue{baseband} signals.
We then exploit the differences in quantization axes to map the spin dephasing times and $g$-factor distributions of an extended 10 quantum dot array, thereby efficiently gathering statistics on relevant metrics in large spin qubit systems.

\subsection*{High-fidelity single-qubit operations and long qubit coherence times at low magnetic field}
\begin{figure*}[htp!]
	\centering
	\includegraphics[width=0.99 \textwidth]{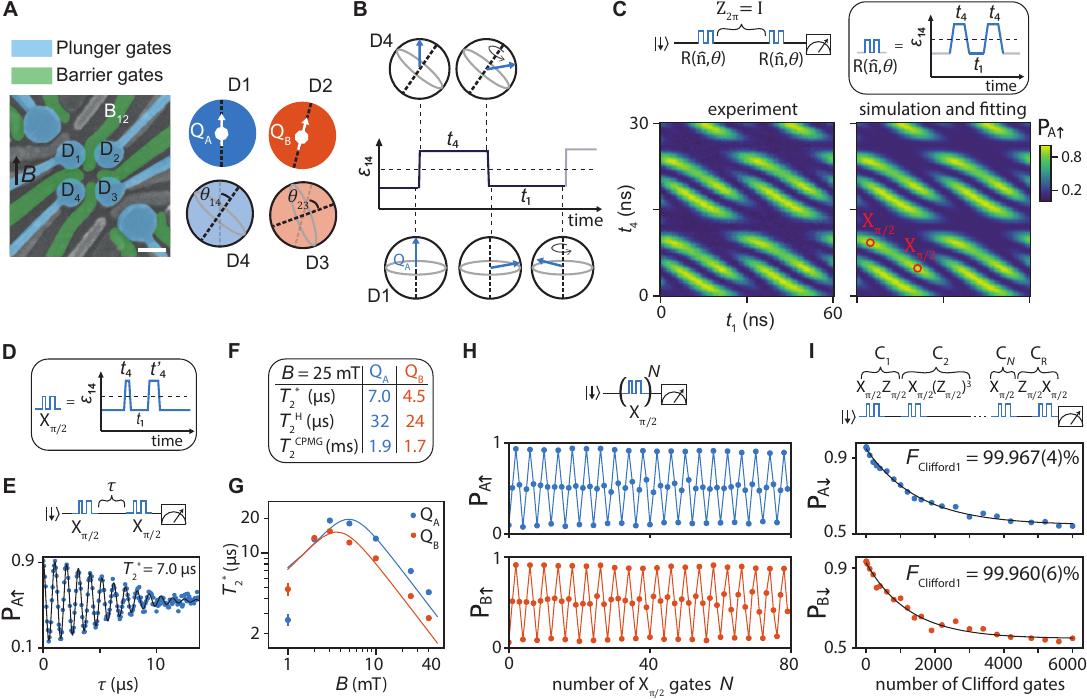}
	\caption{\textbf{High-fidelity hopping-based single-qubit operations and long qubit coherence times at low magnetic field.}   \textbf{(A)} (left) Scanning electron microscopy image of the 2$\times$2 quantum dot array device~\cite{Hendrickx2021AProcessor}, with scale bar of 100 nm, including gate-defined charge sensors at two corners. (right) Schematic of the two spin qubits $\rm Q_A$ and $\rm Q_B$. 
  The black dashed lines mark the relative quantization axis direction in the quantum dot pair D1-D4 (D2-D3), with the angle $\theta_{14}$ ($\theta_{23}$).  \textbf{(B)} Example of a baseband pulse $\epsilon_{14}(t)$ used to manipulate qubit $\rm Q_A$, by shuttling the spin back and forth between quantum dots D1 and D4 and allowing the spin to precess in the individual quantum dots for the time $t_4$ and $t_1$. 
  \textbf{(C)} Tune-up procedure of a four-shuttle pulse for the $\rm X_{\pi/2}$ gate of $\rm Q_A$ at 20~mT. On the top, we display the pulse sequence of the experiment, on the bottom left the measured spin-up probability $P_{\rm A \uparrow}(t_1, t_4)$, and on the bottom right the simulation result. The red markers identify the timings for implementing an $\rm X_{\pi/2,A}$ gate and correspond to the maximal spin-up probability. The markers are periodic in $t_{\rm 1}$ and $t_{\rm 4}$, but for clarity we only plot a few of them. 
  \textbf{(D)} The calibrated pulse for $\rm X_{\pi/2, A}$ gate with unequal wait time $t_4$ and $t_4^\prime$. 
  \textbf{(E)} The free induction decay obtained from Ramsey experiments at 25~mT. 
  \textbf{(F)} The coherence times $T_2^{*}$, $T_2^{\rm H}$ and $T_2^{\rm CPMG-512}$ of both qubits at 25~mT. 
  \textbf{(G)} $T_2^{*}$ as a function of magnetic field. The data points are fitted with an effective model including electric noise and nuclear noise [\cite{Supplementary},~\labelcref{sec:coherence_2qubits}]. 
  \textbf{(H)} The spin-up probability after applying a varying number of $\rm X_{\pi/2}$ gates on each qubit. 
  \textbf{(I)} An example of pulse sequence in $\rm Q_A$ single-qubit randomized benchmarking and the measurement results of both qubits. The uncertainties are obtained from bootstrapping with 95\% confidence intervals.  }
	\label{fig:figure1}
\end{figure*}

A large difference in the orientation of the spin quantization axes between quantum dots is essential for hopping-based qubit operations. 
Holes in planar germanium heterostructures manifest a pronounced anisotropic $g$-tensor, with an out-of-plane $g$-factor $g_\perp$ that can be \txtblue{two orders of magnitude} larger than the in-plane component $g_\parallel$~\cite{Scappucci2021,Jirovec2021b, Wang2022, Hendrickx2024}. 
\txtblue{Consequently}, a small tilt of the applied magnetic field from the in-plane $g$-tensor will lead to a strong reorientation of the spin quantization axis in the out-of-plane direction. Subsequently, when an in-plane magnetic field is applied, the orientation of the spin quantization axis is highly sensitive to the local $g$-tensor, and thus confinement, strain, and electric fields, therefore becoming a site-dependent property~\cite{Hendrickx2024, Abadillo2023, Vanriggelendoelman2023, Corley-Wiciak2023}. 
Here, we exploit this aspect to establish hopping-based quantum operations in two different devices: a four-quantum dot array~\cite{Hendrickx2021AProcessor} arranged in a 2$\times$2 configuration and a 10 quantum dot system arranged in a 3-4-3 configuration. \\
We populate the four-quantum dot array with quantum dots D$m$ with $m$ $\in$ [[1, 4]] with two hole spins $\rm Q_A$ and $\rm Q_B$ which can be shuttled between quantum dots by electrical pulses on the gate electrodes \txtblue{(Fig.~\ref{fig:figure1}A)}. A magnetic field up to 40 mT is applied to split the spin states and positioned in-plane up to sample-alignment accuracy [see Materials and Methods~\cite{Supplementary}]. The relatively small magnetic fields ensure that the maximum qubit frequency (140 MHz) and its corresponding precession period (7 ns) are within the bandwidth of the used arbitrary waveform generators.
In combination with engineered voltage pulses with sub-nanosecond resolution~\cite{Vanriggelendoelman2023} [\cite{Supplementary},~\labelcref{sec:subnanosecond_timing}], we are able to shuttle a spin qubit to an empty quantum dot and thereby accurately change the qubit precession direction several times within one precession period.
Altogether, this enables efficient single-qubit control via discrete voltage pulses (Fig.~\ref{fig:figure1}B). \\
Crucially, the net effect of a multiple-shuttle protocol is a rotation ${\rm R}(\hat{\rm n},\theta )$ of the spin state around an axis $\hat{\rm n}$ and with an angle $\theta$.
To implement a specific rotation such as the quantum gate $\rm X_{\pi/2}$, the number of required shuttling steps depends on the angle between the two quantization axes. 
Due to the large angle between the axes of D1 and D4, $\theta_{14} >90^\circ/4=22.5^\circ$, a pulse consisting of four shuttling steps is sufficient to realize a precise quantum gate $\rm X_{\pi/2, A}$[\cite{Supplementary},~\labelcref{sec:dots_parameters} and~\labelcref{sec:simulations_shuttling_gates}].
As outlined on the top right panel of Fig.~\ref{fig:figure1}C, such a four-shuttle pulse moves the spin between D1 and D4 four times with waiting periods $t_1$ and $t_4$, respectively.
By measuring the spin-flip probability of $\rm Q_A$,  $P_{\rm A \uparrow}$, after two consecutive rotations ${\rm R}(\hat{\rm n},\theta )^2$, we can determine the values of $t_1$ and $t_4$ where $P_{\rm A \uparrow}$ is maximal, which occur when ${\rm R}(\hat{\rm n},\theta )={\rm X_{\pi/2, A}}$.

While this method allows \txtblue{calibration of} the pulse timing to compose an $\rm X_{\pi/2, A}$ gate, it is not necessarily the optimal trajectory. Different choices of $(t_1, t_4)$ are possible (Fig.~\ref{fig:figure1}C), including a composition of four-shuttle pulses with different waiting times in D4. 
The latter implementation allows to construct gates which have a rotation angle $\theta$ less sensitive to Larmor frequency fluctuations in D4. We construct such a gate by fitting the data in Fig.~\ref{fig:figure1}C to an effective model and determine the quantization axes angle $\theta_{14}$ between the quantum dots D1 and D4, the individual Larmor frequencies, and the effective precession time during the ramp.  
Through simulation of the qubit dynamics we design a more noise-resilient $\rm X_{\pi/2, A}$ gate based on four shuttling steps with unequal wait times $t_4$ and $t_4^\prime$ in D4 (Fig.~\ref{fig:figure1}D). 
Following the same approach, we design an $\rm X_{\pi/2, B}$ gate for $\rm Q_B$ that only requires a two-shuttle protocol as the angle of the difference in quantization axes of D2 and D3, $\theta_{23}$, is very close to $45^\circ$ [\cite{Supplementary},~\labelcref{sec:simulations_shuttling_gates}]. 

We further calibrate the pulse timing using repetition sequences as shown in Fig.~\ref{fig:figure1}H and AllXY sequences~\cite{reed2013entanglement} [see~\cite{Supplementary},~\labelcref{sec:simulations_shuttling_gates}].
The $\rm Y_{\pi/2}$ gate in the AllXY sequences is realized by $\rm Y_{\pi/2}=Z_{\pi/2} X_{\pi/2} Z_{3\pi/2}$ and the $\rm Z_{\pi/2}$ gate is implemented by idling the qubit for the time defined by its precession in the lab frame. 
The calibrated $\rm X_{\pi/2}$ gates have a total gate time of 98 (35)~ns for $\rm Q_A$($\rm Q_B$), corresponding to effective qubit rotation frequencies of 2.6 (7.1)~MHz, considerable compared to the Larmor frequencies $f_{\rm A(B)}$ = 42.6 (89.5)~MHz at the in-plane magnetic field of \SI{25}{mT}.

The high ratio between qubit rotation and Larmor frequency results in low power dissipation, which is a critical aspect for scaling up quantum processors~\cite{Vandersypen2017InterfacingCoherent}.
To compare the power consumption of the hopping-based single-qubit control with the electric dipole spin resonance technique, we define the required number of voltage oscillations to flip a qubit, $N_{\rm cycles}$, and the derived energy efficiency $\eta$ =$1/N_{\rm cycles}$, which we find largely determining the power dissipation \txtblue{under the assumption that dielectric losses are dominant over other dissipation mechanisms} [\cite{Supplementary},~\labelcref{sec:power_dissipation}].
For our system, we estimate an efficiency of  $\eta=25 (50)\%$  for qubit A(B).
\txtblue{By comparison,} previous demonstrations of high-fidelity universal qubit logic \txtblue{in silicon exhibited} $\eta $ in the range of 0.04 - 0.07\%~\cite{Xue2022,Undseth2023,Noiri2022}. Moreover, despite applying sizeable amplitudes to move the spins between localized orbitals of adjacent quantum dots, we still obtain \txtblue{a factor of 20} reduction in power dissipation with respect to the electric dipole spin resonance technique [\cite{Supplementary},~\labelcref{sec:power_dissipation}]. Engineering lower required pulse amplitudes and increasing the orthogonality of the spin quantization axes will enable to further reduce the dissipated power. \txtblue{Furthermore, the hopping-based approach can simplify the signal delivery and required control electronics, and alleviate detrimental heating effects. }

Having established universal single-qubit control, we utilize the set of gates \{$X_{\pi/2}$, $Y_{\pi/2}$\} to investigate the qubit coherence times at low magnetic fields.
By \txtblue{using} a Ramsey sequence (Fig.~\ref{fig:figure1}E), we obtain a dephasing time $T_2^{*}$ of 7.0 (4.5) $\mathrm{\mu s}$ at 25 mT for $\rm Q_A$($\rm Q_B$), an order of magnitude larger than measured at 1 T \txtblue{ in the same sample}~\cite{Hendrickx2021AProcessor, Lawrie2023}. 
We can further extend the coherence times using Hahn and CMPG techniques obtaining $T_2^{\rm H} =\SI{32(24)}{\micro s}$ and $T_2^{\rm CPMG-512}=1.9(1.7)$~ms (Fig.~\ref{fig:figure1}F).
The dependence of the dephasing times as a function of magnetic field (Fig.~\ref{fig:figure1}G) indicates that charge noise remains the main cause for decoherence for magnetic fields as low as 5~mT [\cite{Supplementary},~\labelcref{sec:coherence_2qubits}]. 

We characterize the single-qubit gate fidelity using randomized benchmarking (RB) and gate set tomography (GST)~\cite{nielsenProbingQuantumProcessor2020, blume-kohoutDemonstrationQubitOperations2017, dehollainOptimizationSolidstateElectron2016} [with details discussed in~\cite{Supplementary},~\labelcref{sec:RB_method} and~\labelcref{sec:GST_method}]. 
The results of RB with average Clifford fidelity (Fig.~\ref{fig:figure1}I) set the lower bounds of the $\rm X_{\pi/2}$ average gate fidelity $F_{\rm X_{\pi/2},A} \geq$ 99.967(4)\% and $F_{\rm X_{\pi/2},B} \geq$ 99.960(6)\%, consistent with the error modeling [\cite{Supplementary},~\labelcref{sec:model_Xgate_error}].
Using GST we benchmark the $\rm X_{\pi/2}$ and $\rm Y_{\pi/2}$ gates, obtaining an average gate fidelity above 99.9\%. From the GST analysis, we infer that dephasing is the dominant contribution to the average gate infidelity. 
Taking into account the multiple shuttling steps to execute a single gate, we estimate a coherent shuttling fidelity \txtblue{per hop} as high as $F_{\rm shuttle}=99.992\%$[\cite{Supplementary},~\labelcref{sec:shuttling_fidelity}]. 

\subsection*{High-fidelity two-qubit exchange gate}
\begin{figure*}[htp!]
	\centering
	\includegraphics[width=0.99 \textwidth]{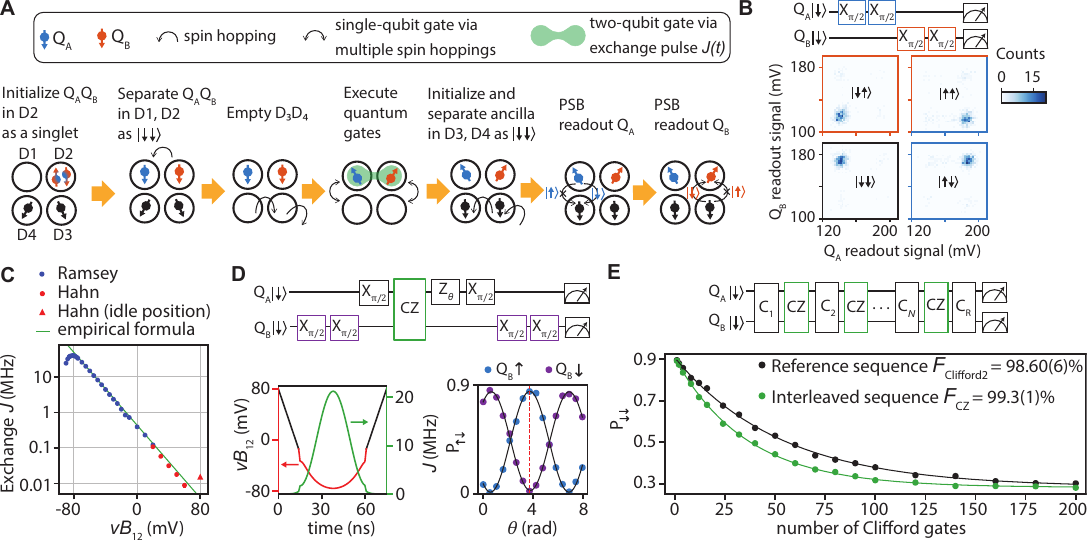}
	\caption{\textbf{High-fidelity two-qubit gate in germanium.}   
	 \textbf{(A)} Schematics of two-qubit initialization, manipulation and individual readout.~\txtblue{$\ket{\rm Q_A Q_B}$ is initialized by relaxing to the singlet ground state in D2 and then adiabatically moving one spin to D1. Quantum circuits consisting of single-qubit gates (spin hoppings) and two-qubit gates (exchange pulse $J(t)$) are performed. The final quantum state is read out by preparing ancillary spins and then performing two PSB readouts. In each readout, the chemical potentials of the quantum dots are pulsed such that the spin can either move to the neighboring dot (indicated by arrows) or stay in the original dot (indicated by arrows with $\times$ markers) with probabilities depending on the spin state $\ket{\rm Q_{A(B)}}$.}  \textbf{(B)} The 2D histograms of the sensor signals formed by 500 single-shot measurements for four different two-qubit states, which are prepared by applying $\rm X_{\pi/2,A(B)}$ gates.   \textbf{(C)} The exchange coupling as a function of virtual barrier gate ${vB}_{12}$, measured by Ramsey (Hahn echo) experiments in the large (small) coupling regime. \txtblue{The idle position corresponds to the barrier voltage where single qubit gates are performed, but at slightly different plunger gate voltage. The empirical formula for mapping ${vB}_{12}$ and $J$ is detailed in~\cite{Supplementary}~\labelcref{sec:calibrate_CZ}.} The bending on the left side of the plot results from the energy level anti-crossing when $J \sim f_{\rm A}$. \textbf{(D)} The voltage pulse of the CZ gate is shaped to have exchange $J(t)$ in the form of a Hamming window, as illustrated in the bottom left. The CZ gate calibration circuit for single-qubit phases is on the top, with the measurement outcome plotted on the bottom right. The target qubit ($\rm Q_A$) phase depends on the control qubit $\rm Q_B$ being in the state $\ket{\downarrow}$ in blue ($\ket{\uparrow}$ in purple). The red dashed line marks the required single-qubit phase of $\rm Q_A$ for the CZ gate. \textbf{(E)} Gate sequence and measurement result of two-qubit interleaved RB.      }
	\label{fig:figure2}
\end{figure*}

We now focus on assessing the single-qubit and two-qubit gate performance in the two-qubit space. 
We implement a two-qubit state preparation and measurement (SPAM) protocol (Figs.~\ref{fig:figure2}A,B).
For the state preparation, we adiabatically convert the two-spin singlet in D2 to the triplet ${\ket{\rm Q_A Q_B}}=\ket{\downarrow \downarrow}$.
For the state measurement, we perform sequential Pauli spin blockade (PSB) readout on $\rm Q_A$ and $\rm Q_B$ by loading ancillary spins from the reservoir \txtblue{and adiabatic conversion to the state $\ket{\downarrow \downarrow}$ in quantum dots D3 and D4}.  
The difference in the effective $g$-factor between the quantum dots D1 and D2 allows for the construction of a \txtblue{controlled-Z (CZ)} gate even at low magnetic fields. 
We do so by pulsing the virtual barrier gate voltage $vB_{12}$, which controls the exchange coupling $J$ between $\rm Q_A$ and $\rm Q_B$ from 10 kHz to 40 MHz (Fig.~\ref{fig:figure2}C) [see~\cite{Supplementary},~\labelcref{sec:measure_residual_exchange} and~\labelcref{sec:model_twoQ} for further details].
\txtblue{Because} the maximum exchange coupling strength is non-negligible compared to the Zeeman energy difference $\Delta E_{\rm Z}$ and the qubit frequency $f_{\rm A}$, pulse shaping is essential to mitigate coherent errors~\cite{Xue2022,rimbach-russSimpleFrameworkSystematic2023}. 
We implement exchange pulses with a Hamming window and perform the \txtblue{CZ} gate calibration (Fig.~\ref{fig:figure2}D)  [\cite{Supplementary},~\labelcref{sec:calibrate_CZ}]. 

\txtblue{We now advance to benchmarking a two-qubit gate in germanium,} by executing two-qubit randomized benchmarking [see~\cite{Supplementary},~\labelcref{sec:RB_method} for further details and~\cite{Supplementary},~\labelcref{sec:GST_method} for two-qubit GST]. Individual Clifford gates are implemented by sequentially applying \txtblue{one or more of} the gates $\rm CZ$, $\rm X_{\pi/2}^{A(B)}$, $\rm Z_{\pi/2}^{A(B)}$, and $\rm I$.
From the fit of the decay constants of the reference and interleaved sequence in Fig.~\ref{fig:figure2}E, we determine the average Clifford gate fidelity $F_{\rm Clifford2}$ = 98.60(6)\% and average CZ gate fidelity $F_{\rm CZ}$ = 99.33(10)\%, consistent with the result of error modeling [\cite{Supplementary},~\labelcref{sec:model_CZ_error}]. 
For the single-qubit gate performance in the two-qubit space, we estimate the lower bound of fidelity, averaged between both qubits, as $\frac{1}{2}(F_{\rm X_{\pi/2},A}+F_{\rm X_{\pi/2},B}) \geq $ 99.90(5)\%. We believe these high fidelities to result from the high driving efficiency and the relatively long $T_2^\star$ at low magnetic field. 

\subsection*{Hopping spins to benchmark large and high-connectivity quantum dot architectures}
\begin{figure*}[htp!]
	\centering
	\includegraphics[width=0.99 \textwidth]{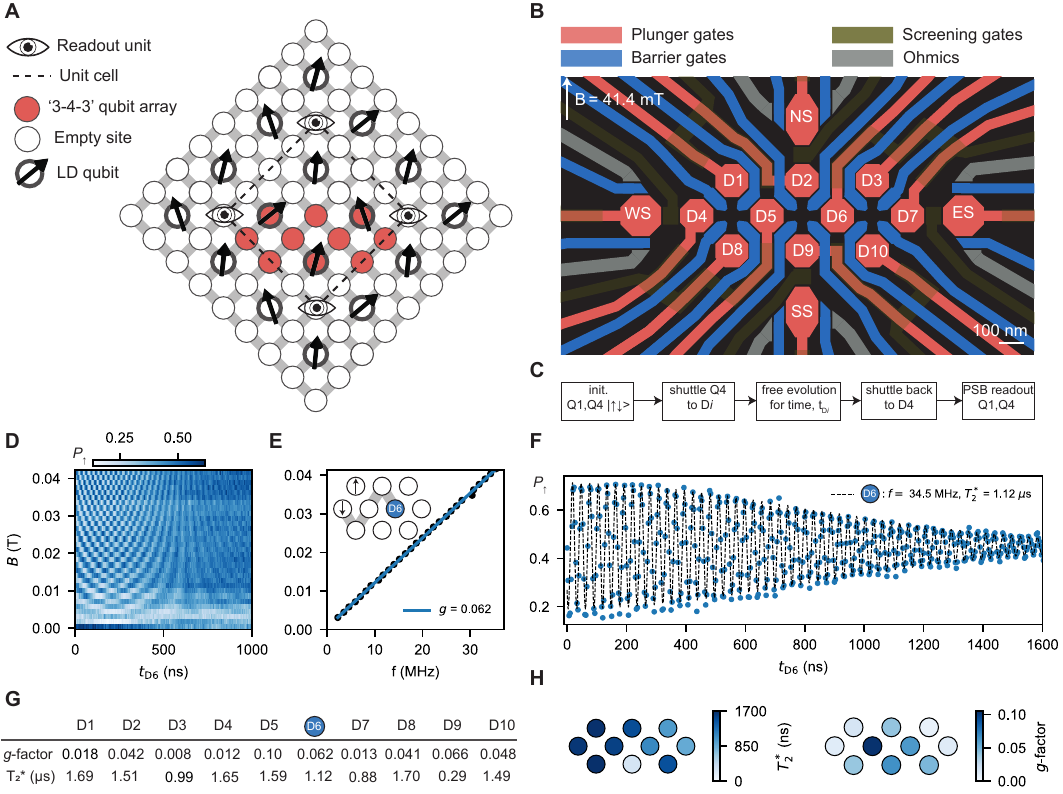}
	\caption{\textbf{Hopping spins to benchmark large and high-connectivity quantum dot architectures.} 
		\textbf{(A)}~Our vision of a semiconductor quantum computing architecture comprising hopping Loss-DiVincenzo (LD) spin qubits (black arrows), readout units (eyes), and empty quantum dot sites for shuttling operations.
	    \textbf{(B)}~Layout of the 10 quantum dot array, with gate-defined charge sensors labelled in analogy of the four cardinal points (NS, ES, WS, SS).
	    \textbf{(C)}~Control sequence employed to characterize the array: a spin originally in D4 is shuttled across the whole array, allowed to evolve at a certain quantum dot, and read out.
		\textbf{(D)}~Qubit rotations induced by the difference in quantization axes as a function of idling time in quantum dot D6 and magnetic field. 
		\textbf{(E)}~D6 Larmor frequency, extracted from the Fourier analysis of (d), versus magnetic field. Linear fit yields an estimated $g$-factor of \txtblue{0.062}. 
		Inset shows the shuttling trajectory of the spin qubit from D4 to D6.
		\textbf{(F)}~Extended time evolution in D6 at \txtblue{B = 41.4} mT, yielding a qubit frequency of 34.51 MHz and a dephasing time of $T^{\ast}_2 = $1.12 $\mu s$.
		The experimental trace is fitted (dashed lines) as described in~\cite{Supplementary}~\labelcref{sec:dephasing_Larmor_10quantumdot}.
		\textbf{(G, H)} Table and visualization of the extracted parameters: $g$-factors, $T^{\ast}_2$, respectively.  
}
	\label{fig:figure3}
\end{figure*}

The presented sparse occupation of a quantum dot array allows to construct high-fidelity hopping-based quantum logic, but it may also facilitate the implementation of quantum circuits with high-connectivity. While two-dimensional quantum circuits with nearest neighbor connectivity can already tolerate high error rates~\cite{Wang2011, Fowler2012, Hetenyi2023}, an increased connectivity may drastically lower the physical qubit overhead and lower the logical qubit error rate~\cite{Bravyi2023}. We therefore envision a qubit architecture with sparse occupation (Fig.~\ref{fig:figure3}A), to be as a potential platform. Here, qubits may be shuttled to remote sites for distant two-qubit logic, while single-qubit logic can be executed during this trajectory. \\
As a first step toward such architectures, we develop and characterize an extended system comprising 10 quantum dots.
The system (Fig.~\ref{fig:figure3}B) consists of a multilayer gate architecture with quantum dots (D$n$ with $n \in [[1, 10]]$) and peripheral charge sensors, which may be integrated within the array through development of vertical interconnects such as in ref.~\cite{Ha2022}.
By exploiting dedicated (virtual) barrier and plunger gate voltages, we prepare the quantum dots D1 and D4 in the single-hole regime, leaving the others empty [\cite{Supplementary},~\labelcref{sec:10_quantum_dot_array} and~\labelcref{sec:shuttling_across_multiple_dots}].
The hopping-based qubit gates \txtblue{are used} to rapidly characterize the different quantum dot $g$-factors and coherence times. After initializing the associated qubit pair Q1,Q4 into its $\ket{\uparrow \downarrow}$ eigenstate, we diabatically shuttle the Q4 spin to another quantum dot site D$n$. We let it precess for a time $t_{\mathrm{D}n}$, after which the spin is shuttled back and read out. 
The misalignment between the spin quantization axes gives rise to spin rotations with the Larmor frequency $f_{\mathrm{D}n}$~\cite{Vanriggelendoelman2023}.
The resulting oscillations \txtblue{are shown} as a function of waiting time in D6, $t_{\mathrm{D6}}$, and magnetic field (Fig.~\ref{fig:figure3}D).
From the linear scaling of the D6 Larmor frequency with the magnetic field, we extract an effective $g$-factor of \txtblue{0.062} (Fig.~\ref{fig:figure3}E), and from the decay of the oscillations a dephasing time of $T^{\ast}_2 = 1.12 \, \mathrm{\mu s}$ (Fig.~\ref{fig:figure3}F).
Repeating this protocol to reach all the quantum dots, we extract the Larmor frequency and dephasing time at each site, as displayed in Figs.~\ref{fig:figure3}G, H.
For the case of Q1 (Q4), we shuttle the spin to D5 (D8) back and forth twice, interleaved by a varying precession time in D1, $t_{\mathrm{Q1}}$ (in D4, $t_{\mathrm{Q4}}$), which we explain in detail in~\cite{Supplementary}~\labelcref{sec:shuttling_occupied_dots}.
Our experiments show an average $T^{\ast}_2$ of $ 1.3 \pm 0.4 \, \mathrm{\mu s}$ at a magnetic field of \txtblue{41.4} mT [\cite{Supplementary},~\labelcref{sec:dephasing_Larmor_10quantumdot}], and
we attribute the fast dephasing of D9 ($T^{\ast}_2$ = \txtblue{290} ns) to charge noise originating from a fluctuator nearby. 
Furthermore, we obtain an average $g$-factor of \txtblue{$0.04 \pm 0.03$}. The observed variability in this distribution is likely a result of multiple factors: the heterogeneity inherent in the shapes of the quantum dots (dot-to-dot variability), the presence of strain gradients in the quantum well arising from the gates above or the SiGe strained relaxed buffer below, and the impact of interface charges. 
Notably, the average $g$-factor is considerably lower than observed in the literature~\cite{Hendrickx2019_2Qubits, Hendrickx2021AProcessor, Jirovec2021b, Hendrickx2024}. 
We \txtblue{suggest} that this reduction is primarily due to two phenomena: a precise in-plane magnetic field configuration and an \txtblue{appreciable} renormalization of the gyromagnetic ratio from the pure heavy-hole value of $\sim 0.18$~\cite{Martinez2022, Abadillo2023, Wang2022}. 
Such renormalization is driven by substantial inter-band mixing between the heavy-hole and the light-hole band, which we attribute to asymmetries in the strain, as simulated in~\cite{Supplementary}~\labelcref{sec:simulations_gfactor}.
Furthermore, these simulations indicate that such a low average effective $g$-factor only occurs when the misalignment of the magnetic field is smaller than $0.1 ^\circ$ with respect to the plane of the $g$-tensors, \txtblue{emphasizing} the importance of accurately controlling the magnetic field orientation when operating with germanium qubits.
\section*{Conclusion}
\txtblue{We have shown} that hopping spin qubits between quantum dots with site-dependent $g$-tensors allows for coherent shuttling with fidelities up to 99.992\% \txtblue{per hop}, single-qubit gate fidelities up to 99.97\%, and two-qubit gate fidelities up to 99.3\%. 
This method allows for efficient control with \txtblue{baseband} pulses only and fast execution of quantum gates even at low magnetic fields where the coherence is high. 
\txtblue{Utilizing this approach for control of dense quantum dot arrays with sparse qubit occupation can alleviate challenges in crosstalk and heating, while providing high connectivity.}
Recent theoretical developments predict that increased connectivity \txtblue{can} substantially improve logical qubit performance and reduce the required overhead on physical qubits~\cite{Bravyi2023}. 
Sparse spin qubit arrays \txtblue{could} be particularly suited for error correction schemes requiring a larger number of nearest neighbors, or requiring coupling beyond nearest neighbors.
A significant challenge remains in addressing the qubit-to-qubit variation. 
Remarkably, this was already highlighted in the original work by Loss and DiVincenzo~\cite{Loss1998QuantumDots}. 
We envision that the characterization of larger qubit arrays and statistical analysis will become pivotal, with the presented 10 quantum dot array already providing a first indication that design considerations can determine relevant qubit parameters. 
\txtblue{Site-dependent quantization axes can be realized by g-tensor engineering for example in elongated quantum dots~\cite{Bosco2021b}, by using nanomagnets, or by applying currents through nanowires above the qubit plane~\cite{Li2018}. The developed control methods for high timing accuracy can also advance exchange-only qubits that are operated using baseband pulses~\cite{Weinstein2023} and impact platforms such as superconducting qubits~\cite{Campbell2020}. We envision establishing high-fidelity quantum operation through low-power control in uniform and large-scale systems to be a critical step in realizing fault-tolerant quantum computing.}


\newpage
%

\bibliographystyle{Science}

\section*{Acknowledgments}
We are grateful to B. Undseth, I. F. de Fuentes, X. Xue, E. Raymenants, C. Ostrove, Y.-M. Niquet, and J. C. Abadillo-Uriel for fruitful discussions. We thank L. M. K. Vandersypen for proofreading. \\
\textbf{Funding:} We acknowledge support by the Dutch Research Council through an NWO ENW grant and by the European Union through ERC Starting Grant QUIST (850641)  and through the IGNITE project of European Union’s Horizon Europe Framework Programme under grant agreement No. 101069515. F.B. acknowledges support from the Dutch Research Council (NWO) via the National Growth Fund programme Quantum Delta NL (Grant No. NGF.1582.22.001). N.W.H. acknowledges support from the European Union through EIC Transition Grant GROOVE (101113173). M.R.-R. acknowledges support from the Dutch Research Council (NWO) under Veni grant (VI.Veni.212.223). This research was sponsored in part by the Army Research Office (ARO) under Awards No. W911NF-23-1-0110 and No. W911NF-17-1-0274. The views, conclusions, and recommendations contained in this document are those of the authors and are not necessarily endorsed nor should they be interpreted as representing the official policies, either expressed or implied, of the Army Research Office (ARO) or the U.S. Government. The U.S. Government is authorized to reproduce and distribute reprints for Government purposes notwithstanding any copyright notation herein.\\
\textbf{Author contributions:} C.-A.W., V.J., H.T., C.X.Y., A.I. and F.B. conducted the experiments.
C.-A.W. and F.B. analyzed the data.  
C.-A.W., C.D., B.D.W., M.F. and M.R.-R. performed the simulations and theoretical analysis.
W.I.L.L. and S.D.O. fabricated the devices.
V.J., C.X.Y., F.B., F.v.R.-D. and N.W.H. contributed to the devices development and measurement setups.
S.L.d.S developed the measurement software.
L.E.A.S., A.S. and G.S. supplied the heterostructures.
C.-A.W., F.B. and M.V. wrote the manuscript with input from all authors.
M.V. and F.B. supervised the project.\\
\textbf{Competing interests:} N.W.H. is also affiliated with Groove Quantum BV and declares equity interest. N.W.H. and M.V. are inventors on a patent application (NL provisional application N2036660) submitted by Delft University of Technology related to controlling semiconductor qubits. The other authors declare no competing interests. \\
\textbf{Data and materials availability:}   All data are available in the manuscript, the supplementary material or deposited at 4TU.ResearchData repository~\cite{Alldata}.


\newpage
\section*{Supplementary materials}
Materials and Methods\\
Supplementary Text, Section 1 to Section 18\\
Figs. S1 to S26\\
Tables S1 to S9\\
References \textit{(47-70)}

\renewcommand{\thepage}{S\arabic{page}} 
\renewcommand{\thesection}{Section }   
\renewcommand{\thesubsection}{Section \arabic{subsection}}   

\setcounter{figure}{0}
\setcounter{page}{1}
\setcounter{section}{0}
\setcounter{subsection}{0}   
\setcounter{table}{0}

\renewcommand{\figurename}{\textbf{Fig.}}
\renewcommand{\thefigure}{\textbf{S\arabic{figure}}}

\renewcommand{\tablename}{\textbf{Table}}
\renewcommand{\thetable}{\textbf{S\arabic{table}}}

\section*{Materials and Methods}
The two devices are fabricated on a Ge/SiGe heterostructure \txtblue{with} a 16 nm germanium quantum well buried 55 nm below the semiconductor/oxide interface~\cite{Sammak2019ShallowTechnology, Lodari2021}.
The devices gate stack is realised using a multilayer of Pd gates and $\mathrm{Al_2 O_3}$ gate oxide, grown by atomic layer deposition. 
Ohmic contacts are made by a thermally-diffused Al and Pt contact layer for the 2$\times$2 and 10 quantum dot devices, respectively.  
Details on the fabrication of the first device can be found in ref.~\cite{Hendrickx2021AProcessor}.
The second device is based on a similar approach, but has an additional layer of gates and gate oxide.
The experiments are performed in two Bluefors dilution refrigerators with an electron temperature lower than $140$ mK~\cite{Borsoi2023}. We estimate a possible misalignment angle between the device plane and the magnetic field axis of $\pm 2^\circ$. \txtblue{We also note that due to an offset in the height position of the 10 quantum dots chip on the cold finger of the cryostat with respect to the center of the field, the effective magnetic field is 69\% of the applied field. We have determined this factor using the Ge-73 gyromagnetic ratio, measured via CPMG sequences on a different device mounted in the same position in a different cool-down. This factor also agrees well to what estimated using simulations of the coil field. }   
In each of the two setups, we \txtblue{utilize} an in-house built battery-powered SPI rack \url{https://qtwork.tudelft.nl/~mtiggelman/spi-rack/chassis.html} to set direct-current (DC) voltages, while we use a Keysight M3202A arbitrary waveform generator (AWG) to apply alternating-current (AC) pulses via coaxial lines.
The DC and AC voltage signals are combined on the printed circuit board (PCB) with bias-tees and applied to the gates. 
\txtblue{ In the individual bias-tee, the DC signal undergoes a resistor of 1 M$\Omega$, and the high-frequency signal undergoes a capacitor of 100 nF. } 
Each charge sensor is galvanically connected to a NbTiN inductor with an inductance of a few $\mathrm{\mu H}$ forming a resonant tank circuit with resonance frequencies of $\sim 100$ MHz. The reflectometry circuit also consists of a directional coupler (ZEDC-15-2B) mounted on the mixing chamber stage. The readout signals are amplified by a cryogenic SiGe amplifier mounted on the 4 K stage (a CITLF3 with gain of 33 dB), by a room-temperature amplifier (a M2j module of the SPI Rack with a gain of 70 dB)  and demodulated with a Keysight M3102A digitizer module with a sampling rate of 500 MSa/s. 

\newpage
\section*{Supplementary Text}
\subsection{Timing precision of shuttling pulses}
\label{sec:subnanosecond_timing}
\begin{figure}
	\centering
	\includegraphics[width=0.99\textwidth]{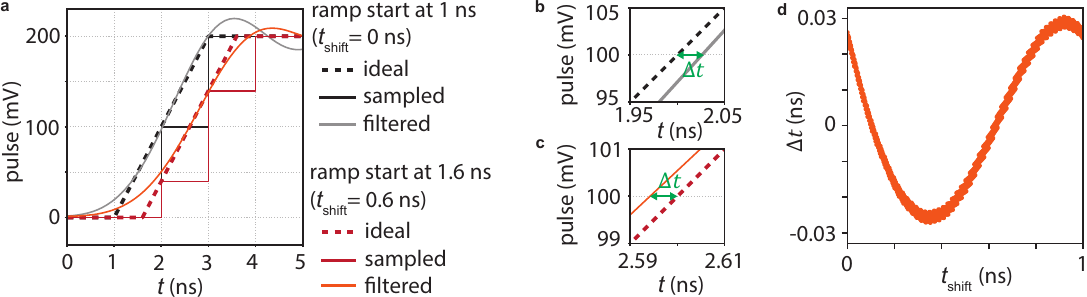}
	\caption{\textbf{Pulse timing.} 
		\textbf{a,} Digital inputs and analogue outputs of the AWG for two pulses with time shifts $t_{\rm shift}=$ 0 and 0.6~ns. The dotted lines are the ideal linear ramps with 0.6~ns time shift that we are aiming for. The solid lines are the digital inputs, represented by the discrete sampling with 1~ns resolution. To produce a time shift of 0.6~ns, a 60~mV decrement of the digital input is made on the rising ramp. The low-pass filtering of the AWG results in the smoothened output voltages represented by the solid curves, as well as the oscillations (ringing) after the ramp. 
  \textbf{b, c,} Zoom-in of the pulses around the middle of the ideal ramps. The deviation in time between the ideal ramps and the analogue outputs at half of the voltage amplitude is denoted as $\Delta t$. 
		\textbf{d,} The deviation $\Delta t$ as a function of time shift $t_{\rm shift}$. The data set is generated with equally distributed time shifts from 0 to 0.999~ns. The mean of the distribution corresponds to $\Delta t=0$. The analogue outputs in all the plots are calculated using the measured step response of the AWG. }
	\label{fig:pulse_timing}
\end{figure}

High fidelity hopping-based gates require a precise timing of shuttling pulses. A qubit fidelity above 99.99\% can be achieved when the rotation has an incoherent error of less than 1.3~degrees. In a simplified example where two quantum dots having quantization axes which are perpendicular, the timing error of ramps for an $\rm X_{\pi/2}$ shuttling gate on a qubit with a Larmor frequency of 40~MHz should be less than 90~ps. 
This timing precision is far below the sample rate of 1~GSa/s of the used AWG. Ramps can be timed with precision higher than the sample rate, because the voltage resolution of the AWG can be used to shift the ramp in time as shown in Fig.~\ref{fig:pulse_timing}a. The time resolution $\Delta t_{\rm res}$ of a ramp with a duration long enough to be not affected by the transients at the start and end of the ramp can be approximated by
$\Delta t_{\rm res} = t_{\rm ramp}  \Delta V / A$, 
where $t_{\rm ramp}$ the duration of the ramp, $\Delta V$ the voltage resolution of the AWG and $A$ the amplitude of the ramp. 
This approximation assumes that the low-pass filter has a cut-off frequency just below the Nyquist frequency. 
Surprisingly, the sample rate has no direct effect on the time resolution of the ramp. A higher sample rate combined with a higher cut-off frequency allows the generation of shorter ramps and shorter ramps have a higher time resolution. The voltage resolution and thus the time resolution effectively decrease when oversampling is used, i.e. when the cut-off frequency is significantly lower than the Nyquist frequency.

We have used AWGs with a voltage resolution of 0.37~mV and pulses with an amplitude on the order of 200 mV at the AWG outputs (this translates to 25.2~mV on the device due to the attenuation on the line) and a ramp time of 2~ns.  This setting gives a time resolution of 3.7~ps, which meets the requirement for high-fidelity gates. However, the ramps for the shuttling pulses are short with respect to the transient response of the low-pass filter. The filter of the AWG adds small wiggles to the short ramps making the timing less precise. This effect is shown in Fig.~\ref{fig:pulse_timing}d, where the time deviation for the ramps with different time shift $t_{\rm shift}$ are plotted. From these calculated ramps we have derived a maximum deviation of 30.4~ps from the average and a standard deviation of 19.4~ps, satisfying the basic requirements for 99.99\% fidelity. We modeled our gate implementation in ~\ref{sec:simulations_shuttling_gates} and estimate the incoherent error due to such timing deviation, as summarized in Table~\ref{tab:model_Xgate_error}.

\newpage

\subsection{Fitting of quantum dot pair parameters \txtblue{for} shuttling \txtblue{gates} }
\label{sec:dots_parameters}

\begin{figure*}[htp!]
	\centering
	\includegraphics[width=0.99 \textwidth]{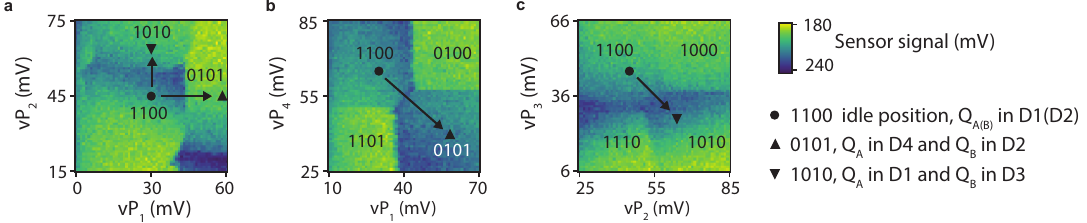}
	\caption{\textbf{Charge stability diagrams of the shuttling gates.} 
	The diagrams in \textbf{a-c} show the voltages of the working points projected on the corresponding virtual plunger gates $\rm vP_i$. The hole occupancies are labelled as $\rm n_1 n_2 n_3 n_4$. The idle position is marked as the circle, in which both qubits $\rm Q_A$ and $\rm Q_B$ are in quantum dots D1 and D2. The $\rm X_{\pi/2}$ on $\rm Q_{A(B)}$ is realized by shuttling between the idle position and the working point marked as the triangle in 0101 (1010) charge occupation. }
	\label{fig:csd_shuttle_gates}
\end{figure*}

\begin{figure*}[htp!]
	\centering
	\includegraphics[width=0.99 \textwidth]{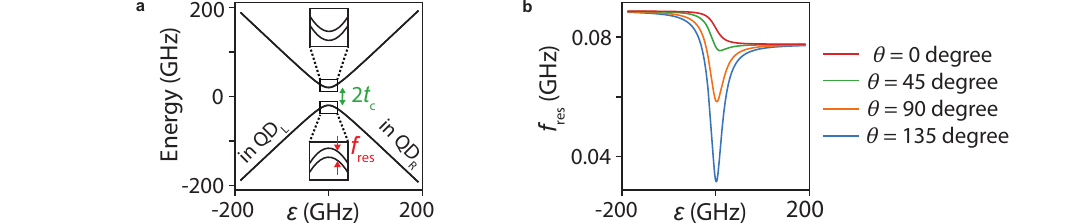}
	\caption{\textbf{Energy levels of a single spin.} 
	\textbf{a,} An example of energy levels of the single spin in double quantum dot given by Eq.~\eqref{Hamiltonian_DQD4x4}. \textbf{b,} The transition frequency between the lowest two states, $f_{\rm res}$, as function of detuning energy $\epsilon$ for the quantization axes angle $\theta$. \txtblue{Parameters used here: $\mu_{\rm B} B g_{\rm L(R)}=0.078(0.089)$~GHz, $t_{\rm c}=20$~GHz.} }
	\label{fig:DQD_H4x4_example}
\end{figure*}

\begin{figure*}[htp!]
	\centering
	\includegraphics[width=0.99 \textwidth]{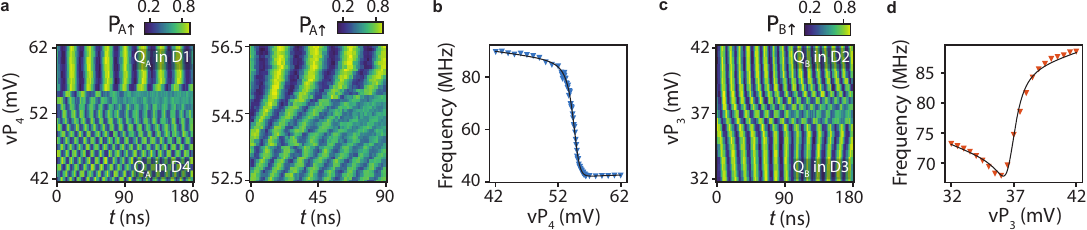}
	\caption{\textbf{Fitting of tunnel coupling at the shuttling gate settings via spin free precession measurement at the magnetic field of 25~mT.} 
	\textbf{a,} The free evolution of $\rm Q_A$ at different detuning across D1-D4 anti-crossing. The panel on the right is the fine scan around the charge anti-crossing where the frequencies changes rapidly. The oscillation frequencies are extracted and plotted in \textbf{b}. \textbf{c,} The free evolution of $\rm Q_B$ at different detuning across D2-D3 charge anti-crossing. The oscillation frequencies are extracted and plotted in \textbf{d}. The black lines are the fits using Eq.~\eqref{freq_anticrossing}, with the lever arm $\frac{\Delta \epsilon_{14(23)}}{\Delta {\rm vP_{4(3)}}}=0.086(0.082)\SI{}{eV/V}$. }
	\label{fig:frequency_fit_tc}
\end{figure*}

Using the Ramsey sequence, we measure the free precession frequency as a function of detuning in the double quantum dot system D1-D4 as well as D2-D3, in order to characterize the tunnel couplings, the position of the anti-crossings, and the relative angle of the quantization axes under the voltage settings used for implementing the hopping-based quantum gates. The corresponding charge stability diagrams are shown in Figs.~\ref{fig:csd_shuttle_gates}b, c. Following the modelling approach in the work~\cite{Vanriggelendoelman2023}, the system is described in the basis \{$\ket{\rm L,\uparrow_{\rm L}}$,$\ket{\rm L,\downarrow_{\rm L}}$,$\ket{\rm R,\uparrow_{\rm L}}$,$\ket{\rm R,\downarrow_{\rm L}}$\}, where `L' or `R' indicates the position of the hole in quantum dot QD$_{\rm L}$ or QD$_{\rm R}$ and $\uparrow_{\rm L}$ or $\downarrow_{\rm L}$ specifies its spin states in the frame of quantum dot L. Its Hamiltonian is written as:

\begin{equation}
H_{\rm 4\times4} = 
  \left( {\begin{array}{cccc}
     \epsilon & 0 & t_{\rm c} & 0 \\
   0 & \epsilon & 0 & t_{\rm c}\\
   t_{\rm c} &  0 & -\epsilon & 0 \\
   0 &  t_{\rm c} & 0 & -\epsilon \\
  \end{array} } \right) +  \frac{1}{2}\mu_{\rm B} B\left( {\begin{array}{cccc}
    g_{\rm L}(\epsilon) & 0 & 0 & 0\\
   0 & -g_{\rm L}(\epsilon) & 0 & 0\\
   0 & 0 & g_{\rm R} (\epsilon) \cos(\theta) & g_{\rm R} (\epsilon)\sin(\theta_{\rm}){\rm e}^{-{\rm i}\phi}\\   
  0 & 0 & g_{\rm R}(\epsilon)\sin(\theta){\rm e}^{{\rm i}\phi} & -g_{\rm R}(\epsilon)\cos(\theta)\\
  \end{array} } \right),
\label{Hamiltonian_DQD4x4}
\end{equation}
where $\epsilon$ is the detuning energy of the double quantum dot system (taken as zero at the charge transition), $\mu_{\rm B}$ is the Bohr magneton and the $g_{i}$ are the $g$-factors of the quantum dot $i$, $\theta$ ($\phi$) is the polar (azimuthal) angle between the two quantization axes. An example of the energy levels is shown in Fig.~\ref{fig:DQD_H4x4_example}. We note that this model is similar to that of a flopping-mode qubit~\cite{Benito2019}. Diagonalizing the Hamiltonian, we obtain the qubit resonance frequency $f_{\rm res}$ (at the limit of small Zeeman energy $\mu_{\rm B} B \ll t_{\rm c}$):

\begin{equation}
f_{\rm res} =   \frac{\mu_{\rm B}B}{h}\frac{\sqrt{(2\epsilon^2+t_{\rm c}^2)(g_{\rm L}(\epsilon)^2+g_{\rm R}(\epsilon)^2)+2\epsilon \sqrt{\epsilon^2+t_{\rm c}^2}(g_{\rm R}(\epsilon)^2-g_{\rm L}(\epsilon)^2)+2g_{\rm L}(\epsilon)g_{\rm R}(\epsilon)t_{\rm c}^2\cos(\theta)}}{2\sqrt{\epsilon^2+t_{\rm c}^2}},
\label{freq_anticrossing}
\end{equation}

Assuming a linear dependence of $g$-factors $g_{\rm L(R)}(\epsilon)$ on the detuning $\epsilon$, we fit the above formula to the data and extract the tunnel coupling $t_{\rm c,14} = 27 \pm 1$~GHz and the angle between quantization axes $\theta_{14}=65\pm 2 ^\circ$ for the quantum dot pair D1-D4. In the quantum dot pair D2-D3 we extract the tunnel coupling $t_{\rm c,23} = 20 \pm 1$~GHz and the angle $\theta_{23}=51\pm 2 ^\circ$. The results are shown in Fig.~\ref{fig:frequency_fit_tc}. We notice that the extracted quantization axis angles are higher than the values extracted from the fitting in Fig.1C of the main text and Fig.~\ref{fig:fit2d_q2}c, where $\theta_{14}=41.5 ^\circ$ and $\theta_{23}=44.7 ^\circ$(see ~\ref{sec:simulations_shuttling_gates}). This discrepancy might be attributed to the adiabaticity of the shuttling process, and the non-linear $g$-factor variation as a function of voltages around the charge anti-crossing.

\subsection{Simulations of the hopping-based single-qubit gates}
\label{sec:simulations_shuttling_gates}
In the lab frame, we have three different models to describe the spin dynamics with decreasing complexity. The first model considers the full $4\times4$ Hamiltonian $H_{\rm 4\times4}$ as shown in Eq.~\eqref{Hamiltonian_DQD4x4}. The second model is a $2\times2$ Hamiltonian $H_{\rm 2\times2}$ where the effective magnetic field experienced by the spin depends on the orbital wave function hybridization in the double quantum dot. It can be obtained by projecting the first model onto the orbital ground state, and can be written as
\begin{equation}
H_{\rm 2\times2} = 
\frac{h f_{\rm L}}{4}  (1-\frac{\epsilon}{ \sqrt{t_{\rm c}^2 + \epsilon^2}}) \left( {\begin{array}{cc}
    1 & 0 \\
   0 & -1 \\ 
  \end{array} } \right)
   +
\frac{h f_{\rm R}}{4}  (1+\frac{\epsilon}{ \sqrt{t_{\rm c}^2 + \epsilon^2}}) \left( {\begin{array}{cc}
    \cos\theta & \sin\theta {\rm e}^{{-\rm i}\phi} \\
   \sin\theta {\rm e}^{{\rm i}\phi} & -\cos\theta \\   
  \end{array} } \right).
\label{Hamiltonian_DQD2x2}
\end{equation}
The third model is derived by the second model, Eq.~\eqref{Hamiltonian_DQD2x2}, by taking the limit $t_{\rm c} \rightarrow 0$

\begin{equation}
H_{\rm dis} = 
\frac{h f_{\rm L}}{2}  \left( {\begin{array}{cc}
    1 & 0 \\
   0 & -1 \\ 
  \end{array} } \right) \Theta(-\epsilon) + 
\frac{h f_{\rm R}}{2}  \left( {\begin{array}{cc}
    \cos\theta & \sin\theta {\rm e}^{-{\rm i}\phi} \\
   \sin\theta {\rm e}^{{\rm i}\phi} & -\cos\theta \\   
  \end{array} } \right) \Theta(\epsilon)
\label{Hamiltonian_DQD_discrete},
\end{equation}
where we have replace the smooth step $\frac{1}{2}  (1\mp\frac{\epsilon}{ \sqrt{t_{\rm c}^2 + \epsilon^2}})$ by the Heaviside step function $\Theta(\mp \epsilon)$. 
Essentially, we discretize the dynamics and consider that the spin precession frequency as well as quantization axis angle only takes two discrete values, $\epsilon<0$ and $\epsilon>0$, instead of a continuous transition through the anti-crossing.

\begin{figure*}[htp!]
	\centering
	\includegraphics[width=0.99 \textwidth]{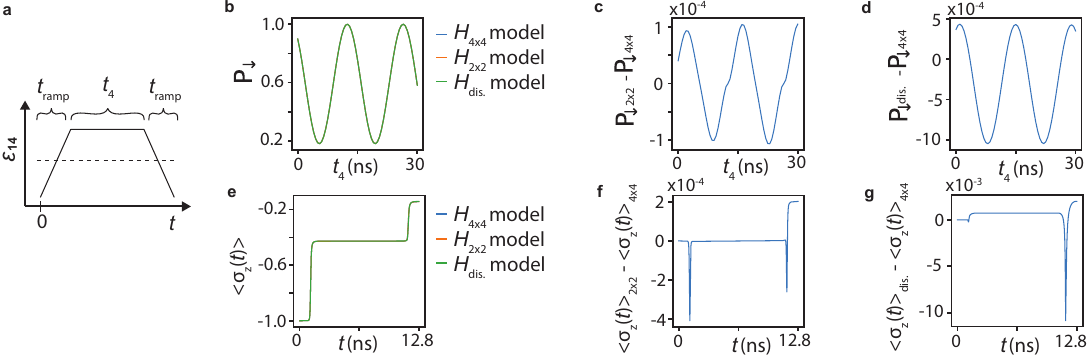}
	\caption{\textbf{Simulation and comparison of different spin shuttling models.} 
	\textbf{a,} The pulse of detuning $\epsilon_{\rm 14}(t)$. \textbf{b,} The spin down probability of the final state $P_{\downarrow}$ as a function of the idle time $t_{\rm 4}$ in quantum dot D4, with the initial state of spin down. The simulation results given by three different models are plotted in blue, orange and green curves. These curves are overlapping because of the small difference between the curves. \textbf{c,} Comparing the $2\times2$ model and the $4\times4$ model by plotting the difference of final state spin down probability. \textbf{d,} Comparing the discrete model and the $4\times4$ model. \textbf{e,} The simulated state evolution under a detuning pulse, given by three different models. The results are presented in the form of spin polarization $\braket{\rm \sigma_{Z}(t)}$ in the instantaneous eigen-basis at time $t$. \textbf{f,} Comparing the $2\times2$ model and the $4\times4$ model by plotting the difference of instantaneous spin polarization.  \textbf{g,} Comparing the discrete model and the $4\times4$ model.  In (e)-(g) the detuning pulse with $t_{\rm 4}=8.8$~ns is used in the simulation to show the maximal discrepancy between the models, based on the observation in (d) that the maximal deviation occurs around $t_{\rm 4}=8.8$~ns.     }
	\label{fig:shuttle_models}
\end{figure*}

\paragraph{Model comparison}
We use QuTiP to compute the final state and the time evolution under the time dependent detuning $\epsilon_{\rm 14}(t) $ as depicted in Fig.~\ref{fig:shuttle_models}a. The detuning is varied linearly from -337~GHz to 226~GHz, corresponding to the virtual plunger gate voltages shown in Fig.~\ref{fig:csd_shuttle_gates}b, within the ramp time $t_{\rm ramp}$ = 2~ns. Other parameters used in the simulations are: the tunnel coupling $t_{\rm c,14} = 27$~GHz, the angle between quantization axes $\theta_{14}=65^\circ$, frequency $f_{\rm L} = 33.8$~MHz, frequency $f_{\rm R} = 71.5$~MHz.

In Fig.~\ref{fig:shuttle_models}c,f the small difference between the 2 by 2 model and the full (4 by 4) model shows that the tunnel coupling is large enough such that the charge degree of freedom is adiabatic. This agrees with the estimation of the vanishing Landau Zener probability of the excited orbital state induced by the detuning ramp, $P_{\rm LZ} = \exp(- 2\pi^2 t_{\rm c,14}^2 / (h \frac{d\epsilon_{14}}{dt}) )=9.9 \times 10^{-23}$. 
In Fig.~\ref{fig:shuttle_models}d the difference between the discrete model and the full model is less than 0.11\%. This good agreement is attributed to the short ramp time, low Larmor frequencies, and the large ratio of the detuning difference over tunnel coupling $\Delta \epsilon_{\rm 14}/t_{\rm c,14}$. These conditions make the description of abrupt change of the spin Hamiltonian a good approximation. We use the discrete model in the manuscript and the rest of the supplementary material to describe the spin dynamics that involves multiple shuttling steps.

\paragraph{Impact of the azimuthal shuttling angle $\phi$}
A two-shuttle process, shuttling to quantum dot 2 and back is described in the following by the time evolution
\begin{align}
        U(\theta,\phi,t) &= \exp{\left[-i\pi f_{\rm B} t \left( {\begin{array}{cc}
    \cos\theta & \sin\theta {\rm e}^{-{\rm i}\phi} \\
   \sin\theta {\rm e}^{{\rm i}\phi} & -\cos\theta \\   
  \end{array} } \right)\right]}.
\end{align}
While the polar shuttling angle $\theta$ is essential for the gate implementation, the azimuthal angle only adds a spin-dependent phase to the double-quantum dot system. This can be easily verified by the transformation
\begin{align}
  U(\theta,\phi,t) &\rightarrow e^{i \beta \sigma_z/2}U(\theta,\phi,t)e^{-i \beta \sigma_z/2}\\
    &= U(\theta,\phi+\beta,t).
\end{align}
Since all remaining gates, the single-qubit z-gate implemented via idling and also the two-qubit CZ gate, commute with the phase gate, we can choose $\beta=-\phi$ allowing us to drop the azimuthal angle.

\begin{figure*}[htp!]
	\centering
	\includegraphics[width=0.99 \textwidth]{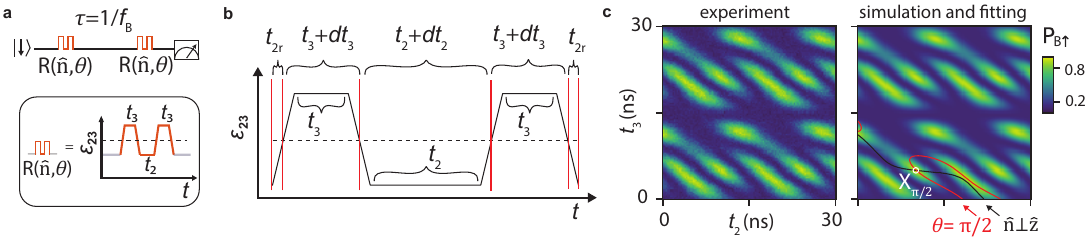}
	\caption{\textbf{Initial tune-up of the single qubit gate on qubit $\mathbf{\rm Q_B}$. } 
	\textbf{a,} The pulse sequence for the experiment. \textbf{b,} The detuning pulse considered in the simulation. The horizontal dashed line marks the D2-D3 charge anti-crossing. The vertical red lines mark the start and end of individual evolution time \{$t_{\rm 2r}$, $t_{3}+dt_{3}$, $t_{2}+dt_{2}$, $t_{3}+dt_{3}$, $t_{\rm 2r}$\}. \textbf{c,} The measured spin-up probability $P_{\rm A \uparrow}(t_2, t_3)$ at magnetic field 20~mT is shown on the left. The simulation result is on the right. The black contour line indicates the wait times ($t_2, t_3$) in which the rotation axis $\hat{\rm n}$ is on the equator of the Bloch sphere, while on the red lines indicate the rotation angle $\theta={\rm \pi}/2$. The intersection, marked in white, is the conditions for $\rm X_{\pi/2}$ gates and corresponds to maximal spin-up probability. The black lines and red lines are periodic in $t_{\rm 2}$ and $t_{\rm 3}$, while for clear illustration we only plot a few of them. }
	\label{fig:fit2d_q2}
\end{figure*}

\paragraph{Gate simulations}
We use the discrete model in the lab frame to simulate an eight-shuttle process as a function of wait time in two double quantum dots, as shown in Fig.~\ref{fig:fit2d_q2}. The process consists of two identical four-shuttle pulses and a wait time in between, $\tau=1/f_{\rm B}$, which is assumed to be an identity operation.  The time evolution of a four-shuttle pulse is a series of free precession for various duration \{$t_{\rm 2r}$, $t_{3}+dt_{3}$, $t_{2}+dt_{2}$, $t_{3}+dt_{3}$, $t_{\rm 2r}$\} around the corresponding quantization axes in \{D2, D3, D2, D3, D2\} with two distinct frequencies \{$f_{\rm D2}$, $f_{\rm D3}$, $f_{\rm D2}$, $f_{\rm D3}$, $f_{\rm D2}$\} as depicted in Fig.~\ref{fig:fit2d_q2}b. For simplicity we assume the Larmor frequencies of the dots to not change with detuning. Fitting to the experimental data gives $t_{\rm 2r}=1.16$~ns, $dt_{3}=1.54$~ns, $dt_{2}=2.29$~ns, $f_{\rm D2}=70.9$~MHz, $f_{\rm D3}=62.0$~MHz, $\theta_{23}=44.7^\circ$ (different than $\theta_{23}=51^\circ$ obtained in ~\ref{sec:dots_parameters} ). Applying the same fitting procedure for the quantum dot pair D1-D4 as shown in the main script Fig.1C, we obtain $t_{\rm 1r}=0.98$~ns, $dt_{4}=1.94$~ns, $dt_{1}=1.94$~ns, $f_{\rm D1}=33.8$~MHz, $f_{\rm D4}=71.5$~MHz, $\theta_{14}=41.5^\circ$ (different than $\theta_{14}=65^\circ$ obtained in ~\ref{sec:dots_parameters} ). For both double quantum dot pairs this effective model fits well to the experimental data. Based on the fitted parameters, we can find the wait times $t_{2}$, $t_{3}$ ($t_{1}$, $t_{4}$) in the individual quantum dot to construct a desired spin state rotation ${\rm R}(\hat{\rm n}, \theta)$ on qubit $\rm Q_A$ ($\rm Q_B$), as shown in the contour lines in Fig.~\ref{fig:fit2d_q2}c. Specifically, the $\rm X_{\pi/2}$ gate is the rotation that satisfies the rotation angle $\theta=90^\circ$ as well as the rotation axis lying on the Bloch sphere equator, $\hat{\rm n} \perp \hat{\rm z}$. The rotation axis can be chosen to point along x-axis by redefining the azimuthal angle of the Bloch sphere, as shown in the previous paragraph. 

\begin{figure*}[htp!]
	\centering
	\includegraphics[width=0.99 \textwidth]{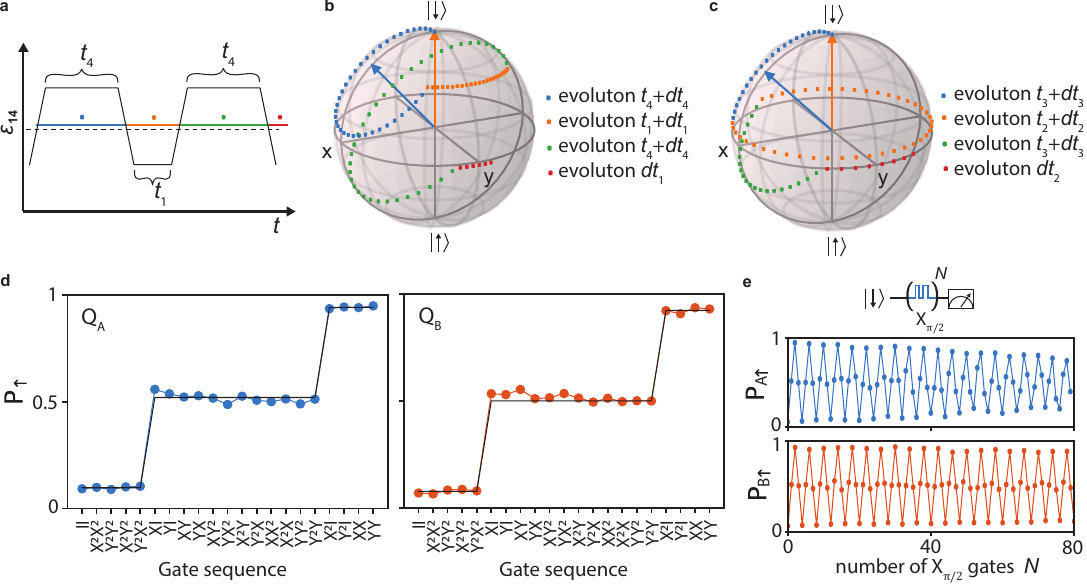}
	\caption{\textbf{Simulation and calibration of $\rm X_{\pi/2}$ gates with equal wait times at 20 mT. } 
	\textbf{a,} The detuning pulse $\epsilon_{14}(t)$ of the gate $\rm X_{\pi/2,A}$ with equal wait times $t_4$ in D4. The horizontal dashed line marks the D1-D4 charge anti-crossing. In different parts of the pulse, marked by \{blue,orange,green,red\} lines, the spin rotates around quantization axes of \{D4,D1,D4,D1\} with frequencies \{$f_{\rm D4},f_{\rm D1},f_{\rm D4},f_{\rm D1}$\}. The simulated state evolution, with the initial state $\ket{\downarrow}$, is plotted in \textbf{b}. The states at time $t$ are plotted as points on the Bloch sphere with time step of 0.3 ns. The quantization axis of D1(D4) is represented by the orange (blue) arrow. \textbf{c,} The simulated state evolution of the gate $\rm X_{\pi/2,B}$ with the initial state $\ket{\downarrow}$.  \textbf{d,} The measured spin-up probability of $\rm Q_A$ and $\rm Q_B$ in AllXY experiments after the gates are calibrated. In each graph, three sections of the black horizontal lines mark the expected outcome of the ideal gates and take into account the state preparation and measurement (SPAM) error. If there is no SPAM error the black horizontal lines are at values of 0, 0.5, and 1. \textbf{e,} The spin-up probability after applying repeated $\rm X_{\pi/2}$ gates on each qubit.   }
	\label{fig:bloch_x90_equal}
\end{figure*}

When the waiting times lead to an exact $\rm X_{\pi/2}$ gate, the spin-up probability shows a local maximum. This property is used for the initial tune-up in the experiment. The subsequent fine-tuning consists of calibrating the rotation axis direction via AllXY sequence~\cite{reed2013entanglement}, as shown in Fig.~\ref{fig:bloch_x90_equal}d. The calibration of the rotation angle is done by applying numbers of $\rm X_{\pi/2}$ gate to amplify over-rotation error. The tuned-up $\rm X_{\pi/2}$ gates are shown in Fig.~\ref{fig:bloch_x90_equal}d,e. The simulation of the state evolution is plotted in Fig.~\ref{fig:bloch_x90_equal}b,c.

\begin{figure*}[htp!]
	\centering
	\includegraphics[width=0.99 \textwidth]{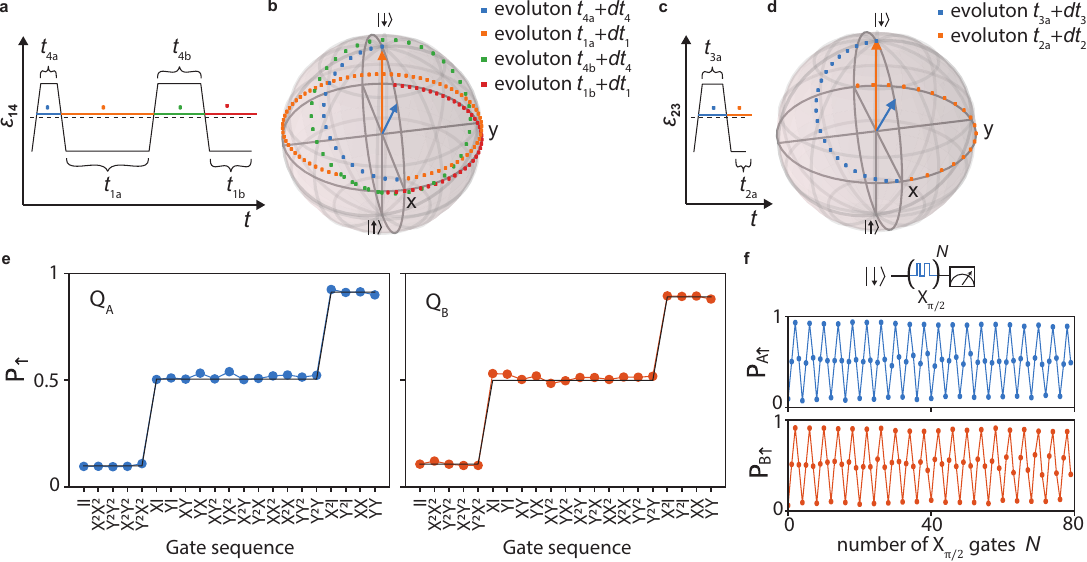}
	\caption{\textbf{Simulation and calibration of $\rm X_{\pi/2}$ gates with unequal wait times at 25 mT.  } 	\textbf{a,} The detuning pulse $\epsilon_{14}(t)$ of the gate $\rm X_{\pi/2,A}$ with unequal wait times ($t_{\rm 1a}$, $t_{\rm 1b}$) in D1 and ($t_{\rm 4a}$, $t_{\rm 4b}$) in D4. The horizontal dashed line marks the D1-D4 charge anti-crossing. In different parts of the pulse, marked by \{blue,orange,green,red\} lines, the spin rotates around quantization axes of \{D4,D1,D4,D1\} with frequencies \{$f_{\rm D4},f_{\rm D1},f_{\rm D4},f_{\rm D1}$\}. The simulated state evolution, with the initial state $\ket{\downarrow}$, is plotted in \textbf{b}. The states at time $t$ are plotted as points on the Bloch sphere with time step of 0.3 ns. The quantization axis of D1(D4) is represented by the orange (blue) arrow. \textbf{c,} The detuning pulse $\epsilon_{23}(t)$ of the gate $\rm X_{\pi/2,B}$ with two shuttles. \textbf{d,} The simulated state evolution of the gate $\rm X_{\pi/2,B}$ with the initial state $\ket{\downarrow}$.     \textbf{e,} The measured spin-up probability of $\rm Q_A$ and $\rm Q_B$ in AllXY experiments after the gates are calibrated. In each graph, three sections of the black horizontal lines mark the expected outcome of the ideal gates and take into account the state preparation and measurement (SPAM) error. If there is no SPAM error the black horizontal lines are at values of 0, 0.5, and 1.  \textbf{f,} The spin-up probability after applying repeated $\rm X_{\pi/2}$ gates on each qubit. Here we replot the data in Fig.1H of the main text for easier comparison with Fig.~\ref{fig:bloch_x90_equal}e. 
	}
	\label{fig:bloch_x90_unequal}
\end{figure*}

\paragraph{Alternative pulse scheme }
We further consider the pulse with unequal wait times for the gate on qubit $Q_{\rm A}$. We design the pulse such that the first rotation in D4 is $\pi$ and the subsequent rotations in D1 and D4 are either close to $\pi$ or $2 \pi$. The intuition is that, in this scheme the polarization $\braket{\rm \sigma_Z}$ of the final state evolved from the initial spin-down state might have weaker dependence on the frequency fluctuations in D1 and D4, which \txtblue{may result in a} gate rotation angle $\theta$ \txtblue{that is} more robust against noise. We use the discrete model and the fitted parameters obtained above to compute the required waveform of the detuning pulse, as shown in  Fig.~\ref{fig:bloch_x90_unequal}a. It gives $t_{\rm 4a}=3.65$~ns, $t_{\rm 1a}=19.75$~ns, $t_{\rm 4b}=10.51$~ns, $t_{\rm 1a}=8.17$~ns. In the experiment we start with this set of wait times and further fine-tune the wait times using AllXY sequence and the repetition sequence $X_{\rm \pi/2}^N$. The parameters after calibration experiments are $t_{\rm 4a}=3.747$~ns, $t_{\rm 1a}=19.33$~ns, $t_{\rm 4b}=10.17$~ns, $t_{\rm 1a}=9.4$~ns, which are close to the initial values predicted by the model. The AllXY and repetition sequences of a calibrated $\rm X_{\pi/2,A}$ gate are shown in Fig.~\ref{fig:bloch_x90_unequal}e,f. 
When comparing to Fig.~\ref{fig:bloch_x90_equal}e, the extended decay time in Fig.~\ref{fig:bloch_x90_unequal}f might be explained by the pulse designed to be more robust in rotation angle.
Further discussion and estimation are in Table~\ref{tab:gate_sensitivity_q1} and the corresponding paragraph.

For the  $\rm X_{\rm \pi/2}$ on qubit $\rm Q_{B}$, we design a two-shuttle pulse because the quantization axis angle $\theta_{23}=44.7^\circ$ is very close to $45^\circ$.
In theory, the angle $45^\circ$ can realize a gate with only two shuttles and at the same time have rotation angle insensitive to frequency fluctuations in both quantum dots.
\txtblue{We therefore} implement the two-shuttle gate in our experiment, even though in theory it will not make an exact $\rm X_{\pi/2}$.
Following similar procedure as described above, we start from the predicted values $t_{3}=4.91$ ns and $t_{2}=3.35$ ns (assume $\theta_{23}=45^\circ$), perform calibration experiments and determine $t_{3}=4.86$~ns and $t_{2}=3.42$~ns, which only differ slightly from the initial predictions. 
The AllXY and repetition sequences of a calibrated $\rm X_{\pi/2,B}$ gates are shown in Fig.~\ref{fig:bloch_x90_unequal}e,f. 
These results show that the gate we created is very close to $\rm X_{\pi/2}$.
In particular from the repetition sequence in Fig.~\ref{fig:bloch_x90_unequal}f we estimate a small rotation angle error ${\rm \sigma}_{\theta,{\rm rep}}=0.2^\circ$.
An alternative estimation using gate set tomography (GST) (Table~\ref{tab:1qGST_raw_estimate}) shows a small rotation angle error ${\rm \sigma}_{\theta,{\rm GST}}=0.29^\circ$. 
Combining the rotation angle error ${\rm \sigma}_{\theta,{\rm GST (rep)}}$ and the values $\Delta\theta_{\rm rot},\Delta\theta_{23}$ in Table~\ref{tab:gate_sensitivity_q2}, we can estimate the lower bound $\theta_{23,{\rm GST (rep)}} \geq 44.86(44.9)^\circ$. 

The drift or fluctuation in the Larmor frequency, in the quantization axis angle, and in the timing of individual shuttling event can contribute to the gate rotation error. Using the model and parameters described above, we can estimate the corresponding variations of gate rotation angle $\theta_{\rm rot}$ and the polar angle of the rotation direction $\theta_{\hat{\rm n}}=\arccos(\hat{\rm n}\cdot\hat{\rm z})$. 
We denote such variations as $\Delta\theta_{\rm rot}$ and $\Delta\theta_{\rm \hat{n}}$.
We consider the timing error of the shuttling events caused by the fluctuations in the position of the charge anti-crossing $\Delta \epsilon_{\rm ij,AC}$. 
It is estimated to be $t_{\rm ramp} \Delta \epsilon_{\rm ij,AC}/\Delta \epsilon_{ij} $. 
The estimation is summarized in Table~\ref{tab:gate_sensitivity_q1} and Table~\ref{tab:gate_sensitivity_q2}. From the estimation we observe the rotation angle $\theta_{\rm rot}$ of the modified gates is more robust against fluctuations on most of the parameters. On the other hand, the rotation axis direction becomes more sensitive to certain parameters. 

\begin{table}
    \centering
    \begin{tabular}{|c|c|c|c|c|}
    \hline
         & \multicolumn{2}{|c|}{  \begin{tabular}[x]{@{}c@{}} four-shuttle $\rm X_{A}$ \\ with unequal wait times \end{tabular}  } &  \multicolumn{2}{|c|}{  \begin{tabular}[x]{@{}c@{}} four-shuttle $\rm X_{A}$ \\ with equal wait times \end{tabular}   }  \\\hline
         &  $\Delta\theta_{\rm rot} (^\circ)$   & $\Delta\theta_{\rm \hat{n}} (^\circ)$  &  $\Delta\theta_{\rm rot} (^\circ)$  & $\Delta\theta_{\rm \hat{n}} (^\circ)$  \\\hline
        $\Delta f_{\rm D1}=100$ kHz & -0.22 & 1.44 & -0.11  & 1.22 \\\hline
        $\Delta f_{\rm D4}=100$ kHz &  0.019 & 0.44 & -0.53 & 0.34 \\\hline
        $\Delta \theta_{\rm 14}=0.1^\circ$ & 0.21  & \txtblue{0.034} & \txtblue{0.17}  & \txtblue{0.12} \\\hline
        $\Delta \epsilon_{\rm 14,AC} =\SI{10}{\micro eV}$ & -0.037  & -0.49  & 0.67  & -0.16   \\\hline
    \end{tabular}
    \caption{\textbf{Estimation of}     $\boldmath{\rm X_{A}}$    \textbf{gate sensitivity to the fluctuations of the system parameters.}  
    Here we assume both gates are operated at 25 mT. The values of uncertainty are not the measured values. They are chosen to make the calculation easier.   
    }
    \label{tab:gate_sensitivity_q1}
\end{table}

\begin{table}
    \centering
    \begin{tabular}{|c|c|c|c|c|}
    \hline
         & \multicolumn{2}{|c|}{  \begin{tabular}[x]{@{}c@{}} two-shuttle $\rm X_{B}$  \end{tabular}  } &  \multicolumn{2}{|c|}{  \begin{tabular}[x]{@{}c@{}} four-shuttle $\rm X_{B}$ \\ with equal wait times \end{tabular}   }  \\\hline
         &  $\Delta\theta_{\rm rot} (^\circ)$   & $\Delta\theta_{\rm \hat{n}} (^\circ)$  &  $\Delta\theta_{\rm rot} (^\circ)$  & $\Delta\theta_{\rm \hat{n}} (^\circ)$  \\\hline
        $\Delta f_{\rm D2}=100$ kHz &  0.0036   &  0.51   &   -0.46  &  0.58   \\\hline
        $\Delta f_{\rm D3}=100$ kHz &  0.0037   &  0.16   &   -0.29  &  0.24   \\\hline
        $\Delta \theta_{\rm 23}=0.1^\circ$ &  0.2   &  -0.0023    &  \txtblue{0.11}    &  \txtblue{0.14}  \\\hline
        $\Delta \epsilon_{\rm 23,AC} =\SI{10}{ \micro eV}$  &  $-8.5\times10^{-3}$   &   -0.095   &   0.070    &  -0.27     \\\hline

    \end{tabular}
    \caption{\textbf{Estimation of $\rm X_{B}$ gate sensitivity to the fluctuations of the system parameters.} 
    Here we assume both gates are operated at 25 mT. The values of uncertainty are not the measured values. They are chosen to make the calculation easier.  }
    \label{tab:gate_sensitivity_q2}
\end{table}

\newpage
\subsection{Power dissipation and scaling advantages of shuttling-based control} 
\label{sec:power_dissipation}
To execute the shuttling operations, trapezoidal voltage pulses are applied on the gates. To achieve high-fidelity single qubit control a handful of such shuttling pulses are required, each with ramp times of a few nanoseconds between two discrete voltage levels. This stands in stark contrast with state-of-the-art electron dipole spin resonance (EDSR) control where typically high frequency, sinusoidal pulses are applied, and many oscillations of the driving signal are needed to achieve the desired gate fidelity~\cite{Xue2022,Noiri2022}. This gives an advantage to a shuttling\txtblue{-}based architecture \txtblue{considering} energy dissipation, crosstalk and \txtblue{} complexity of \txtblue{the required} control electronics.\txtblue{}

\begin{figure}[htp!]
	\centering
	\includegraphics[width=0.8\textwidth]{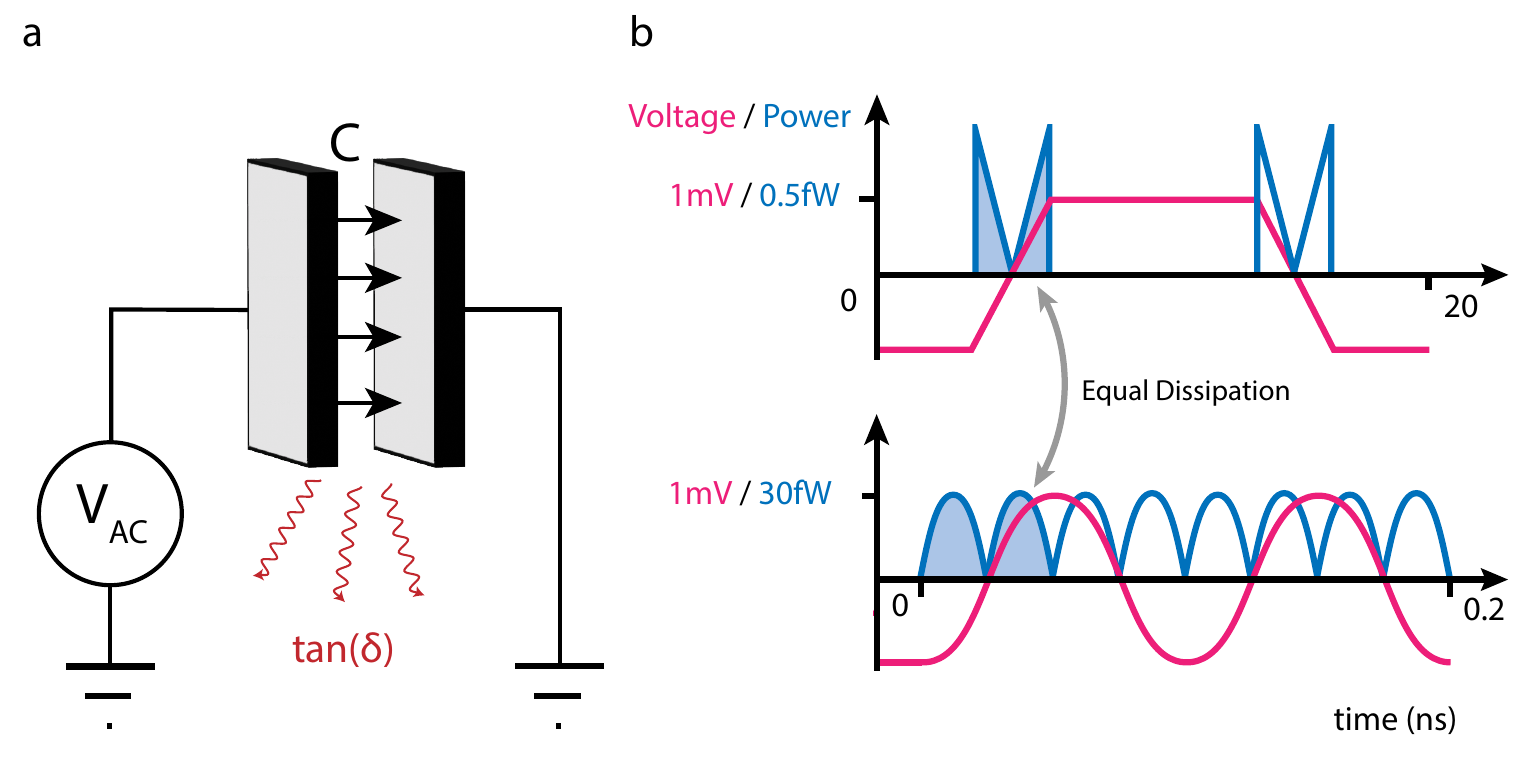}
	\caption{\textbf{Heat dissipation.} \textbf{a,} Schematic of the model of the heat dissipation, due to some capacitor $C$ with loss tangent $\tan(\delta)$. \textbf{b,} For equal pulse amplitude and DC-offset, the heat dissipated per cycle is the same independent of the pulse shape. $C\cdot\tan(\delta)=10^{-18}\SI{}{F}$ was assumed in this example.}
	\label{fig:heat_dissipation}
\end{figure}

Already at the current system sizes, EDSR-based devices experience a drift in qubit frequency that is linked to heat dissipation of the signal~\cite{Undseth2024}. When resistive losses are limited, this heat-dissipation is believed to result from a dielectric loss of energy is stored in the electric field around the signal line. Effectively the system is a capacitor with some loss tangent $\tan(\delta)$ \txtblue{, defined as $\tan(\delta)=\epsilon''/\epsilon'$ in a non-conductive system, with $\epsilon''$ and $\epsilon'$ the imaginary and real part of the electric permittivity~\cite{Pozar_1998}}. During each charging cycle, \txtblue{a fraction proportional to} $\tan(\delta)$ of the stored \txtblue{electric} energy is lost \txtblue{as depicted in Fig.~\ref{fig:heat_dissipation}}. With a DC bias around zero the total capacitive energy stored and discharged by the signal line is proportional to $C V_{\rm AC}^2$, where $C$ is the capacitance and $V_{\rm AC}$ the EDSR amplitude, with which the capacitor is charged. The total energy lost is proportional to $E_{\rm Loss} = N_{\rm cycles}\tan(\delta) C V_{\rm AC}^2$, where $N_{\rm cycles}$ gives the \txtblue{number} of oscillations required to perform a qubit operation. \txtblue{In a simplified model, we can take the electric permittivity and with it the loss tangent of silicon and germanium to be largely independent of frequency in the relevant frequency range~\cite{A_C_Baynham_1960,Krupka_2015}. In this model, for} an identical geometry and driving amplitude, the energy dissipation \txtblue{is assumed to} solely depends on the \txtblue{number} of cycles \txtblue{of} the operation and not on the pulse-shape, as \txtblue{indicated} in Fig.~\ref{fig:heat_dissipation}. Hence $1/N_{\rm cycles}$ is a measure of the efficiency of the operation. 

For an EDSR-based $\rm X_\pi$-gate the \txtblue{number} of cycles is given by $N_{\rm cycles,EDSR}=\frac{f_{\rm Larmor}}{2f_{\rm Rabi}}$, which is exactly the inverse of the efficiency $\eta$ as defined in the main text. The driving efficiency is inherently limited by the relatively small Rabi frequency $f_{\rm Rabi}\ll f_{\rm Larmor}$ \txtblue{when operating in} the weak-driving (adiabatic) regime, in which the rotating-wave approximation holds. We note that while faster driving is possible, it requires complex amplitude and phase modulation for high-fidelity implementations~\cite{theisCounteractingSystemsDiabaticities2018,rimbach-russSimpleFrameworkSystematic2023} which also dissipate additional heat. An experimental demonstration \txtblue{of high-fidelity qubit logic is given} by Xue et al. \txtblue{} operated with Rabi (Larmor) frequencies of $f_{\rm Rabi}=\SI{2}{MHz}$ ($f_{\rm Larmor}=\SI{12}{GHz}$)~ \cite{Xue2022,Undseth2023}. This corresponds to an efficiency of $\eta=2f_{\rm Rabi}/f_{\rm Larmor} \approx 1/3000$. Similarly Noiri et al. demonstrated $\eta\approx 1/1500$~\cite{Noiri2022}. For the prior device, an EDSR amplitude of $V_{\rm AC}\approx \SI{5}{mV}$ is reported at the bond pads of the chip~\cite{Undseth2023}. This corresponds to an energy dissipation of $E_{\rm Loss}\approx 0.075 \tan(\delta)C~\SI{}{V^2}$ per $\rm X_\pi$-gate for high-fidelity EDSR control.

Shuttling based gates do not face a similar inherit efficiency limitation, instead being limited by the relative tilt in quantization axis. In the main part of the paper we demonstrate that to perform an $\rm X_\pi$-gate using shuttling, the hole is shuttled two to four times back and forth depending on the angle between the quantization axes of the quantum dot pairs. With periodic pulse timings and negligible ramp times \txtblue{this corresponds to} $N_{\rm cycles}=1/\eta$. This is done with a typical amplitude $V_{\rm AC}=\SI{20}{mV}$. Using the $N_{\rm cycles}=4$ figure, this corresponds to a heat-dissipation corresponding to $E_{\rm Loss}= 2\cdot0.0016 \tan(\delta)C~\SI{}{V^2}=0.0032\tan(\delta)C~\SI{}{V^2}$, where the additional factor of two accounts for the two plunger gates on which the voltage is applied.

Crosstalk, like heat dissipation, is a problem observed in the current spin-qubit devices and is expected to become more significant as the number and density of qubits increase~\cite{Philips2022}. This crosstalk can originate from close spacing of signal lines, both on and off the qubit chip. \txtblue{As} the density of the quantum dots increases, the capacitance between the gates is expected to grow accordingly, increasing the crosstalk further.  
Since the admittance between \txtblue{signal lines} is directly proportional to the signal frequency, the capacitive crosstalk will be less for low-frequency\txtblue{} shuttling\txtblue{-}based pulses, compared to high-frequency EDSR experiments which face challenges similar to conventional high-frequency integrated circuits~\cite{Mbairi2008}. In integrated circuits design, a rule of thumb is to keep the distance between traces to three times the trace width ~\cite{Mbairi2008}. This might pose a significant limitation for qubit routing, \txtblue{especially} for larger 2D arrays. An architecture based on the demonstrated high-fidelity shuttling gates is thus expected to be less sensitive to crosstalk, which will be advantageous in scaling to large qubit counts.

In large spin systems consisting of many hundreds or thousands of qubits\txtblue{,} the scalability of control electronics is a major consideration. The electronic hardware required to generate the \txtblue{IQ modulated} sinusoidal EDSR pulses need high analog voltage resolution, which is significantly more involved than the shuttling pulses consisting of two voltage levels. The lower required voltage accuracy and precision of the shuttling based control allows scaling the qubit count while limiting the electronic overhead, cost and energy consumption. The required timing resolution of shuttling based control should be below \SI{90}{ps} for a \SI{40}{MHz} Larmor frequency (\labelcref{sec:subnanosecond_timing}), higher than the sampling resolution of the IQ modulated EDSR driving. However, EDSR signals need to control the qubit phase with a similar precision as the shuttling pulses, thus requiring a similarly high resolution. 

\newpage
\subsection{Coherence \txtblue{times} of the \txtblue{individual} qubits }
\label{sec:coherence_2qubits}
\begin{figure*}[htp!]
	\centering
	\includegraphics[width=0.99 \textwidth]{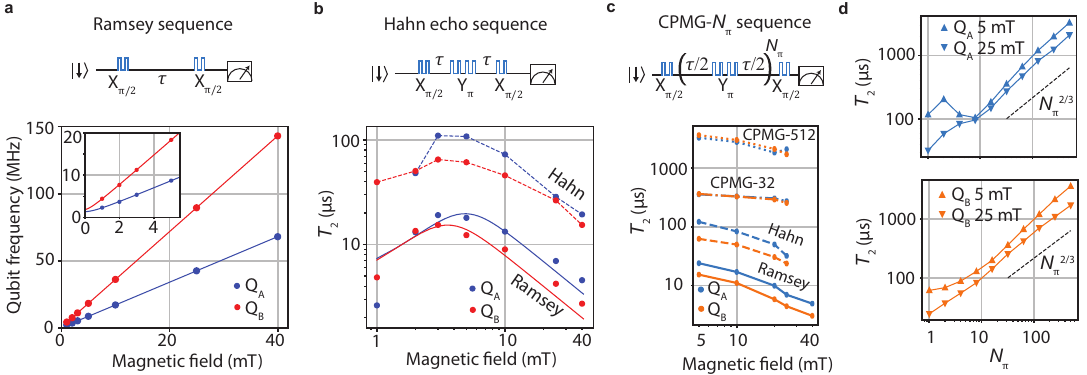}
	\caption{\textbf{Coherence time and dependence on magnetic field strength at the idle position of the qubits.}	\textbf{a,} The frequencies of qubits $\rm Q_A$ and $\rm Q_B$ as a function of external magnetic field, measured by the Ramsey sequence consisting of shuttling gates as shown on the top. The inset is the zoom-in at low field regime, where a non-linear behavior is observed. The fitting method is described in the text. Here the superconducting magnet is in the driven mode. In this mode, the power supply is galvanically connected to its power supply. It introduces extra noise in the system compared to the normal operation mode. Note that field below 5 mT can only be reached with the magnet in driven mode. \textbf{b,} The $T_2^\star$ and $T_2^{\rm H}$ as a function of external magnetic field when the magnet is at the driven mode. The Hahn echo sequence consists of shuttling gates is plotted at the top. Here a $\rm Y_{\pi}$ is realized by two $\rm Y_{\pi/2}$ shuttling gates. The $T_2^\star$ is extracted from the Ramsey measurement with an average of 10 traces and the experimental time 12-19 minutes. Here we replot the data in Fig.1G of the main text for easier comparison. \textbf{c,} The coherence time as a function of magnetic field above 5 mT when magnet is in the normal operation mode. The longest coherence time is obtained at 5 mT, with $T_2^\star= \SI{24.1}{\micro s}$, $T_2^{\rm H}= \SI{122}{\micro s}$ and $T_2^{\rm CPMG-512}>\SI{3}{}$~ms. The $T_2^\star$ is extracted from the Ramsey measurement with an average of 10 traces and the experimental time 12-19 minutes. \txtblue{When fitting $T_2^{\rm CPMG-512}$ of $\rm Q_A$, we disregard data points corresponding to total evolution time $\tau N_{\pi} > 4$~ms that are influenced by the reservoir-induced decay. Exemplary CPMG datasets are shown in  Fig.\ref{fig:2x2_CPMG_5mT}.} \textbf{d,} The $T_2^{\rm CPMG}$ as function of number of $\rm \pi$-pulses for both qubits at two different magnetic fields. 	}
	\label{fig:2x2_T2_idle}
\end{figure*}

Because the $g$-tensor and hyperfine interaction for heavy hole qubits are expected to be highly anisotropic, a small magnetic field offset pointing towards an out-of-plane direction can change the dephasing time significantly.
For our device, we find that the measured qubit frequencies are not completely linear in magnetic field for field strength of 1 mT. Therefore, we can estimate the magnetic field offset for in-plane and out-of-plane direction by fitting the measured qubit frequency to $hf(B_{\rm ext}) = \sqrt{(g_\parallel\mu_B (B_{\rm ext}+B_0^{\parallel}))^2+ (g_{\perp} \mu_B B_0^{\perp})^2}$ (Fig.~\ref{fig:2x2_T2_idle}a and inset).
Our best fits show perpendicular magnetic field offsets $g_{\perp} \mu_B B_0^{\perp}=1.4(1)$~MHz for $\rm Q_A$, 1.8(2)~MHz for $\rm Q_B$ and parallel offsets $B_0^{\parallel}=0.08(3)$~mT for $\rm Q_A$, 0.13(2)~mT for $\rm Q_B$.
The perpendicular offsets are $\SI{10}{}$ and $\SI{13}{\micro T}$ assuming an out-of-plane $g$-factor $g_{\perp}=10$.
The offsets might originate from magnetic materials on the sample board, trapped flux in superconducting magnet, polarized nuclear spins, or the Earth magnetic field.

To estimate the magnetic field dependence of the dephasing time, we consider a simplified model assuming Gaussian quasi-static fluctuations of the qubit frequency originating from nuclear spin noise and quasi-static fluctuations of the $g$-factor caused by charge noise. The qubit frequency for an external applied magnetic field $B_{\rm ext}$ is given by 
\begin{equation}
f(B_{\rm ext}, \delta g, \delta f_{\rm n}) = \frac{1}{h}\sqrt{((g_\parallel+\delta g)\mu_B (B_{\rm ext}+B_0^{\parallel}))^2+ (\delta f_{\rm n}+g_{\perp} \mu_B B_0^{\perp})^2}.
\label{freq_noise_1q}
\end{equation}
In linear order, the in-plane $g$-factor fluctuation $\delta g$ gives rise to qubit frequency fluctuation $\delta f_{\delta g}=f(B_{\rm ext}, \delta g,0 )-f(B_{\rm ext}, 0,0)$ with standard deviation $\sigma_{f,\delta g}$ and the out-of-plane hyperfine field fluctuations $\delta f_{\rm n}$ give rise to qubit frequency fluctuation $\delta f_{\rm n}=f(B_{\rm ext}, 0, \delta f_{\rm n})-f(B_{\rm ext},0, 0)$ with standard deviation $\sigma_{f,\delta f_{\rm n}}$. 
Assuming both noise sources to be independent and uncorrelated, the standard deviation of the total qubit frequency fluctuation at $B_\text{ext}$ is $\sigma_{f}=\sqrt{\sigma_{f, \delta g}^2 + \sigma_{f, \delta f_{\rm n}}^2}$ giving rise to a coherence time $T_2^\star=\frac{1}{\sqrt{2}\pi \sigma_f}$.
From our fit in Fig.1G (replotted in Fig.~\ref{fig:2x2_T2_idle}b), we extract an effective hyperfine noise $\delta f_{\rm n}=52(7)$~kHz for $\rm Q_A$ and 78(8)~kHz for $\rm Q_B$, corresponding to the coherent time $T_2^\star=4.3(6)$ and $\SI{2.9(3)}{\micro s}$.
This result is larger than $\delta f_{\rm n}=34.4$~kHz reported in Ref.~\cite{LawrieThesis2022} in D3 of the same device and significantly smaller than $\delta f_{\rm n}=250$~kHz reported in Ref.~\cite{Hendrickx2024}.
The difference could arise from microscopic details in the device, the simplicity of the model, as well as the complexity of the nuclear spin noise at low magnetic fields, where the $\rm ^{73}Ge$ nuclear spins have a quadrupolar splitting caused by strain which has a similar magnitude as the precession frequency.

\begin{figure*}[htp!]
	\centering
	\includegraphics[width=0.99 \textwidth]{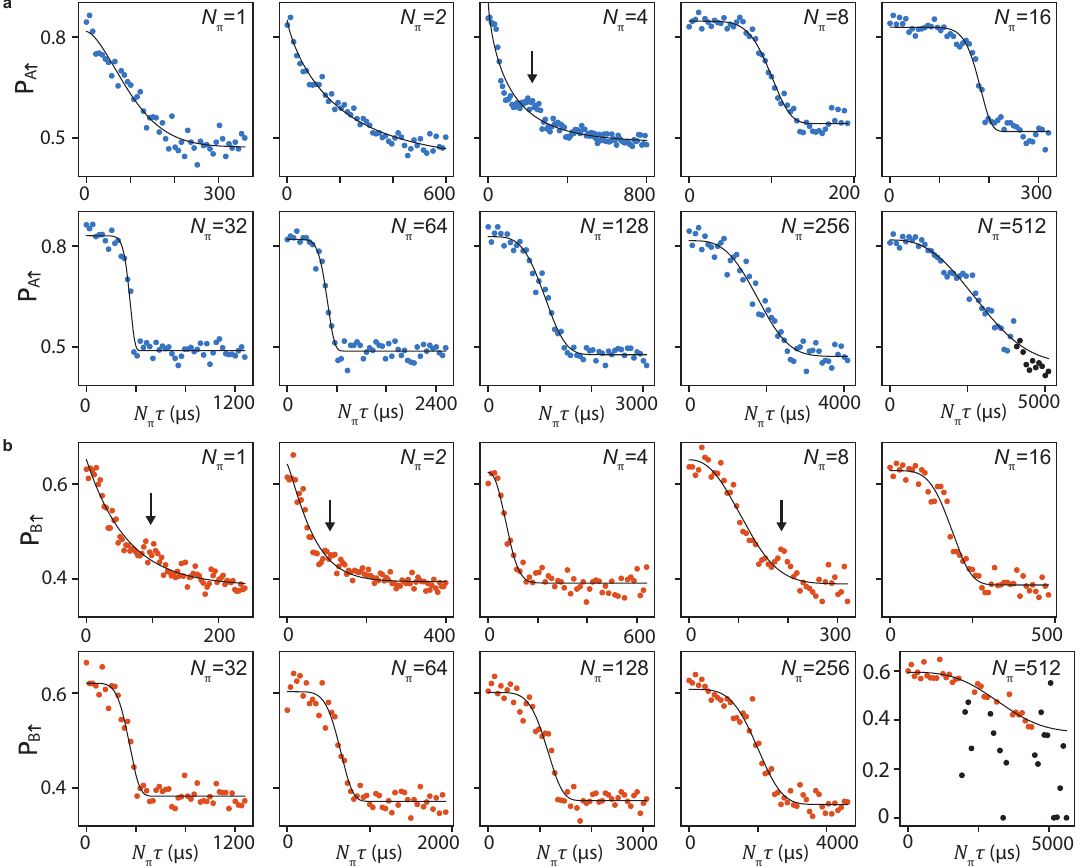}
	\caption{\textbf{Coherence time with dynamical decoupling pulses at magnetic field of 5~mT.} \textbf{a,} The coherence of qubit A  and \textbf{b,} the coherence of qubit B  as a function of total evolution time $N_{\rm \pi} \tau$ with CPMG dynamical decoupling sequence (schematics depict in top panel of Fig.~\ref{fig:2x2_T2_idle}c). The collapse and revival of coherence (peaks marked by black arrows in the plots $N_{\pi}\leq$8) should be attributed to hyperfine noise of $\rm ^{73}$Ge nuclear spin. We notice that at such low magnetic field the expected linewidth of hyperfine noise becomes comparable to the nuclear spin precession frequency, which might explain the observed smoother features compared to the work \cite{Hendrickx2024}. Despite the collapse-and-revival features, we still fit the data to the formula $P(t)=A \exp{(-(t/T_2)^\alpha)}+B$ to extract coherence time $T_2$. We also notice the coherence at $N_{\rm \pi} \tau=0$ almost stay the same for the plots from $N_{\rm \pi}=1$ to $N_{\rm \pi}=512$, which implies the spin states do not have noticeable decay with numbers of shuttles up to 4096(2048) times for qubit $\rm Q_{A(B)}$. We remark that the black data points in the plots $N_{\rm \pi}=512$ are removed from the coherence time fitting, due to the decay induced by tunnel coupling to the reservoir in (a), and due to the charge jumps of the sensor in (b). In both cases, the fitted $T_2$ should still be valid because the fitting curves agree with the data, and the fitted $T_2$ fall on the trend of $T_2$-$N_{\rm \pi}$ data in Fig.~\ref{fig:2x2_T2_idle}d.  }
	\label{fig:2x2_CPMG_5mT}
\end{figure*}

\newpage
\subsection{Randomized benchmarking}
\label{sec:RB_method}
\paragraph{Experiment implementation} 
In the single-qubit randomized benchmarking (RB), the sequence lengths are varied from \{ 1, 3, 10, 30, ..., 6000 \}, in total 25 different lengths. 
We execute sequences of different lengths once in a random order.
After going over all the 25 sequences, we repeat a random execution again with different random order. 
In total we repeat this execution 32 times. 
For every sequence we perform 400 single-shot readout.
The final spin-down probability $P_{\rm A(B), \downarrow}$ of the RB sequences on qubit A(B) with the idled qubit B(A) is obtained by averaging over 400 single-shot readout and tracing out the qubit B(A) from the two-qubit state probability $\rm P_{\sigma \sigma^\prime}$. 
An experiment takes 4.5 hours to complete, with no re-calibration within the individual experiment.

In the two-qubit interleaved randomized benchmarking (IRB), the sequence lengths are \{ 1, 2, 4, 8, ..., 200 \}, in total 20 different lengths.
The order of sequence execution is similar to the single-qubit RB.
We execute a reference sequence and right afterward an interleaved sequence with the same length, and then continue on the sequences with different lengths in a random order. 
After going over all the 20 sequences, we repeat a random execution again with different random order.
In total we repeat 128 times.
For every sequence we perform 200 single-shot readout.
An experiment takes 7.5 hours to complete, with no re-calibration during the individual experiment. 

In both single- and two-qubit RB, we observe the 2D histograms of the charge sensor signal are well-separate even at the maximal sequence lengths, while they have an overall shift which gradually increases for the longer sequence. 
We speculate that the intensive pulsing locally heats up the two-level fluctuators and the high-kinetic inductors, shifting chemical potential of the single-hole charge sensor and the impedance of the LC circuits, respectively~\cite{Undseth2024}. 
For the single-qubit RB and the first two-qubit RB ($\rm IRB_{1}$), we apply adaptive thresholding on the histograms to obtain the two-qubit state probability $\rm P_{\sigma \sigma^\prime}$. 
For the other two-qubit RB experiments ($\rm IRB_{2,3}$), we add an extra wait time of $\SI{300}{\micro s}$ before reloading the ancilla qubits for readout. 
This amount of wait time is sufficient to reduce the sensor signal shift and we are able to use pre-defined constant thresholds to obtain the two-qubit state probability $\rm P_{\sigma \sigma^\prime}$. 

\paragraph{Fidelity extraction} 
In single-qubit RB, the single-qubit Cliffords consist of the gates $\rm X_{\pi/2}$, $\rm Z_{\pi/2}$, and the idle gate $\rm I$. We measure the  final state probability of the sequences containing $m$ Clifford gates and a recovery Clifford gate which is the inverse of the corresponding $m$-Clifford sequence. The spin-down probability averaged over 32 random sequences is fitted to  $P_{\downarrow}(m)=A p^m+B$, where $p$ is the decay rate of the sequence, $m$ is the number of Cliffords, $A$ and $B$ are the parameters absorbing SPAM errors. The average Clifford fidelity is related to the decay rate by $F_{\rm Clifford1}=1-\frac{1}{2}(1-p)$. The measurements in Fig.1I of the main text shows the average Clifford fidelity $F_{\rm Clifford1,A}=$ 99.967(4)\%  and $F_{\rm Clifford1,B}=$ 99.960(6)\%. The uncertainties are obtained from bootstrapping re-sampling with 95\% confidence intervals.  The average number of gates for single-qubit Clifford is 1.0 $\rm X_{\pi/2}$, 2.42 $\rm Z_{\pi/2}$ and 0.04 $\rm I$. Defining the infidelity of gate $i$ as $r_{i}=1-F_{i}$ and assuming the Clifford gate infidelity equals to the sum of the primitive gate infidelity weighted by the average composition, $r_{\rm Clifford}= r_{\rm X_{\pi/2}} + 2.42 r_{\rm Z_{\pi/2}} + 0.04 r_{\rm I}$,  the average Clifford fidelity sets the lower bounds of the $\rm X_{\pi/2}$ average gate fidelity $F_{\rm X_{\pi/2},A} \geq F_{\rm Clifford1,A}$ and $F_{\rm X_{\pi/2},B} \geq F_{\rm Clifford1,B}$.

In two-qubit RB, the two-qubit Cliffords consist of the gates  $\rm CZ$, $\rm X_{\pi/2}^{A(B)}$, $\rm Z_{\pi/2}^{A(B)}$, and $\rm I$.
Similar to the single-qubit RB, we measure the final state probability of the sequences containing $m$ Clifford gates and a recovery Clifford gate.
The return probability of the reference sequence (interleaved sequence) is fitted to  $P_{\rm \downarrow \downarrow, ref(int)}(m)=A p_{\rm ref(int)}^m+B$, where $p_{\rm ref(int)}$ is the decay rate of the sequence, $m$ is the number of Cliffords, while $A$ and $B$ are the parameters absorbing the SPAM errors. From the reference sequence decay curve in main text Fig.2E, we determine the average Clifford gate fidelity $F_{\rm Clifford2} \equiv F_{\rm ref}=1-\frac{3}{4}(1-p_{\rm ref})=$ 98.60(6)\%. The uncertainties are obtained from bootstrapping re-sampling with 95\% confidence intervals. The average number of gates for two-qubit Clifford is 1.63 $\rm CZ$, 1.60 $\rm X_{\pi/2}^{A(B)}$, 2.68 $\rm Z_{\pi/2}^{A(B)}$, and 0.00009 $\rm I$. This implies the relation between gate errors, $r_{\rm Clifford2} \equiv r_{\rm ref}=1.63 r_{\rm CZ} + {\rm \Sigma}_{i={\rm A,B}} 1.60 r_{\rm X_{\pi/2},i} + 2.68 r_{\rm Z_{\pi/2},i}$. From this relation we find the average Clifford gate fidelity sets the lower bound of CZ gate fidelity $F_{\rm CZ} = 1-r_{\rm CZ} \geq 1-\frac{r_{\rm ref}}{1.63}=$ 99.14(4)\%, which is consistent with the IRB result $F_{\rm CZ}=1-\frac{3}{4}(1-p_{\rm int}/p_{\rm ref})=$ 99.33(10)\%. We estimate the lower bound of single qubit gate fidelity in the two-qubit subspace, average between both qubits, as $\frac{1}{2}(F_{\rm X_{\pi/2},A}+F_{\rm X_{\pi/2},B}) = 1-\frac{1}{2}(r_{\rm X_{\pi/2},A}+r_{\rm X_{\pi/2},B}) \geq 1-\frac{1}{2}\frac{r_{\rm ref}-1.63 r_{\rm CZ}}{1.60}=$ 99.90(5)\%. 

We perform additional check for the potential echoing effect in two-qubit RB/IRB experiments, by fitting the data with super-exponential formula. 
As shown in Table~\ref{tab:2qRB_result}, the exponents are in the range of  0.86 - 1.05, \txtblue{showing small deviations from a pure exponential decay}.

\begin{table}[]
\begin{tabular}{|c|c|c|c|c|}
\hline
                Fitting formula  & Results &  $\rm IRB_1$  &  $\rm IRB_2$   &  $\rm IRB_3$   \\ \hline
\multirow{4}{*}{\begin{tabular}[x]{@{}c@{}} Fit with super-exponent model \\ $P(m)=A p^{(m^\alpha)} + B$ \end{tabular}} & $\alpha_{\rm ref}$ &  0.862 $\pm$ 0.046 & \txtblue{1.050 $\pm$ 0.058} & \txtblue{0.988 $\pm$ 0.057}  \\ \cline{2-5} 
                  & $\alpha_{\rm int}$  & 0.867 $\pm$ 0.041 & \txtblue{0.946 $\pm$ 0.047} & \txtblue{0.954 $\pm$ 0.047} \\ \cline{2-5} 
                  & $r_{\rm ref}$ (\%) & 2.55 $\pm$ 0.40 &  \txtblue{1.17 $\pm$ 0.25} & \txtblue{1.55 $\pm$ 0.31}  \\ \cline{2-5} 
                  & $r_{\rm CZ}$ (\%) & \txtblue{1.20} $\pm$ 0.68 &  \txtblue{1.36 $\pm$ 0.49} & \txtblue{1.21 $\pm$ 0.56}  \\ \hline
\multirow{2}{*}{\begin{tabular}[x]{@{}c@{}} Fit with single-exponent model \\ $P(m)=A p^{m} + B$ \end{tabular}} &  $r_{\rm ref}$ (\%) & 1.56 $\pm$ 0.07 & 1.40 $\pm$ 0.06 & 1.48 $\pm$ 0.06 \\ \cline{2-5} 
                  &  $r_{\rm CZ}$ (\%) & 0.79 $\pm$ 0.11 & 0.67 $\pm$ \txtblue{0.10} & 0.86 $\pm$ 0.11  \\ \hline
\end{tabular}
    \caption{\textbf{Two-qubit interleaved randomized benchmarking results for three individual runs.} The parameter settings are identical to two-qubit GST experiments, where magnetic field $B=25$~mT and the CZ gate has maximum exchange coupling $J \approx 21$~MHz. The infidelity of the two-qubit Clifford $r_{\rm ref}$ is related to the decay rate of the reference RB sequence by $r_{\rm ref}=1-F_{\rm ref}=\frac{3}{4}(1-p_{\rm ref})$. The infidelity of the interleaved CZ gate $r_{\rm CZ}$ is related to the decay rates of the reference sequence and interleaved sequence by $r_{\rm CZ}=1-F_{\rm CZ}=\frac{3}{4}(1-p_{\rm int}/p_{\rm ref})$.  The uncertainty represents the 95\% confidence interval.      }
    \label{tab:2qRB_result}
\end{table}

\subsection{Gate set tomography and \txtblue{comparison} with two-qubit randomized benchmarking}
\label{sec:GST_method}
\paragraph{Gate set tomography implementation}
We carried out gate set tomography (GST) experiments using the python package pyGSTi~\cite{nielsenProbingQuantumProcessor2020}. 
For single-qubit GST, we use the default gateset \{$\rm I$, $\rm X$, $\rm Y$\}, where $\rm I$ is the idle gate of $\frac{5}{f_{\rm A}}\approx$ 118 ns ($\frac{9}{f_{\rm B}}\approx$ 102 ns), $\rm X(Y)$ stands for of $\rm X(Y)_{\pi/2}$. The six fiducials for state preparation and measurements are \{ $\rm null$, $\rm X$, $\rm Y$, $\rm XX$, $\rm XXX$, $\rm YYY$ \}, where $\rm null$ is the gate with zero idle time. The five germs are \{$\rm I$, $\rm X$, $\rm Y$, $\rm XY$, $\rm XXY$ \}. The circuit length are power of two from 1 up to 128, resulting in total 1120 sequences, which takes 17 minutes to complete in the experiment. In every sequence, the spin-up probability $P_{\rm A(B), \uparrow}$ of qubit A(B) with the idled qubit B(A) are obtained by averaging over 500 single-shot readout and tracing out the qubit B(A) state from the two-qubit state probability $\rm P_{\sigma\sigma^\prime}$.

For two-qubit GST, we use the default gateset \{$\rm I$, \rm $X_A$, $\rm X_B$, $\rm Y_A$, $\rm Y_B$, $\rm CZ$\}. 
Here the idle gate takes 100~ns. 
The 11 measurement fiducials are \{$\rm null$, $\rm X_B$, $\rm Y_B$, $\rm X_B X_B$, $\rm X_A$, $\rm Y_A$, $\rm X_A X_A$, $\rm X_A X_B$, $\rm X_A Y_B$, $\rm Y_A X_B$, $\rm Y_A Y_B$ \}. 
The 16 preparation fiducials are measurement fiducials plus the gates \{$\rm X_A X_B X_B$, $\rm Y_A X_B X_B$, $\rm X_A X_A X_B$, $\rm X_A X_A Y_B$, $\rm X_A X_A X_B X_B$   \}.
The 16 germs are \{ $\rm I$, $\rm X_A$, $\rm Y_A$, $\rm X_B$, $\rm Y_B$, $\rm CZ$, $\rm X_A Y_A$, $\rm X_B Y_B$, $\rm X_A X_A Y_A$, $\rm X_B X_B Y_B$, $\rm X_B Y_B CZ$, $\rm CZ X_A X_A X_A$, $\rm X_A X_B Y_B X_A Y_B Y_A$, $\rm X_A Y_B X_B Y_A X_B X_A$, $\rm CZ X_B Y_A CZ Y_B X_A$, $\rm Y_A X_A Y_B X_A X_B X_A Y_A Y_B$  \}. The circuit length are \{1,2,4,8\}, resulting in total 1702 sequences, which takes 18 minutes to complete in the experiment. In every sequence the two-qubit state probability $\rm P_{\sigma\sigma^\prime}$ is obtained by averaging over 500 single-shot readout. 

The measurement outcome of the gate sequence is analyzed in the python package pyGSTi with CPTP model, which considers the gates, the state preparation and measurement as completely positive trace-preserving processes. 
The corresponding process matrices are estimated and multiple derived quantities can be computed.
In the case of single-qubit GST, the estimated process of the single qubit gates can be projected and decomposed into rotation operators as listed in Table~\ref{tab:1qGST_raw_estimate}. 
For both single-qubit and two-qubit GST, we report gate errors metrics (Table~\ref{tab:1qGST_result},~\ref{tab:2qGST_result}) and SPAM error (Table~\ref{tab:1qGST_SPAM},~\ref{tab:2qGST_SPAM}). 
The tables include the averaged gate infidelity $1-\frac{{\rm tr}({G_{\rm exp}^{-1}G_{\rm ideal}}) + d}{d(d+1)} $, non-unitary averaged gate infidelity $\frac{d-1}{d}(1-\sqrt{u({G_{\rm exp}^{-1} G_{\rm ideal}}) }$, 1/2 trace distance $\frac{1}{2} \lVert J_a(G_{\rm ideal})-J_a(G_{\rm exp}) \rVert$, and 1/2 diamond-distance  $\frac{1}{2}{\rm max}_\rho \lVert (G_{\rm ideal} \otimes I)\rho - (G_{\rm exp} \otimes I)\rho \rVert$.
Here $d=2^{\rm N_{qubits}}$ is the dimension of the Hilbert space, $G_{\rm exp}$ is the process of the gate in the GST experiment in the form of Pauli transfer matrix (PTM), $G_{\rm ideal}$ is the PTM of the ideal gate, $u(M)=\text{tr}(J_a(M)^2)$ is the unitarity of the matrix $M$, $J_a(M)$ is the Jamiolkowski isomorphism map between the matrix $M$ \txtblue{and} the corresponding Choi Matrix, $\lVert . \rVert$ denotes the trace norm, and $\rho$ is a density matrix of dimension  $n^2$~\cite{nielsenProbingQuantumProcessor2020,BlumeKohoutTaxonomy2022}.

\paragraph{Discrepancy between RB and GST in two-qubit gate benchmarking} 
The different benchmarking results obtained by GST and interleaved RB may stem from the presence of low-frequency noise. 
In GST, the CZ gate is repeated to amplify and extract the single-gate dephasing error $r_{\rm s}$. Similar to the Ramsey dephasing, repeating the CZ gate $N$-times results in an error  $r(N)=r_{\rm s}^{(N^\alpha)}$ where $\alpha=1$ if the error is Markovian, or $\alpha\approx2$ if the dephasing error is dominated by the energy level fluctuations with $1/f$ noise spectrum~\cite{Martinis2003,Epstein2014}. 
In the latter case, the errors of the CZ gates in different position within a repeated CZ gate sequence (e.g. the first CZ gate and the second CZ gate) are correlated.
This type of error with temporal correlation is non-Markovian.
Analyzing the decay $r(N)=r_{\rm s}^{(N^2)}$ using a Markovian error model can result in deviations of estimated single-gate errors from the actual error. 
The outcome of our GST experiments always shows model violations, which is in line with this hypothesis. 
\txtblue{On the other hand, in RB the CZ gates are placed between Cliffords that reduce the correlation of the CZ gate errors at different position of a sequence.
According to the numerical study~\cite{Epstein2014}, under the $1/f$ noise the RB provides better than a factor-of-2 estimate of the gate error. 
We believe this worse-case deviation of the error estimate (a factor of 2) is smaller than the one from GST, in view of the $1/f$ noise and gate implementation in our system.   }
Therefore, we consider the results of the interleaved RB to be more representative for the average gate fidelity, while GST is used to access the full tomographic reconstruction of the quantum processes.

\begin{table}
    \centering
    \begin{tabular}{|c|c|c|}
    \hline
        Gate 
        & \begin{tabular}[x]{@{}c@{}} Rotation axis \\ $\hat{n}=(n_x, n_y, n_z)$ \end{tabular} 
        & \begin{tabular}[x]{@{}c@{}} Rotation angle \\ $\theta_{\rm rot} (\pi)$ \end{tabular}  \\ \hline 
        $\rm I_{A}$    & $(0.038, 0.027, 0.999 )$ & 0.0038 \\ \hline
        $\rm X_{\rm A}$    & \txtblue{$(1, 1\times10^{-3}, -1.7\times10^{-6})$} & 0.5018 \\\hline
        $\rm Y_{\rm A}$    & \txtblue{$(1\times10^{-3},1, 2\times10^{-7})$} & 0.5019 \\ \hline
        $\rm I_{\rm B}$    & $(-0.0057, 0.014 ,1 )$ & 0.0051 \\ \hline
        $\rm X_{\rm B}$    & $(1,-1\times10^{-4},-2\times10^{-7})$ & 0.5015 \\ \hline
        $\rm Y_{\rm B}$    & $(-1\times10^{-4},1,-4\times10^{-7})$ & 0.5016 \\\hline         
    \end{tabular}
    \caption{\textbf{Single qubit gate parameters determined from GST.}         }
    \label{tab:1qGST_raw_estimate}
\end{table}

\begin{table}
    \centering
    \begin{tabular}{|c|c|c|c|c|c|}
    \hline
        Gate 
        & \begin{tabular}[x]{@{}c@{}}Avg. gate\\infidelity (\%)\end{tabular}
        & \begin{tabular}[x]{@{}c@{}}Non-unitary\\ avg. gate \\infidelity (\%)\end{tabular}  
        & \begin{tabular}[x]{@{}c@{}} 1/2 trace\\ distance (\%)\end{tabular}
        & \begin{tabular}[x]{@{}c@{}} 1/2 diamond-\\ distance (\%)\end{tabular}
        & \begin{tabular}[x]{@{}c@{}}Eigenvalues \\ 1/2 diamond-\\ distance (\%)\end{tabular}    \\ \hline 
        $\rm I_{\rm A}$  & 0.38 $\pm$ 0.02 & 0.38 $\pm$ 0.02 &  0.82$\pm$0.03  & 0.83$\pm$0.05  & 1.22$\pm$0.05  \\ \hline 
        $\rm X_{\rm A}$  & 0.061 $\pm$ 0.008 &  0.061 $\pm$ 0.008 & 0.33$\pm$0.02 &  0.34$\pm$0.07  & 0.44$\pm$0.03  \\ \hline 
        $\rm Y_{\rm A}$  & 0.058 $\pm$ 0.008 &  0.057 $\pm$ 0.008 & 0.35$\pm$0.02 & 0.35$\pm$0.05  & 0.45$\pm$0.02 \\ \hline 
        $\rm I_{\rm B}$  & 0.71 $\pm$ 0.03 & 0.70 $\pm$ 0.03 & 1.32$\pm$0.06 & 1.33$\pm$0.09  & 1.97$\pm$0.09  \\ \hline 
        $\rm X_{\rm B}$  & 0.019 $\pm$ 0.007 &   0.019 $\pm$ 0.007 & 0.24$\pm$0.02 & 0.25$\pm$0.03  & 0.36$\pm$0.03 \\ \hline 
        $\rm Y_{\rm B}$  & 0.023 $\pm$ 0.007 &  0.022 $\pm$ 0.007 & 0.25$\pm$0.02 & 0.26$\pm$0.04  & 0.37$\pm$0.02 \\\hline 
    \end{tabular}
    \caption{\textbf{Single-qubit GST gate fidelity.} The single-qubit GST is performed under the same setting as single-qubit RB and two-qubit IRB and GST, where residual exchange coupling $J \approx 10-15$~kHz. The uncertainty represents the 95\% confidence interval.       }
    \label{tab:1qGST_result}
\end{table}

\begin{table}
    \centering
    \begin{tabular}{|c|c|c|c|c|c|c|}
    \hline
        Gate 
        & \begin{tabular}[x]{@{}c@{}}Avg. gate\\infidelity (\%)\end{tabular}
        & \begin{tabular}[x]{@{}c@{}}Non-unitary \\avg. gate \\infidelity (\%)\end{tabular}   
        & \begin{tabular}[x]{@{}c@{}} 1/2 trace\\ distance (\%)\end{tabular}
        & \begin{tabular}[x]{@{}c@{}} 1/2 diamond-\\ distance (\%)\end{tabular}
        & \begin{tabular}[x]{@{}c@{}}Eigenvalues \\ 1/2 diamond-\\ distance (\%)\end{tabular}    \\ \hline 
        $\rm I_{\rm A} \otimes I_{\rm B}$ & 0.36 $\pm$ 0.27 & 0.36 $\pm$ 0.26 & 0.9$\pm$1.5 & 1.0$\pm$2.4  & 1.4$\pm$0.6   \\ \hline 
        $\rm X_{\rm A} \otimes I_{\rm B}$ & 0.46 $\pm$ \txtblue{0.28} & 0.43 $\pm$ 0.28 & 2.0$\pm$0.9 & 2.7$\pm$2.4 & 3.6$\pm$1.6   \\ \hline 
        $\rm Y_{\rm A} \otimes I_{\rm B}$ & 0.82 $\pm$ 0.35 & 0.78 $\pm$ 0.35 & 2.7$\pm$1.2 & 3.5$\pm$4.5 & 4.4$\pm$2.4   \\ \hline 
        $\rm I_{\rm A} \otimes X_{\rm B}$ & 0.33 $\pm$ 0.27 & 0.32 $\pm$ 0.27 & 0.8$\pm$0.9 & 1.2$\pm$1.7 & 0.7$\pm$1.2   \\ \hline 
        $\rm I_{\rm A} \otimes Y_{\rm B}$ & 0.51 $\pm$ 0.39 & 0.49 $\pm$ \txtblue{0.38} & 1.7$\pm$0.9 & 2.4$\pm$2.5 & 2.4$\pm$1.6  \\ \hline 
        $\rm CZ$                          & 1.87 $\pm$ 0.52 & 1.78 $\pm$ 0.50 & 4.4$\pm$0.7 & 6.2$\pm$3.8 & 8.1$\pm$0.9  \\\hline        
    \end{tabular}
    \caption{\textbf{Two-qubit GST gate fidelity.} The parameter settings are identical to two-qubit IRB experiments, where magnetic field $B=25$~mT and the CZ gate has maximum exchange coupling $J \approx 21$~MHz. The uncertainty represents the 95\% confidence interval.           }
    \label{tab:2qGST_result}
\end{table}

\newpage
\begin{table}[hbt!]
    \centering
    \begin{tabular}{|c|c|c|c|c|c|}
    \hline
    \multirow{2}{*}{Qubit} & \multirow{2}{*}{Readout probability} & \multicolumn{2}{|c|}{Single-qubit GST experiment} &  \multicolumn{2}{|c|}{Two-qubit GST experiment}  \\
    \cline{3-6}
         & &  Prepare $\ket{\downarrow }$  &  Prepare $\ket{ \uparrow}$   
        &  Prepare $\ket{\downarrow }$  &  Prepare $\ket{ \uparrow}$  \\ \noalign{\hrule height2.0pt}
        \multirow{2}{*}{$\rm Q_A$} &  $\rm P_{\downarrow }$  (\%) & \textbf{96.9}  & 8.6    & \txtblue{\textbf{97.3}}  & 10.0 \\ \cline{2-6}
                                                 &  $\rm P_{ \uparrow}$    (\%) &  3.1 &  \textbf{91.4}    & \txtblue{2.7}  & \textbf{90.0} \\ \cline{1-6}
                      \multirow{2}{*}{$\rm Q_B$} &  $\rm P_{\downarrow }$  (\%) & \textbf{95.0}  & 8.0    & \textbf{95.1}  & 7.2 \\ \cline{2-6}
                                                 &  $\rm P_{ \uparrow}$    (\%) & 5.0  &  \textbf{92.0}    &  4.9 & \textbf{92.8} \\ \hline                    
    \end{tabular}
    \caption{\textbf{Estimation of SPAM fidelity in single-qubit space based on single-qubit GST and two-qubit GST experiments.}         }
    \label{tab:1qGST_SPAM}
\end{table}

\begin{table}[hbt!]
    \centering
    \begin{tabular}{|c|c|c|c|c|c|}
    \hline
        Readout probability
        &Prepare $\ket{\downarrow \downarrow}$ 
        &Prepare $\ket{\downarrow \uparrow}$ 
        &Prepare $\ket{\uparrow \downarrow}$    
        &Prepare $\ket{\uparrow \uparrow}$     \\ \hline 
          $\rm P_{\downarrow \downarrow}$  (\%) &  \textbf{94.0}  &  6.2  &  8.6  &  1.5   \\ \hline
          $\rm P_{\downarrow \uparrow}$    (\%) &  3.7  &  \textbf{90.7}  &  1.3  &  8.5  \\ \hline
          $\rm P_{\uparrow \downarrow}$    (\%) &  2.1  &  0.7  &  \textbf{85.4}  &  6.0 \\ \hline
          $\rm P_{\uparrow \uparrow}$      (\%) &  0.2  &  2.4  &  4.7  &  \textbf{84.0}  \\\hline        
    \end{tabular}
    \caption{\textbf{Estimation of SPAM fidelity based on two-qubit GST results.}  We use the SPAM operations estimated by GST, including the initial state (a density matrix) and the positive operator-valued measure (POVM), to compute the expected readout probability when preparing specific computational states. The computational states are prepared using the imperfect initialization of $\ket{\downarrow \downarrow}$ and the perfect single-qubit gates.        }
    \label{tab:2qGST_SPAM}
\end{table}

\subsection{Error modeling of \txtblue{the} hopping-based single-qubit gate }
\label{sec:model_Xgate_error}
\begin{figure*}[htp!]
	\centering
	\includegraphics[width=0.99 \textwidth]{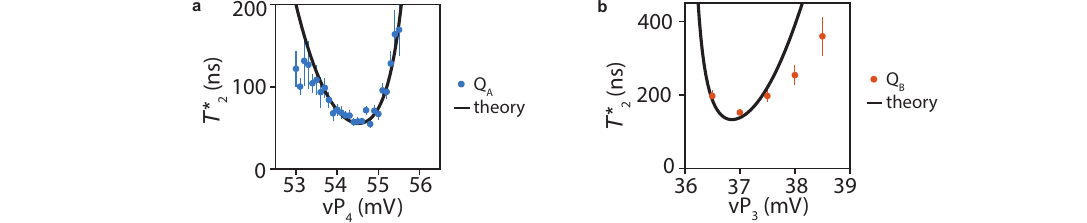}
	\caption{\textbf{Qubit coherence times near the charge anticrossings.} The coherence time for qubit A \textbf{(a)} and qubit B \textbf{(b)} extracted from Fig.~\ref{fig:frequency_fit_tc} by fitting to the formula $P_{\uparrow}(\tau)=A \exp(-(\tau/T_2^\star)^2) + B$. The black curves are the expected coherence time assuming quasi-static electric noise on the gates $\rm vP_4$($\rm vP_3$), $T_2^\star = \frac{1}{\sqrt{2}\pi\sigma_{f}}$ and the voltage-dependent qubit frequency fluctuation is $ \sigma_{f} \approx \frac{\partial f}{\partial \rm{vP_i}} \delta\rm{vP_i} + \frac{1}{2} \frac{\partial^2 {\it f}}{\partial \rm{vP_i}^2} \delta\rm{vP_i}^2$ ~\cite{russAsymmetricResonantExchange2015,Benito2019}. We estimate the effective electric noise $\delta\rm{vP_4}(\delta\rm{vP_3})=0.19(0.14)$~mV, which minimize the square sum of dephasing rate difference $\Delta \frac{1}{T^{\star}}$  between the measured values and the theoretical values.   	}
	\label{fig:2x2_T2_anticrossing}
\end{figure*}

\paragraph{Noise estimation}
We model incoherent error originating from (1) fluctuations in Larmor frequencies of the individual quantum dot, (2) fluctuations in detuning energies, (3) waveform uncertainty, and (4) thermalization processes near the charge anti-crossing.
First we estimate the noise strength of individual error sources.
From the $T_2^\star$ of the static qubits as shown in the main Fig.1F, we estimate Larmor frequency fluctuation $\sigma_{f}=\frac{1}{\sqrt{2}\pi T_2^\star}=32$~kHz for $\rm Q_A$ and $\sigma_{f}=50$~kHz for $\rm Q_B$. 
For Larmor frequency fluctuations in D3 and D4, we assume that they are uncorrelated and have equal magnitude as $\rm Q_B$.
From the fitting of the coherence times in Fig.~\ref{fig:2x2_T2_anticrossing}, we obtain the effective electric noise $\delta\rm{vP_4}(\delta\rm{vP_3})=0.19(0.14)$~mV, which is equivalent to the fluctuations in the position of the charge anti-crossing $\Delta \epsilon_{\rm 14(23),AC}=\SI{17(12)}{\micro eV}$ and creates the timing fluctuation of 14(23)~ps for shuttling operations of $\rm Q_A$ ($\rm Q_B$). 
For the errors from waveform uncertainty (Fig.~\ref{fig:pulse_timing}d), we compute the expected waveforms of the gates $\rm X_{\pi/2,A(B)}$ for the time shifts $t_{\rm shift}$ ranging from 0 to 0.99~ns. 
Each waveform results in slightly different timing of shuttling, and therefore contributes to incoherent error.

\paragraph{Error model}
We use the discrete model Eq.~\eqref{Hamiltonian_DQD_discrete} to compute the unitary matrices of the target gates and noisy gates, and estimate the incoherent error.
The method is described as follow.
The unitary of a gate $U$ is a series of free precession for various duration around the corresponding quantization axes of the quantum dots with different frequencies as depicted in Fig.~\ref{fig:fit2d_q2}b. 
A noise source can either change the duration or change the precession frequencies, resulting in a slightly different gate unitary.
Averaging over the distribution of the noise parameter gives average gate infidelity, similar to the method used in~\ref{sec:model_CZ_error}. 
For the calculation of errors caused by waveform uncertainty, instead of using a single target unitary, we use a set of target unitaries generated by the waveforms with uniformly distributed time shift $t_{\rm shift}$. 
This treatment results in a range of infidelity rather than a single value. 
We also estimated infidelity caused by $T_1$-like processes, where the qubits are thermalized to 50-50 population around the charge anti-crossing with the time scale $1/\Gamma(\epsilon)$.
This time scale depends on the detuning $\epsilon$ and has a minimum value around $300 \mathrm{\mu s}$.
The corresponding infidelity per gate is therefore estimated by integrating the transition rates over the ramp time and multiplying the number of shuttles,  $\frac{N_{\rm shuttle}}{3} \int_0^{t_{\rm ramp}} \Gamma(\epsilon(t)) dt$.

As summarized in Table~\ref{tab:model_Xgate_error}, the results show that a large portion of errors arises from the waveform uncertainty.
The relative impact of the detuning noise and Larmor frequency fluctuations depends on the details of the pulses and quantization axes angle.
The thermalization process has little contribution, because of the extended thermalization time at low field and the short ramp time we use.
The estimated infidelity of both qubits are on the same order as the measured infidelity, $r_{\rm X,A(B)} \approx$ 0.03 (0.04)\% given by randomized benchmarking (RB) and $r_{\rm X,A(B)} \approx$ 0.06 (0.02)\% given by gate set tomography (GST).
The deviations can arise from unaccounted error sources as well as the robustness of the benchmarking protocols under realistic experimental conditions.

\begin{table}
    \centering
    \begin{tabular}{|c|c|c|}
    \hline
        Error source 
        & \begin{tabular}[x]{@{}c@{}} $\rm X_{\pi/2,A}$  infidelity     ($\times10^{-5}$) \end{tabular} 
        & \begin{tabular}[x]{@{}c@{}} $\rm X_{\pi/2,B}$  infidelity     ($\times10^{-5}$)  \end{tabular}  \\ \hline 
        Larmor frequency fluctuations    &   3.1       &    0.8   \\ \hline
        detuning noise      &   7.2       &    0.13  \\ \hline
        waveform uncertainty&   4.0 - 14.6     &   5.1 - 17.2  \\ \hline
        thermalization      &   0.04           &    0.05 \\ \noalign{\hrule height2.0pt} 
        total infidelity    &   14.3  - 25.0   &    6.0 - 18.1 \\\hline         
    \end{tabular}
    \caption{\textbf{Incoherent error estimation.} Here we present the error metric in terms of average gate infidelity in single-qubit space. }
    \label{tab:model_Xgate_error}
\end{table}

\newpage
\subsection{Evaluation of the shuttling fidelity}
\label{sec:shuttling_fidelity}
In this section we show the connection between shuttling fidelity $F_{\rm shuttle}$ and the gate fidelity extracted from single-qubit randomized benchmarking.
The $\rm X_{\pi/2,A}$ gate is composed of four shuttling ramps of 2~ns and some idle periods. 
Because the spin state rotates during the 2~ns-ramp in a predictable way, we consider the 2~ns-ramp as a quantum gate.
The average gate fidelity of this single-shuttle gate is taken as shuttling fidelity $F_{\rm shuttle}$.
In principle, the deterministic part of the gate can be compensated by applying a calibrated rotation after the ramp.
The stochastic part of the gate (incoherent error) that cannot be compensated contributes to the shuttling infidelity.

\txtblue{In Table~\ref{tab:model_Xgate_error} we list the error sources and find that the wave function uncertainty due to pulse timing is the major error source. The non-integer waiting time between each shuttling step, as well as the differences in execution times of the Clifford gates, result in randomization of this error. We therefore consider the errors as uncorrelated, consistent with the assumptions of randomized benchmarking, and use the relation $r_{\rm X_{\pi/2,A}} = 4 r_{\rm shuttle} + r_{\rm idle}$, where $r_{\rm X_{\pi/2,A}}$ is the infidelity of $\rm X_{\pi/2,A}$, $r_{\rm shuttle}=1-F_{\rm shuttle}$ is the shuttling infidelity and  $r_{\rm idle}$ is the infidelity \txtblue{that} accounts for all the idling operations. } 
This relation gives the lower bound of the shuttling fidelity,  $F_{\rm shuttle}=1- r_{\rm shuttle} \geq 1- \frac{1}{4} r_{\rm X_{\pi/2,A}}$.
Based on the single-qubit RB fidelity $F_{\rm X_{\pi/2},A} \geq$ 99.967(4)\%, we calculate the shuttling fidelity $F_{\rm shuttle}\geq$ 99.992(1)\%. From the gate $X_{\pi/2,B}$ we estimate the shuttling fidelity $F_{\rm shuttle}\geq$ 99.980(3)\%. However, we remark that the quantization axis \txtblue{of qubit B is very close to 45$^\circ$, which may result in decoupling, and therefore an underestimation of $r_{\rm idle}$ and possibly $r_{\rm shuttle}$.}

\newpage
\subsection{Measurement protocol for residual exchange couplings }
\label{sec:measure_residual_exchange}
\begin{figure*}[htp!]
	\centering
	\includegraphics[width=0.99 \textwidth]{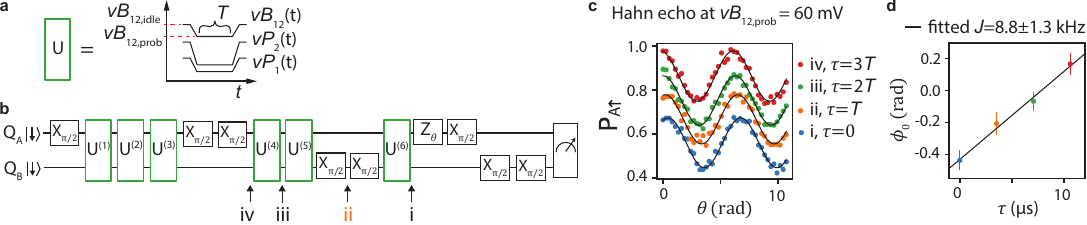}
	\caption{\textbf{Measurement of exchange coupling using a Hahn echo sequence at magnetic field of 25~mT. }	\textbf{a,} Illustration of a pulse used to probe the exchange coupling at $vB_{\rm 12,probe}$ starting from the idle point $vB_{\rm 12,idle}$ where the single-qubit gates are performed. The virtual gate voltages $vP_{1,2}$ are ramped to the values used for the Ramsey experiments (Fig.~\ref{fig:2x2_T2_exchange_on}) as well as the GST and RB experiments. \textbf{b,} The circuits for Hahn echo measurement, probing the difference of accumulated phases on qubit A induced by the flipped state of qubit B. Echo fringes of qubit A are measured in \textbf{c} by inserting $\rm X_{\pi/2,B}^2$ at various positions of the circuits \{i,ii,iii,iv\}, which lets $\rm Q_A$ interact with flipped $\rm Q_B$ for various amount of time $\tau=nT$, $n=\{0,1,2,3\}$. (b) shows the gates $\rm X_{\pi/2,B}^2$ inserting at the position ii. \textbf{c,} The fringes of the Hahn echo measurement. The data sets are shifted vertically for clearer display. The fringes are fitted to $A \cos(\theta + \phi_0) + B$ as black lines and the extracted phase offsets $\phi_0$ are plotted in \textbf{d}. The linear fit of the phase offsets $\phi_0$  as a function of evolution time $\tau$ gives the $\rm Q_B$-state-dependent frequency variation of $\rm Q_A$. The phase accumulation during the ramp and the idle time before and after the pulses $\rm X_{\pi/2,B}^2$ are corrected  by the residual exchange at the idle position, which is 15(1) kHz measured via the same method. We note that the measurement displayed in this figure are taken after a charge jump of $vB_{12}$, similar to the situation described in the caption of Fig.~\ref{fig:2x2_T2_exchange_on}.  }
	\label{fig:residual_exchange_hahn}
\end{figure*}

\newpage
\subsection{Measurement and simulations of the two-qubit energies and coherence time  }
\label{sec:model_twoQ}
\begin{figure*}[htp!]
	\centering
	\includegraphics[width=0.99 \textwidth]{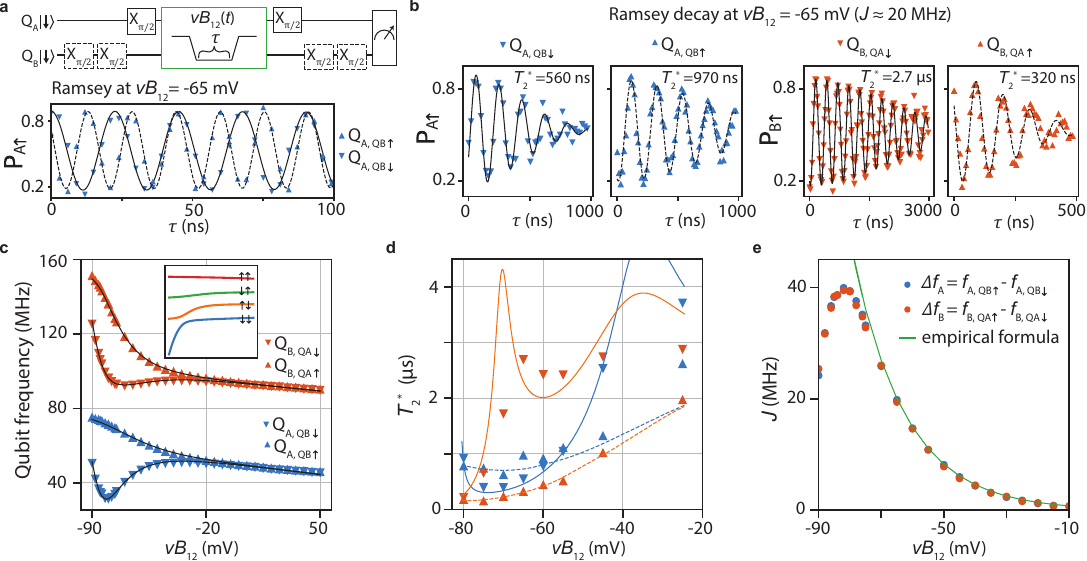}
	\caption{\textbf{Qubit frequencies and coherence time as a function of virtual barrier voltage at magnetic field of 25~mT.}	\textbf{a,} The Ramsey experiments for measuring qubit frequencies as well as the free evolution decay time $T_2^\star$ at various virtual barrier gate voltage $vB_{12}$. The circuits here is an example of qubit A frequency measurement conditioned on qubit B state. The pulse on $vB_{12}$ is trapezoidal with linear ramp times of 80~ns to avoid diabatic state transitions. \textbf{b,} Free induction decay of individual qubit conditioned on the other qubit at $vB_{12}=-65$~mV. The data are fitted to $P(\tau)=A \exp(-(\tau/T_2^\star)^2) + B$ to extract $T_2^\star$. \textbf{c,} The state-dependent qubit frequencies. The fitting results are plotted in black lines, with the energy diagram in the inset. \textbf{d,} The $T_2^\star$ measurement and the fitting curves. The sampling time and numbers of sample points are chosen to adapt for the qubit frequencies and decay rates that depends on $vB_{12}$, resulting in the $T_2^\star$ experiment time of 18-58 seconds for $\rm Q_A$ and 38-133 seconds for $\rm Q_B$. \textbf{e,} The exchange couplings $J=\Delta f_{\rm A(B)}$. The exchange couplings predicted by empirical formula \txtblue{$J=J_0 \exp(-\kappa\ {(vB_{12}-\Delta vB_{12})})$} is plotted, where $J_0=0.24$~MHz, $\kappa=0.059~{\rm mV}^{-1}$, and \txtblue{$\Delta vB_{12}=10$~mV}. One set of the data  $\Delta f_{\rm A}$ is also plotted in Fig.2C of the main text.  We note that the data displayed in this figure \txtblue{and in Fig.2C of the main text} are taken after a charge jump that shifts $vB_{12}$ by about \txtblue{$\Delta vB_{12}=10$~mV}. As an example, the measurement taken at $vB_{12}=-65$~mV in this figure should be considered as the measurement taken at $vB_{12} \approx -75$~mV in other parts of the paper.  }
	\label{fig:2x2_T2_exchange_on}
\end{figure*}

We measure the qubit energies and the coherence times of the two-qubit system as shown in Fig.~\ref{fig:2x2_T2_exchange_on}.
We observe several features, such as the non-monotonic dependence of qubit energies as function of barrier gate voltages.  
To explain this result, we model the two-qubit system using an extended minimal-size Fermi-Hubbard model with the six basis states \{ $S(2,0)$, $S(0,2)$, $T^{+}(1,1)$, $S(1,1)$, $T^0(1,1)$, $T^{-}(1,1)$\}. The Hamiltonian is written as~\cite{Mutter2021,geyerTwoqubitLogicAnisotropic2022,zhangUniversalControlFour2023}
\begin{equation}
\begin{aligned}
H_{2Q} & = 
\left( {\begin{array}{cccccc}
    U+\epsilon_{\rm 2Q} & 0 & - t_{\rm y} + i t_{\rm x}  & \sqrt{2} t_{\rm c}  & - i \sqrt{2} t_{\rm z}  &   - t_{\rm y} - i t_{\rm x}  \\
    0 & U-\epsilon_{\rm 2Q} & - t_{\rm y} + i t_{\rm x}  & \sqrt{2} t_{\rm c}  & - i \sqrt{2} t_{\rm z}  &   - t_{\rm y} - i t_{\rm x}  \\
  - t_{\rm y} - i t_{\rm x} &   - t_{\rm y} - i t_{\rm x}   &  h f_{\rm +}   & 0 & 0 & 0 \\
 \sqrt{2} t_{\rm c} &  \sqrt{2} t_{\rm c}       &  0 & 0 & h f_{\rm -} & 0 \\
 i \sqrt{2} t_{\rm z} &  i \sqrt{2} t_{\rm z}   &  0 & h f_{\rm -} & 0 & 0 \\
 - t_{\rm y} + i t_{\rm x} & - t_{\rm y} + i t_{\rm x}      &  0 & 0 & 0 & - h f_{\rm +}\\
  \end{array} } \right).
\end{aligned}
\label{Hamiltonian_twoQ}
\end{equation}
The charging energy takes the value $U$ = 2.56 meV~\cite{John2023}. The detuning energy of the two-spin system is $\epsilon_{\rm 2Q}$ (which is different than the single-spin system discussed in ~\ref{sec:simulations_shuttling_gates}). The Zeeman interactions are included in $h f_{\rm \pm}=\frac{1}{2}(g_{\rm A} \pm g_{\rm B})\mu_{\rm B} B$. The hopping between the quantum dots is modelled through a spin-probability conserving tunnel coupling $t_{\rm c}+i t_{\rm z}$ and a spin-probability non-conserving tunnel coupling $t_{\rm x}+i t_{\rm y}$. The impact of a magnetic field is described by the Zeeman interaction Hamiltonian, where we use a local spin basis such that the two spins are aligned. Consequently, this redefines the spin-conserving and spin-non-conserving tunnel couplings.

In the experiments, we change the voltage $vB_{\rm 12}$  at constant detuning to tune the tunnel couplings ($t_{\rm c}$, $t_{\rm x}$, $t_{\rm y}$, $t_{\rm z}$) and the resulting exchange coupling. 
We assume that all the tunnel couplings change exponentially as a function of the barrier gate~\cite{geyerTwoqubitLogicAnisotropic2022} $\exp{(-\frac{1}{2}\kappa {\rm vB_{12}})}$ with identical $\kappa=0.059$ $\rm mV^{-1}$ and estimate the prefactors by fitting the parameters to our measurements. 
This assumption also implies that the ratios $t_{\rm x,y,z}/t_{\rm c}$  remain constant. Since the eigenenergies of Hamiltonian~\eqref{Hamiltonian_twoQ} only depend on the absolute value of $t_{\rm c}+i t_{\rm z}$ and $t_{\rm x}+i t_{\rm y}$ and not on their complex argument (can be easily verified by computing the characteristic polynomial), the phases cannot be estimated by analyzing the eigenenergies. 
For the Zeeman interactions, we assume the $g$-factors depend linearly on the gate voltage, $g_{\rm A(B)}(vB_{12})=g_{\rm A(B)}^{(0)}+g_{\rm A(B)}^{(1)}vB_{12}$. 
Finally, we set the detuning $\epsilon_{\rm 2Q}$ to a fixed value of zero, because we operate at fixed plunger gate voltages (vP1, vP2) close to the symmetry point for all the two-qubit experiments.

We fit the qubit frequencies in Fig.~\ref{fig:2x2_T2_exchange_on}c to the eigenenergies of Eq.~\eqref{Hamiltonian_twoQ}. Our fit shows a good agreement between the model and the experiments. We find the relative strength between spin-dependent tunnel couplings to be $\frac{t_{\rm x}^2+t_{\rm y}^2}{t_{\rm c}^2+t_{\rm z}^2}=0.11$. The corresponding energy levels are plotted in the inset of Fig.~\ref{fig:2x2_T2_exchange_on}c, where we identify the anti-crossing between $\ket{\uparrow\downarrow}$ and $\ket{\downarrow\downarrow}$ as the cause of the bending of exchange coupling around $vB_{\rm 12}=-85$~mV.

Based on this model, we  estimate the dephasing of the two-spin system by considering qubit frequency fluctuations due to three noise sources: the effective electric noise on $vB_{\rm 12}$ and fluctuations of the $g$-factors $g_{\rm A(B)}^{(0)}$~\cite{Xue2022}. 
Assuming $1/f$ noise dominates qubit dephasing, the coherence time reads $T_2^\star= \sqrt{2/(S_{1/f} \ln \frac{0.401}{t_{\rm e}/t_{\rm m}})} $~\cite{Martinis2003,Cywinski2008}, where we define the evolution time $t_{\rm e}$ as the high-frequency cutoff and the total measurement time $t_{\rm m}$ as the low-frequency cutoff, $S_{1/f}$ is the strength of the single-sided spectral density of the qubit \txtblue{angular} frequency\txtblue{. The strength is related to the noise spectrum of} a particular noise source \txtblue{$x \in$ \{$vB_{12}$, $g_{\rm A}$, $g_{\rm B}$\} by $S_{1/f}=(\frac{\partial \omega}{\partial x})^2 S^x_{1/f}$, where $\frac{\partial \omega}{\partial x}$ is the sensitivity of the qubit angular frequency and the strength of the $1/f$ noise $S^x_{1/f}$ is defined by $S^x(\omega)=\int_{0}^{\infty}S^x(t) e^{i \omega t} dt = 2\pi S^x_{1/f}/\omega$ with the autocorrelation function $S^x(t)=\braket{x(t)x(0)}$}. 
Here we choose $t_{\rm e}=T_2^\star$ which is the evolution time relevant for a $T_2^\star$ measurement. 
We assume that the three noise sources are independent and their fluctuations uncorrelated, giving rise to a total dephasing time ${T_{\rm 2,total}^\star}=1/\sqrt{{T_{ 2,vB_{12}}^\star}^{-2}+{\rm T_{\rm 2,g_{\rm A}^{(0)}}^\star}^{-2}+{\rm T_{\rm 2,g_{\rm B}^{(0)}}^\star}^{-2}}$.
For the transition between two energy levels $i$ and $j$, we use the derivatives of the transition angular frequency $\omega_{ij}$ with respect to the voltage fluctuations to compute theoretical predictions of the coherence time. We pay close attention to the different bandwidths ($t_{\rm m}$, $t_{\rm e}$) in the respective measurements. 
For example, the gate voltage noise $S_{1/f}^{ vB_{12}}$ yields ${T_{ 2,vB_{12}}^\star}^{-2}=\frac{1}{2} \ln \frac{0.401}{t_{\rm e}/t_{\rm m}} (\frac{\partial \omega_{ij}}{\partial vB_{\rm 12}})^2 S_{1/f}^{ vB_{12}}$.
We now use the the fitting parameters obtained in Fig.~\ref{fig:2x2_T2_exchange_on}c to fit the noise strength $S_{1/f}^{ vB_{12}, g_{\rm A}, g_{\rm B}}$ to the coherence time for all the transitions. 
We estimate the noise strengths by minimizing the square sum of the dephasing rate differences $\Delta \frac{1}{T_2^\star}$ between theoretical and measurement values.
Fig.~\ref{fig:2x2_T2_exchange_on}d shows the fitting results, having qualitative agreement between the model and the experiment. 
The model reproduce the trend and several features of $T_2^\star(vB_{12})$, and also predicts the relative dephasing time of different qubit transitions.
We find the noise strength $S_{1/f}^{ vB_{12}}=\SI{0.031}{mV^2}$, which is equivalent to $\sigma_{vB_{12}}=0.78$~mV if integrating from $\SI{1}{\micro s}$ to 1000 seconds, a typical time scale for Ramsey measurement, and on the same order as the results reported in Ref.~\cite{Xue2022}.
The noise strength of $S_{1/f}^{ g_{\rm A} (g_{\rm B})}$ at this magnetic field is equivalent to the qubit frequency noise $S_{1/f}^{ f_{\rm QA} (f_{\rm QB})} = (\mu_B B)^2 S_{1/f}^{ g_{\rm A} (g_{\rm B})}= \SI{130(200)}{kHz^2}$, which translates to $\sigma_{f_{\rm QA(QB)}}=50(63)$~kHz and $T_2^\star= 4.5(3.5) \mathrm{\mu s} $ if integrating the noise from   $\SI{1}{\micro s}$ to 1000 seconds.

\newpage
\subsection{Calibration of \txtblue{the pulse-shaped} CZ gates}
\label{sec:calibrate_CZ}
\begin{figure*}[htp!]
	\centering
	\includegraphics[width=0.99 \textwidth]{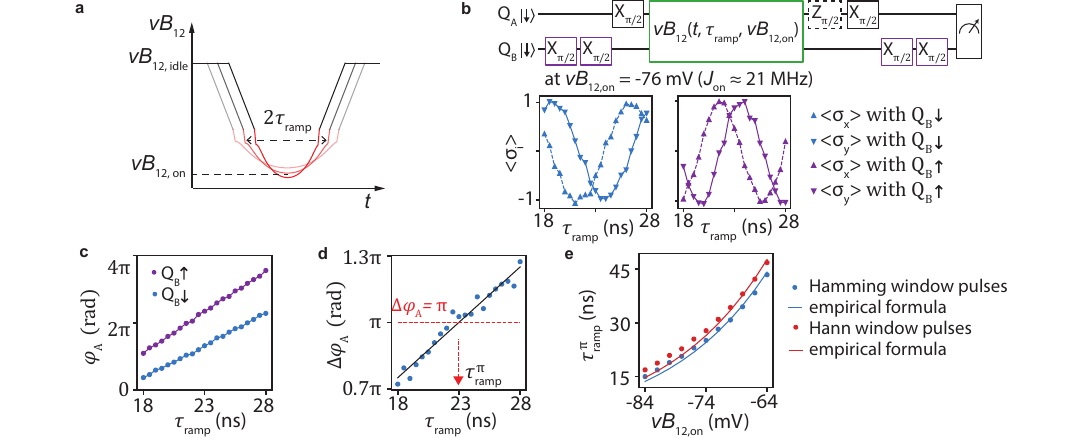}
	\caption{\textbf{Calibration of the conditional phase for \txtblue{the pulse-shaped} CZ gates.}	\textbf{a,} The illustration of virtual barrier gate voltage $vB_{12}(t)$ with two pulse parameters  $vB_{\rm 12,on}$ and $\tau_{\rm ramp}$ . The pulse $vB_{12}(t)$ generates Hamming window waveform $J(t)=J_{\rm on}(0.54 - 0.46 \cos(\pi t/\tau_{\rm ramp})$. The maximum exchange coupling $J_{\rm on}$ is predicted by empirical formula $J_{\rm on}=J_0 \exp(-\kappa\ {vB_{\rm 12,on}})$, where $J_0=0.24$~MHz and $\kappa=0.059~{\rm mV}^{-1}$. \textbf{b,} The normalized $\braket{\sigma_{\rm x(y)}}$ of qubit A depending on the state of qubit B, as a function of $\tau_{\rm ramp}$ at a certain gate voltage $vB_{\rm 12,on}$. The values $\braket{\sigma_{\rm x(y)}}$  are measured by the $\rm X_{ \pi/2}$ without (with) $\rm Z_{ \pi/2}$ before the readout, normalized with the Ramsey amplitudes of a reference experiments without the exchange pulse. Here is an example of $vB_{\rm 12,on}=-76$~mV. \textbf{c,} The state-dependent phases of the qubit A as a function of the ramp time $\tau_{\rm ramp}$. \textbf{d,} The ramp time for the state-dependent $\pi$ phase shift, $\tau_{\rm ramp}=\tau_{\rm ramp}^{\pi}$, is determined by linear interpolation and finding the point where the state-dependent phase shift $\Delta \varphi_{\rm A}=\varphi_{\rm A,B\uparrow}-\varphi_{\rm A,B\downarrow}=\pi$. \textbf{e,} The ramp time $\tau_{\rm ramp}^{\pi}$ that results in CZ gate at various gate voltages $vB_{\rm 12,on}$. We also tune up the CZ gates with Hann window pulses using the same method. The predictions are based on the analytical formula $t_{\rm ramp}=0.25/(a_0 J_0 \exp(-\kappa {{vB}_{\rm 12,on}}))$, where $a_0=0.54(0.5)$ for Hamming (Hann) window.  }
	\label{fig:2x2_CZ_tune_up}
\end{figure*}

We implement exchange pulses with a Hamming window $ J(t)=J_{\rm on} (0.54-0.46 \cos(\pi t/\tau_{\rm ramp}))$, using an empirical relation between the exchange coupling and the gate voltage $vB_{12}$ , $J({vB_{12}})=J_0 \exp(-\kappa\ {vB_{12}})$ where $J_0$ = 0.24 MHz and $\kappa$ = 0.059 ${\rm mV^{-1}}$. 
The CZ gate calibration is performed in the following order: \\
\begin{enumerate}
    \item Conditional phase calibration: for a given pulse amplitude $vB_{\rm 12,on}$, we measure the accumulated state-dependent phases as function of the ramp time $\tau_{\rm ramp}$, as described in Fig.~\ref{fig:2x2_CZ_tune_up}bc. We find the ramp time $\tau_{\rm ramp}=\tau_{\rm ramp}^\pi$ that allows the state-dependent phase difference of $\pi$ (Fig.~\ref{fig:2x2_CZ_tune_up}d). The pulse amplitudes and ramp times allowing conditional phase of $\pi$ are measured and plotted in Fig.~\ref{fig:2x2_CZ_tune_up}e. 
    \item Single-qubit phase correction: as described in Fig.2D of the main text, after applying an exchange pulse with a given pulse amplitude and the ramp time, the target qubit $\rm Q_A$ picks up a phase that should be calibrated to zero if the control qubit $\ket{\rm Q_B}=\ket{\downarrow}$, and to $\pi$ if the control qubit $\ket{\rm Q_B}=\ket{\uparrow}$. The same correction needs to apply to both qubits.
    \item GST calibration: we fine-tune the ramp time $\tau_{\rm ramp}$ and the single-qubit phase correction with the error reports from gate set tomography (GST)~\cite{Xue2022,BlumeKohoutTaxonomy2022}. 
\end{enumerate}
We measure the non-adiabatic transitions of the implemented exchange pulses in Fig.~\ref{fig:2x2_CZ_tune_up}. 
We observed the gate is sufficiently adiabatic when maximum exchange is below 20~MHz, motivating the choice of CZ gate parameter \txtblue{$vB_{\rm 12,on}=-76{\rm mV}$ for two-qubit RB and GST experiments}.

\begin{figure*}[htp!]
	\centering
	\includegraphics[width=0.99 \textwidth]{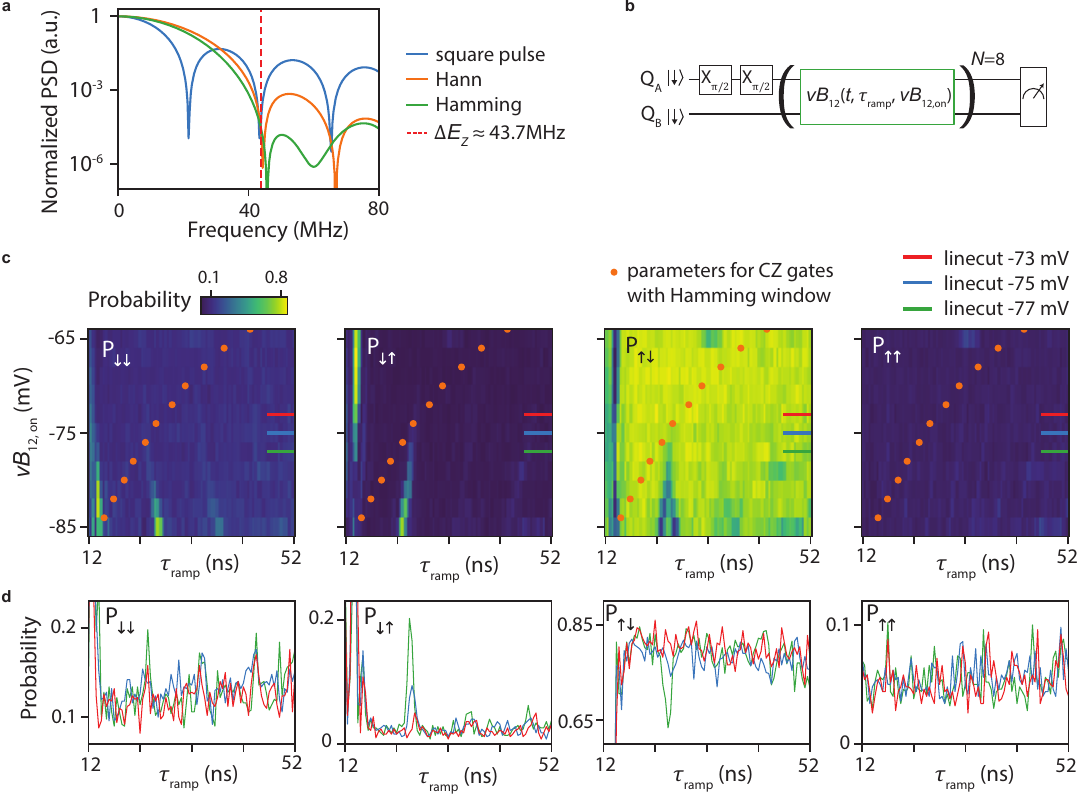}
	\caption{\textbf{Exchange pulse shapes and the resulting non-adiabatic state transitions. } \textbf{a,} The power spectrum density (PSD) of the exchange pulse shapes, indicating the energy emission that can drive non-adiabatic state transitions. Hamming (Hann) window functions are $J(t)=J_{\rm on} (a_{\rm 0} -(1-a_{\rm 0}) \cos(\pi t/\tau_{\rm ramp}))$, where $a_{\rm 0}=0.54(0.5)$. All the shapes have the same pulse time of 46~ns, close to the value used in the RB and GST experiments.                     \textbf{b,} The circuit for measuring state transitions induced by the exchange pulses. We use eight exchange pulses to amplify the transition probability. The pulses with the Hamming window shape parameters $(\tau_{\rm ramp}, vB_{\rm 12,on})$ are applied on the state $\ket{\uparrow \downarrow}$, and the full two-qubit state is readout at the end. \textbf{c,}  The probability $\rm P_{\sigma \sigma^\prime}$ that indicates non-adiabatic state transitions are measured at 25~mT ($\Delta E_Z \approx 43.7$~MHz). The parameters $(\tau_{\rm ramp}, vB_{\rm 12,on})$ for CZ gates, taken from Fig.~\ref{fig:2x2_CZ_tune_up}e, are marked in orange circles. The linecuts at $vB_{\rm 12, on}=-73,-75,-77$~mV (corresponding $J_{\rm on} \approx 18, 20, 23$~MHz) are displayed in \textbf{d}.	An onset of SWAP transition, $\ket{\uparrow \downarrow} \rightarrow \ket{\downarrow \uparrow}$, is observed as the emerging dip (peak) around $\tau=25$~ns in the plot of $P_{\uparrow \downarrow (\downarrow \uparrow)}$ when $vB_{\rm 12, on}$ becomes more negative. In the nearby parameter space we observe another transition dip (peak) $\ket{\uparrow \downarrow} \rightarrow \ket{\downarrow \downarrow}$. Combining with other measurement data (not showing here), we interpret this transition as $\rm Q_B$-state-dependent $\rm Q_A$ transition. }
	\label{fig:2x2_CZ_leakage}
\end{figure*}

\newpage
\subsection{Error modeling of \txtblue{the} two-qubit gate }
\label{sec:model_CZ_error}

In this section we estimate the average gate infidelity of the CZ gate due to the coherent error and incoherent error.
In a d-dimensional Hilbert space, for a unitary operation implemented in the experiment, $U_{\rm exp}$, the corresponding average fidelity is~\cite{Pedersen2007}
\begin{equation}
F= \frac{  |{\rm tr}({U_{\rm ideal}^{-1}U_{\rm exp}})|^2 + d}{d(d+1)}.
\label{fidelity_unitary}
\end{equation}

\paragraph{Coherent error}
To evaluate coherent errors, we compute the time evolution of the two-qubit state under the influence of the gate voltage pulse $vB_{12}(t)$ with a pulse shape matching a Hamming window~\cite{rimbach-russSimpleFrameworkSystematic2023} as depicted in Fig.2D of the main text  by solving the time-dependent Schrödinger equation numerically~\cite{johanssonQuTiPPythonFramework2013}.
If the system evolves adiabatically, the final state only acquires one conditional two-qubit phase and two single\txtblue{-}qubit phases.
These phases can be calibrated in the experiment by fine-tuning the time and amplitude of the pulse~\cite{rimbach-russSimpleFrameworkSystematic2023}.
On the other hand, non-adiabatic state transitions, as shown in Fig.~\ref{fig:2x2_CZ_leakage}cd, result in errors that cannot be simply calibrated.
In our simulation, we fine-tune the voltage pulses $vB_{12}(t)$ to achieve a conditional phase of $\pi$, compute the unitary time evolution operator of the quantum process without noise, and compensate for the single qubit Z rotations.
We find the resulting unitary evolution has an average gate infidelity 0.089\%.
Additionally, we decompose the error in the Pauli basis and express the simulated unitary by the dominant terms, $U_{\rm exp}=e^{-i{\rm (-0.010YI-0.021XY+0.021YX )} } U_{\rm ideal}$.
This result is in good agreement with the fact that the implemented pulse shape is designed to suppress the transition $\ket{\uparrow \downarrow} \rightarrow \ket{\downarrow \uparrow}$ while the transitions induced by spin-non-conserving tunneling are not fully suppressed.
We believe that a further reduction of non-adiabatic transitions can be achieved by incorporating Eq.~\eqref{Hamiltonian_twoQ} directly into the optimization process for finding the pulse.

\paragraph{Incoherent error}
Incoherent errors are dominantly caused by the 1/f-type low-frequency fluctuations in $vB_{12}$ and $g$-factors $g_{\rm A,B}$, which result in the random deviations of the unitary operation $U_{\rm exp}$ from the ideal operation $U_{\rm ideal}$. We can now write the unitary operation $U_{\rm exp}(x)$ that is dependent on a stochastic parameter $x$ of the noise source. While this can be straightforward generalized to multiple sources, we consider for simplicity only fluctuations of the accumulated phases and neglect fluctuations of the transition matrix elements caused by the non-adiabatic time evolution discussed in the previous paragraph. This allows us to further approximate the 1/f spectral noise with quasistatic fluctuations by integrating over the corresponding frequencies
$\sigma^2=2\int_{t_{\rm m}^{-1}}^{t_{\rm e}^{-1}} \frac{S_x}{f} df$.
Assuming $x$ to be a stochastic variable drawn from a Gaussian distribution with zero mean and standard deviation of $\sigma$, we can replace the quantity $|{\rm tr}({U_{\rm ideal}^{-1}U_{\rm exp}})|$ in Eq.~\eqref{fidelity_unitary} with the expectation value~\cite{Green2012,vanDijk2019},
\begin{equation}
\braket{|{\rm tr}({U_{\rm ideal}^{-1}U_{\rm exp}})|^2} = \int_{-\infty}^{\infty} |{\rm tr}({U_{\rm ideal}^{-1}U_{\rm exp}(x)})|^2 \frac{1}{\sqrt{2\pi}\sigma} e^{-\frac{x^2}{2\sigma^2}} dx.
\label{fidelity_random_avg}
\end{equation}
We estimate the accumulated phases by integrating the qubit frequencies $f_{\rm Qi, Qj}(t,x)$ over time under the influence of the voltage pulse $vB_{12}(t)$ and the noise amplitude $x$. The corresponding (stochastic) unitary matrix in the basis $\ket{\downarrow \downarrow}$, $\ket{\uparrow \downarrow}$, $\ket{\downarrow \uparrow}$, $\ket{\uparrow \uparrow}$ is then given by
\begin{equation}
U_{\rm exp}(x) = 
\left( {\begin{array}{cccc}
   1 & 0 & 0  &  0 \\
   0 & e^{-2\pi i \int f_{\rm QB, QA\downarrow}(t,x) dt } & 0  &  0  \\
   0 &  0 & e^{-2\pi i \int f_{\rm QA, QB\downarrow}(t,x) dt } & 0 \\
   0 &  0 & 0  &  e^{-2\pi i \int f_{\rm QA, QB\downarrow}(t,x) + f_{\rm QB, QA\uparrow}(t,x) dt } \\
  \end{array} } \right).
\label{unitary_twoQ_phase}
\end{equation}
The standard deviation of the noise $\sigma$ is estimated in a way similar to the $T_2^\star$ fitting in Fig.~\ref{fig:2x2_T2_exchange_on}d and depends on the low(high)-frequency cutoff $t_{\rm m}^{-1}$($t_{\rm e}^{-1}$) as $\sigma \propto \frac{1}{T_2^\star} \propto \sqrt{\ln \frac{0.401}{t_{\rm e}/t_{\rm m}}}$~\cite{Martinis2003,Cywinski2008}.
In the case of two-qubit IRB experiments, the total experimental time is $t_{\rm m}=2680$~s and $t_{\rm e}$ is chosen as the total gate time of 108 ns (including padding time).
Based on these experimental conditions and the results of the $T_2^\star$ fitting in ~\labelcref{sec:model_twoQ}, we estimate the effective standard deviations $\sigma_{vB_{12}}=0.88$~mV,  $\sigma_{f_{\rm QA}}=57$~kHz and $\sigma_{f_{\rm QB}}=72$~kHz during the IRB experiments. 
Taking the above considerations, we obtain an average gate infidelity 0.23\%, where the main contribution from the noise is caused by fluctuations of $vB_{12}$ accounting for an error of 0.19\%.

In summary, we find that incoherent error caused by dephasing are dominant over coherent errors for the average gate fidelity.
The total average gate infidelity from the models is equal to 0.32\%, which is on the same scale as the estimated value of $0.67\pm0.09\%$ extracted from the IRB experiment, while it significantly differs from the estimated value of $1.87\pm0.52\%$ extracted from the GST experiment (Table~\ref{tab:2qRB_result} and Table~\ref{tab:2qGST_result}).
The deviations can arise from unaccounted error sources as well as the robustness of the benchmarking protocols under realistic experimental conditions.

\break
\newpage
\subsection{Charge tuning and virtual \txtblue{gate} control of the 10 quantum dot array}
\label{sec:10_quantum_dot_array}
\begin{figure*}[htp!]
	\centering
	\includegraphics{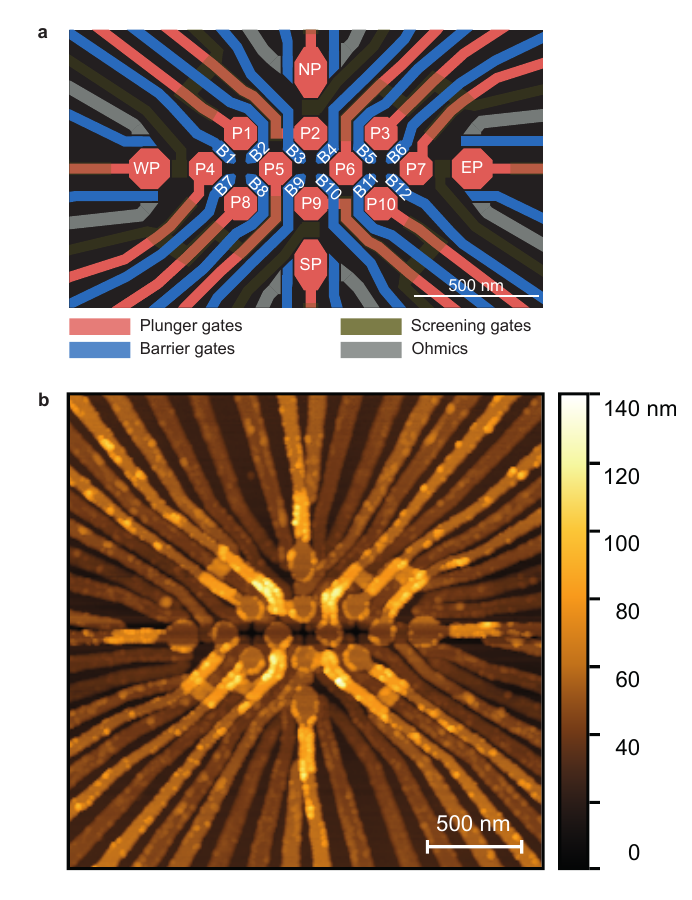}
	\caption{\textbf{The 10 quantum dot array device.}
  \textbf{a,}~Layout of the device indicating the names of the relevant gates.
  Plungers and barriers are labelled with P and B, respectively. In analogy to the cardinal coordinates, the sensors plunger gates are labelled as NP (north), EP (east), SP (south), and WP (west). \txtblue{\textbf{b,} Atomic force microscopy image of the device.} }
	\label{fig:gate-layout}
\end{figure*}

\begin{figure*}[htp!]
	\centering
	\includegraphics[width=0.99 \textwidth]{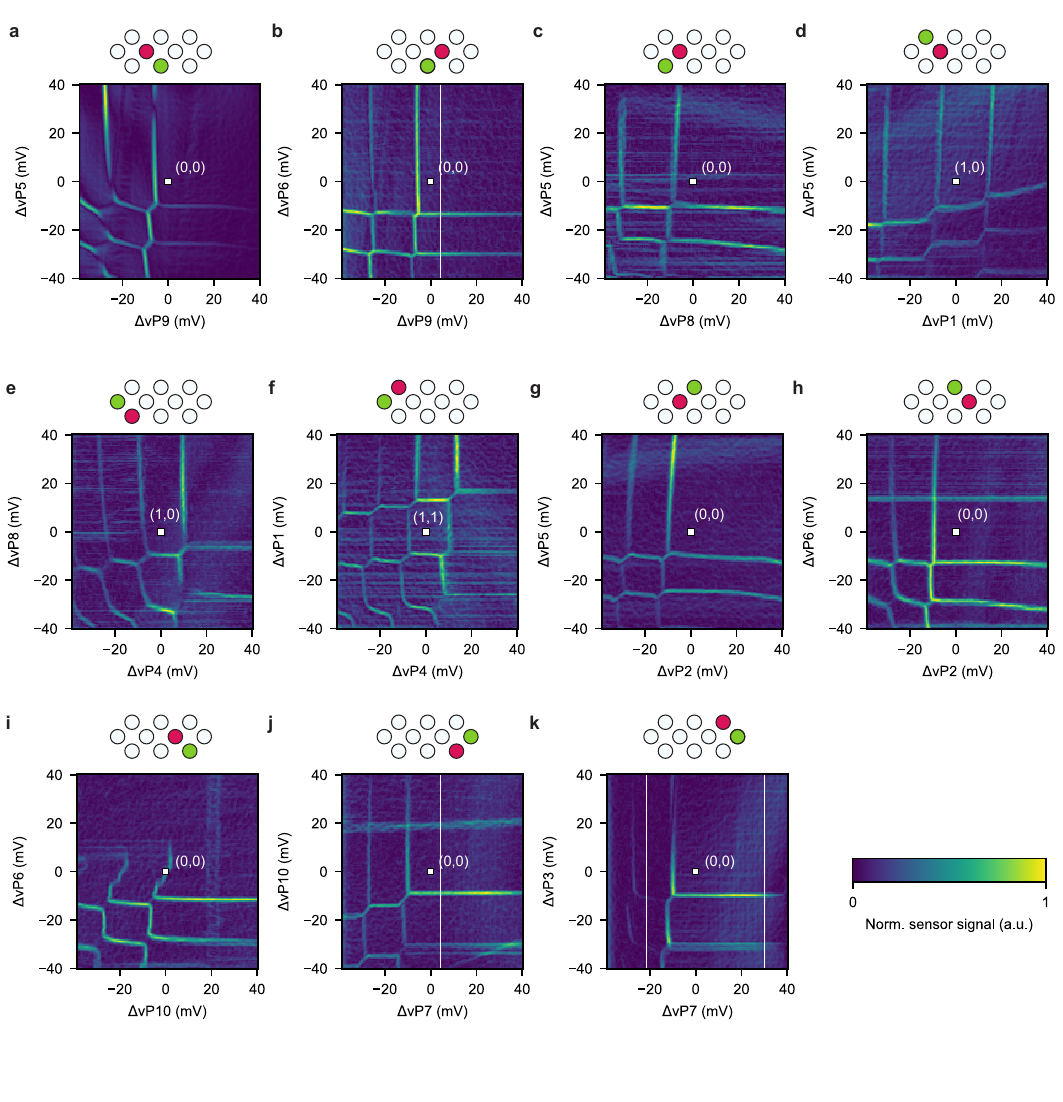}
	\caption{\textbf{Charge tune-up in the 10 quantum dot array.}
	\textbf{a-k,} Charge stability diagrams of the 10 quantum dot system showcasing the transition lines of each quantum dot. 
	At the centre of each map (white square), quantum dots D1 and D4 are prepared in the single-hole regime, while all the other quantum dots are in an empty state.
	In the schematic above each map, the green quantum dot is tuned by the gate swept on the x axis, and the red dot by the gate swept on the y axis.
	The horizontal lines at $\Delta$vP6 $\sim 15$ mV in panel \textbf{h} and at $\Delta$vP10 $\sim 20$ mV are spurious quantum dots transition lines, while the deformation of the vertical transition lines in \textbf{i} is due to charge latching effects. $\Delta$vP$i$ indicates a relative voltage swing with respect to the dc voltage point.}
	\label{fig:charge-tuning}
\end{figure*}

\begin{figure*}[htp!]
	\centering
	\includegraphics{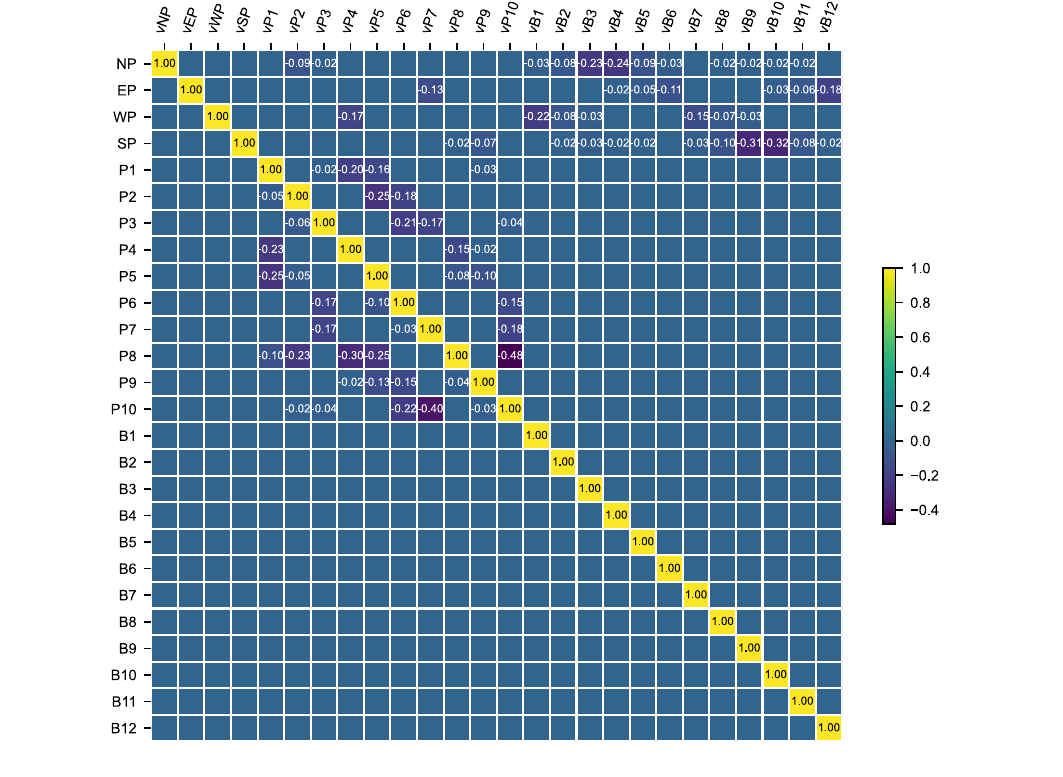}
	\caption{\textbf{Virtual gate matrix.}
		~Virtual gates are defined to compensate the crosstalk to the charge sensors and to obtain independent control of the chemical potential of each quantum dot via virtual plungers (vP1-vP10).}
	\label{fig:virtual_matrices}
\end{figure*}

\begin{figure*}[htp!]
	\centering
	\includegraphics{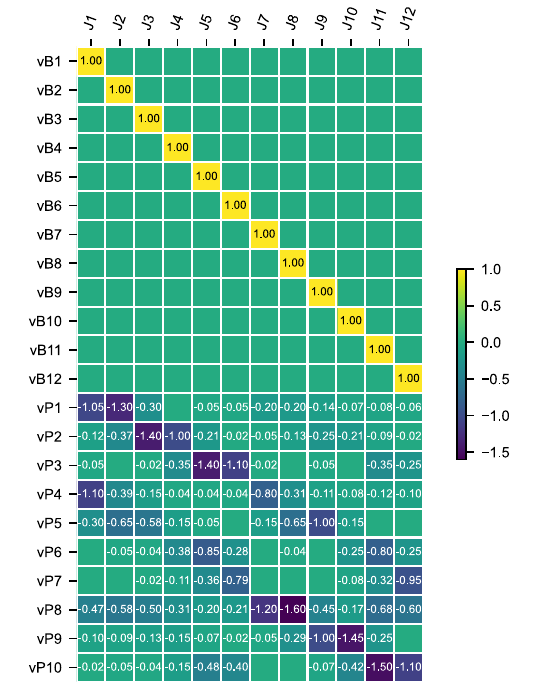}
	\caption{\textbf{Virtual barriers matrix.}
		J barriers are built as control parameters defined as linear combinations of the virtual barriers and virtual plungers. They are defined to obtain independent control over the interdot couplings, while leaving the quantum dots chemical potentials unaffected. 
	}
	\label{fig:virtual_barrier_matrices}
\end{figure*}

We prepare the 10 quantum dot system shown in Fig.~\ref{fig:gate-layout}  in the charge configuration with D1 and D4 in the single-hole regime, and the others in the empty charge regime.
Figs.~\ref{fig:charge-tuning}a-k display the charge stability diagrams acquired via charge sensing as a function of virtual plunger gates.  
At first, a virtual gate framework, with virtual matrix shown in Fig.~\ref{fig:virtual_matrices}, is defined in software to:
\begin{itemize}
    \item compensate the cross-capacitance of each gate with fast (ac) control to the four charge sensors;
    \item achieve independent control of the quantum dots chemical potentials via virtual plunger gates vP1-vP10.
\end{itemize}
A second matrix, shown in Fig.~\ref{fig:virtual_barrier_matrices}, is used for the definition of virtual barriers J1-J12, as a linear combinations of vB1-vB12 and vP1-vP10. J1-J12 serve to independently control the interdot tunnel couplings, without changing the quantum dots chemical potentials.

\newpage
\subsection{Shuttling across multiple quantum dots: detuning and barrier \txtblue{voltage} dependence}
\label{sec:shuttling_across_multiple_dots}
\begin{figure*}[htp!]
	\centering
	\includegraphics[width=0.99 \textwidth]{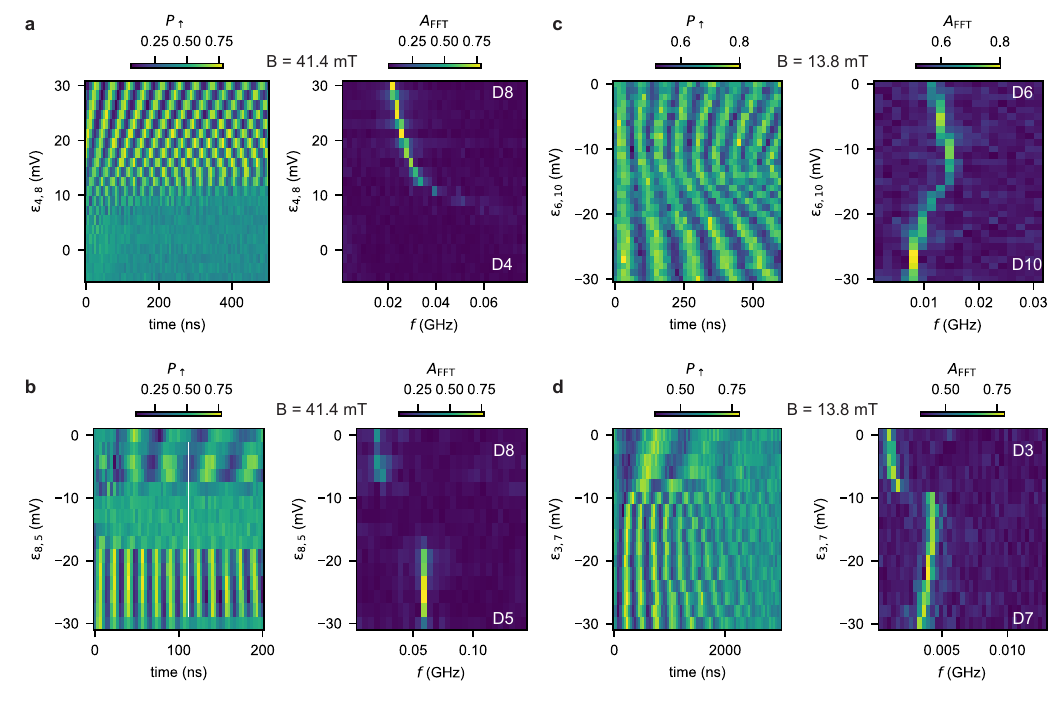}
	\caption{\textbf{Detuning dependence of the hopping-induced spin oscillations.}
	\textbf{a,} We vary the detuning gate voltage of the (0, 1) set point, corresponding to the shuttling sequence that moves the single spin from D4 to D8, i.e., from the (1,0) to the (0,1) charge state, across the charge interdot.
	Similarly to ref.~\cite{Vanriggelendoelman2023}, oscillations arise when the spin is transferred from one dot to the other. We observe that the onset of the oscillations corresponds to the charge interdot point.
	The panel on the right shown the FFT of the data.
	In \textbf{b, c, d,},  we illustrate similar measurements taken for spin shuttling from D8 to D5, from D6 to D10, from D3 to D7, respectively.
	}
	\label{fig:detuning_dependence}
\end{figure*}

\begin{figure*}[htp!]
	\centering
	\includegraphics[width=0.99 \textwidth]{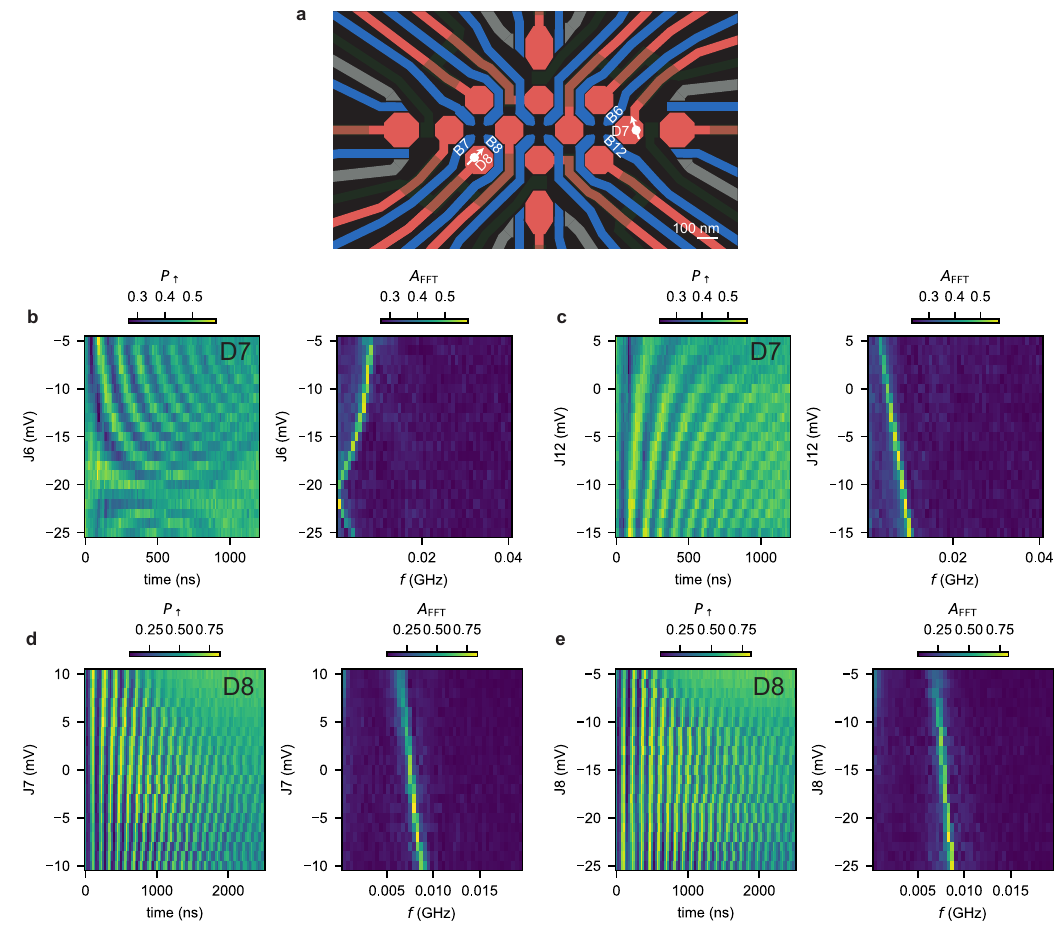}
	\caption{\textbf{Barrier gate dependence of the hopping-induced spin oscillations.}
		\textbf{a,}~Device layout indicating the two quantum dots D8 and D7, together with the surrounding barrier gates. 		\textbf{b, c}~D7 Larmor frequency evolution while sweeping the J6 and J12 voltages, respectively. 
		\textbf{d, e}~D8 Larmor frequency evolution as a function of J7 and J8. Small changes in the barrier voltages induce a linear shift of the D8 frequency.
	}
	\label{fig:barrier_dependence}
\end{figure*}

We probe the oscillations induced by differences in quantization axes as a function of detuning and barrier voltages. In practice, to shuttle from D4 to D8, we follow this protocol:
\begin{enumerate}
    \item initialize the D1, D4 double quantum dot system in the $\ket{\uparrow \downarrow}$;
    \item ramp the gate voltages from the set point defined as (1,0) to the (0,1), passing through the (1,0)-(0,1) charge anticrossing (AC). Here, the first number defines the filling of D4, and the second of D8. Ramp times in between these points are of $\sim$ 10 ns;
    \item wait in the (0,1) point for a varying free-precession time;
    \item pulse back to the AC, and to the (1,0) setpoint;
    \item readout the spin via Pauli spin blockade.    
\end{enumerate}
To probe the dependence of the D8 Larmor frequency, we sweep the detuning of the (0,1) set point. The results of this measurement are shown in Fig.~\ref{fig:detuning_dependence}a. Oscillations starts to arise when the gate voltage overcomes the charge anticrossing, that is found at $\epsilon_\mathrm{4, 8} = 10$ mV. For lower detuning voltages, the spin remains in D4, and therefore oscillations are not present. The Fast Fourier Transform of the data shows well the dependence of the Larmor frequency in the detuning voltage window. Similar measurements are shown for the case of a spin transfer from D8 to D5 (Fig.~\ref{fig:detuning_dependence}b), from D6 to D10 (Fig.~\ref{fig:detuning_dependence}c) and from D3 to D7 (Fig.~\ref{fig:detuning_dependence}d). 
We observe that, except for the region around the charge anticrossing, the qubit frequencies are not strongly affected by the detuning voltages. 
Rather, barrier gates do have a much stronger effect on the qubit frequencies, which mostly shift linearly, as illustrated in Fig.~\ref{fig:barrier_dependence}.
Interestingly, the D7 Larmor frequency crosses zero as a function of J6, suggesting a change of sign in the $g$-factor of the qubit.

\subsection{Hopping-induced oscillations in occupied quantum dots}
\label{sec:shuttling_occupied_dots}
\begin{figure*}[htp!]
	\centering
	\includegraphics[width=\textwidth]{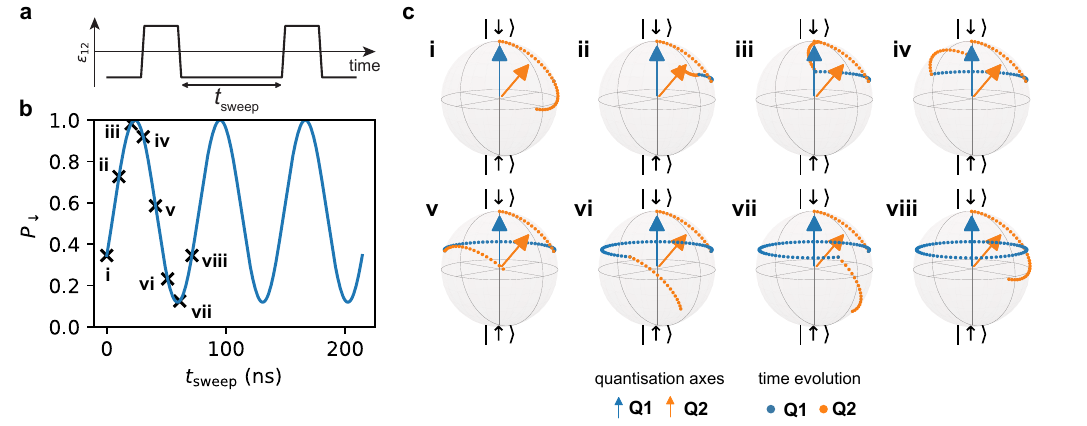}
	\caption{\textbf{Hopping-induced oscillations in occupied dots. a,} Shuttling sequence that moves Q1 from D1 to D2 and back to D1, twice. In the experiment, we vary the time in Q1, indicated as $t_{\textrm{sweep}}$ \textbf{b,} Calculated spin down probability as a function of sweep time in D1. The visibility is dependent on the waiting times in D2 and the difference in quantization axis. \textbf{c,} State evolution during the shuttling sequence for different waiting times in D1, as indicated in \textbf{b}. The final time evolution around the z-axis is not displayed for clarity.
	}
	\label{fig:double_shuttling}
\end{figure*}

Obtaining shuttling-induced oscillation in occupied quantum dots (as for the case of the filled quantum dots D1 and D4 of the main text) requires shuttling the spin back and forth between the corresponding quantum dot and an empty neighboring dot. In this section we motivate our procedure and explain why shuttling two times is required. \\
We assume to have two quantum dots D1 and D2 with a spin qubit Q1 in D1, and D2 empty. 
For simplicity, both sites have a $g$-factor of 0.05 and have a quantisation angle difference of $0.3 \pi$. 
If we want to obtain shuttling-induced oscillations of Q1 in D1, it is not sufficient to shuttle Q1 using the sequence D1 $\rightarrow$ D2 $\rightarrow$ D1, since the rotation in D1 needs to be projected onto another quantisation axis. 
Hence, we require shuttling the spin Q1 using this sequence: D1 $\rightarrow$ D2 $\rightarrow$ D1 $\rightarrow$ D2 $\rightarrow$ D1, as displayed in Fig.~\ref{fig:double_shuttling}a. 
Here, we vary the second time in D1 and wait 10 ns between all shuttle events. This protocol enables to convert the free evolution in D1 around the z axis to a rotation around a different axis of the D1 Bloch sphere. 
The resulting oscillation is shown in Fig.~\ref{fig:double_shuttling}b. The corresponding state evolution in the Bloch sphere for the points labelled as i-viii in Fig.~\ref{fig:double_shuttling}b, are shown in Fig.~\ref{fig:double_shuttling}c.

\newpage
\subsection{Dephasing times and Larmor frequencies in the 10 quantum dot array}
\label{sec:dephasing_Larmor_10quantumdot}
\begin{figure*}[htp!]
	\centering
	\includegraphics{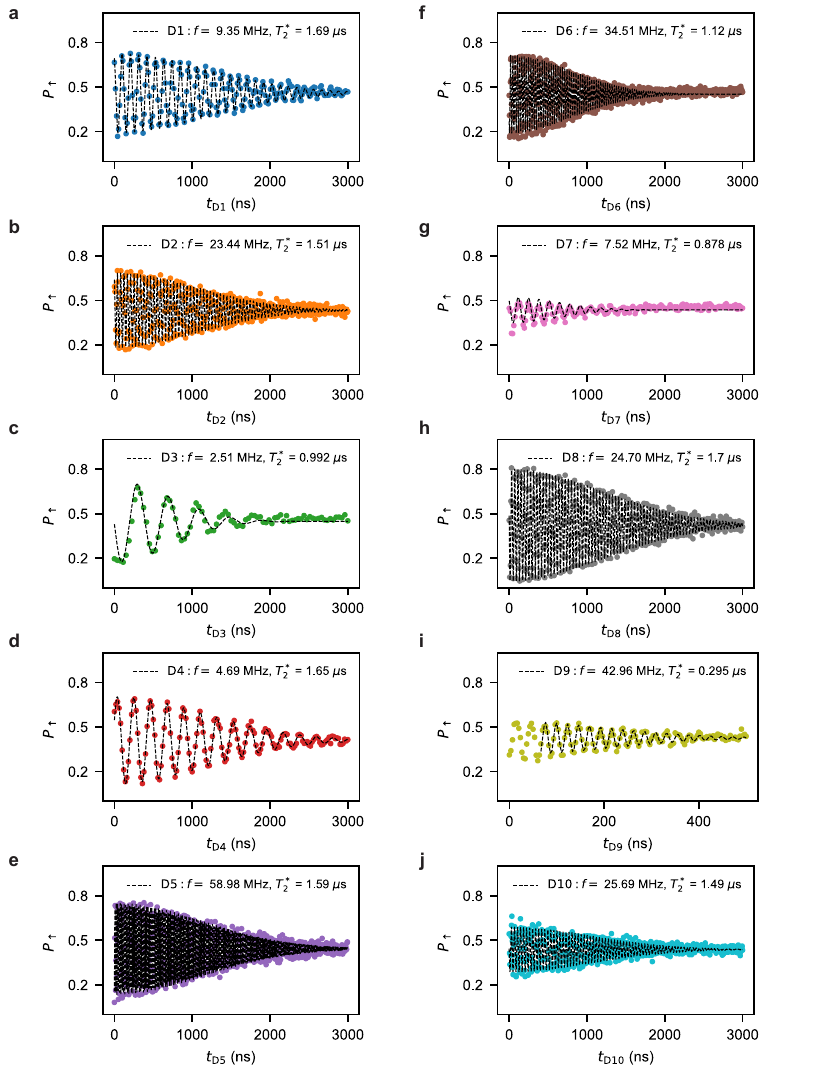}
	\caption{\textbf{$T^{\ast}_2$ of the 10 quantum dot array at \txtblue{41.4} mT.}
		\textbf{a-j,}~Each panel is measured using the same method as presented in the main text Fig.~3. \txtblue{We fit the dataset of D9 from 68 ns onward as we observe a frequency shift in the first $\sim 100$ ns possibly due to a delay in the electrical response.}}
	\label{fig:decoherence}
\end{figure*}

\begin{figure*}[htp!]
	\centering
	\includegraphics[width=0.85 \textwidth]{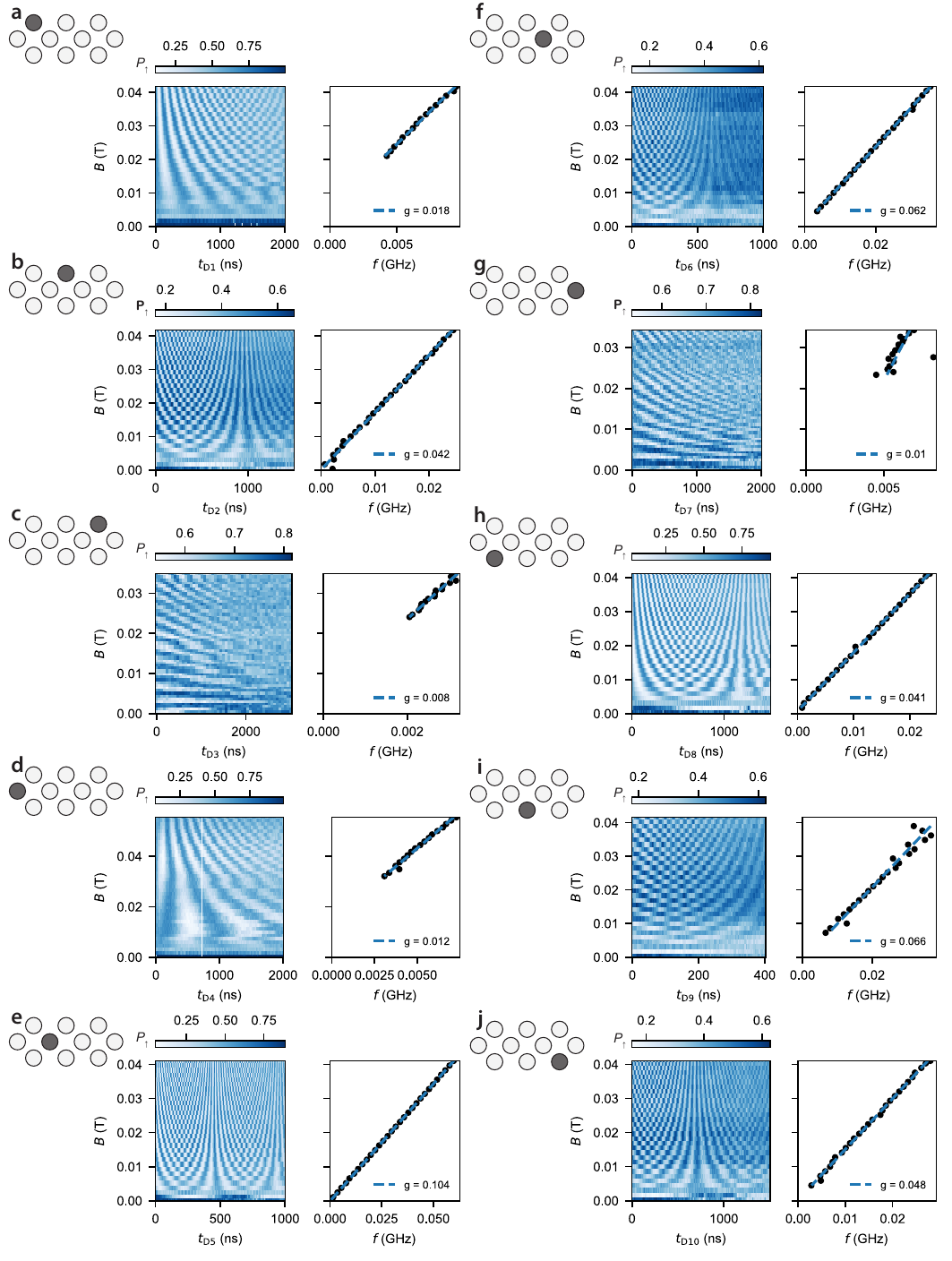}
	\caption{\textbf{Single-qubit rotations as a function of in-plane magnetic field for the 10 quantum dots.}
		\textbf{a-j,} We repeat the experiments shown in main text Fig.~3 and Fig.~\ref{fig:decoherence} as a function of magnetic field to estimate the $g$-factors. We linearly fit the oscillation frequencies as a function of the magnetic field. We observe that all qubits but Q1 and Q4 display a Larmor frequency that intersects zero at zero magnetic field.}
	\label{fig:field_dependence}
\end{figure*}

We study the dephasing times ($T^{\ast}_2$) of the 10 quantum dots by shuttling a spin diabatically from the double quantum dot system D1, D4 to each of the quantum dots, and let it evolve for a varying idle time. We measure the decay of the oscillations as a function of the time spent in each site by fitting the data shown in Fig.~\ref{fig:decoherence} and main text Fig.~3F using the equation: $A \cdot \sin(2  \pi  f  t + \phi)  \exp{ \left(  - (t / T^{\ast}_2)^2 \right) } + C$. Here, $2 \cdot A$ is the visibility, $f$ the Larmor frequency, $t$ the free precession time, $\phi$ the starting phase, and $C$ the oscillations offset. \\
The Larmor frequency of an isolated Loss-diVincenzo spin qubit satisfies the relation: $ f = \frac{g \mu_{\rm{B}} B}{h}$, with $g$ the $g$-factor, $\mu_{\rm{B}}$ the Bohr magneton, $B$ the applied magnetic field and $h$ the Planck constant. From the measurements of the oscillations as a function of magnetic field, we extract the $g$-factor for all the 10 quantum dots (Fig.~\ref{fig:field_dependence}). We find that except for the tunnel coupled Q1, Q4 qubits, $f$ shows a linear dependence to the magnetic field. The deviation from the linear trend can be explained from the coexistence of finite exchange coupling and non-parallel quantization axes. 

\txtblue{In general, the lower-than-unity and varying visibilities of the hopping-induced oscillations (Figs.~S21, S22, S24, S25) are caused by both SPAM errors and by the non-orthogonality of the quantization axes of adjacent quantum dots. 
As the estimated SPAM fidelities are typically in the range of 80-95\% (details for qubits A, B in Tables~\ref{tab:1qGST_SPAM} and ~\ref{tab:2qGST_SPAM}), we speculate that the origin of oscillation amplitudes below $\sim 0.8$ and their variability are mainly due to unfavourable spin alignment. In the current approach, we adopted a simple and sequential tuning approach, which can result in reduced rotations in the Bloch sphere. However, we could envision more involved tuning protocols that would lead to a higher contrast if desired, such as further optimization of the time spent in each dot and possibly additional shuttling steps to ensure that a phase rotation in a dot leads to a full amplitude rotation.}

\newpage

\subsection{Variability of the $g$-factors and quantization axes differences}
\label{sec:simulations_gfactor}
The semiconductor hetorostructure hosting our qubits is prone to imperfections, giving rise to a variability of the $g$-tensor. There are two dominant mechanisms: first, variability of the electrostatics from variations in the confinement through charged defects or neighboring gate electrodes, and second, variability of the strain in the quantum well through defects in the lattice and differences in the thermal expansion coefficients of the composite materials. 

Since our quantum dot structures are large compared to the inter-atom distances and operated at low densities $\rho\sim 10^{10}\,\text{cm}^{-2}$ (single hole regime), their dynamics is captured well in the standard $4\times 4$ Luttinger-Kohn-Bir-Pikus Hamiltonian. In the basis of total angular momentum eigenstates $\ket{j,m_j}=\lbrace\ket{\frac{3}{2},\frac{3}{2}},\ket{\frac{3}{2},-\frac{3}{2}},\ket{\frac{3}{2},\frac{1}{2}},\ket{\frac{3}{2},-\frac{1}{2}}\rbrace$ the Luttinger-Kohn Hamiltonian in $[001]$ growth direction reads as
\begin{align}
    H_\text{LKBP} = \left(\begin{array}{cccc}
        P+P_\varepsilon+Q+Q_\varepsilon & 0 & S+S_\varepsilon & R+R_\varepsilon  \\ 
        0 & P+P_\varepsilon+Q+Q_\varepsilon & R^\dagger+R^\dagger_\varepsilon & -S^\dagger-S^\dagger_\varepsilon  \\
        S^\dagger+S^\dagger_\varepsilon & R+R_\varepsilon & P+P_\varepsilon-Q-Q_\varepsilon & 0  \\
        R^\dagger+R^\dagger_\varepsilon & -S-S_\varepsilon & 0 & P+P_\varepsilon-Q-Q_\varepsilon  
    \end{array}\right).
    \label{eq:HLKBP}
\end{align}
The upper-left 2x2 block describes the energy of the spin-$\frac{3}{2}$ heavy hole state, the lower-right 2x2 block describes the energy of the spin-$\frac{1}{2}$ light hole state. The remaining terms describe the heavy-light hole coupling. 
The momentum operators read as
\begin{align}
    P&= \frac{\hbar^2}{2m_0}\gamma_1 (k_x^2+k_y^2+k_z^2),\\
    Q&= \frac{\hbar^2}{2m_0}\gamma_2 (k_x^2+k_y^2-2k_z^2),\\
    R&= \sqrt{3}\frac{\hbar^2}{2m_0}\left[-\gamma_2 (k_x^2-k_y^2)+i\gamma_3 k_x k_y+i\gamma_3 k_y k_x\right],\\
    S&= -\sqrt{3}\frac{\hbar^2}{2m_0}\gamma_3 \left[(k_x-i k_y)k_z+k_z(k_x-i k_y)\right],
    \label{eq:momentumLKBP}
\end{align}
where $\hbar k_{x,y,z}=-i \hbar \partial_{x,y,z}$ is the x,y,z momentum operator, $\hbar$ the reduced Planck constant, $m_0$ the bare electron mass, and $\gamma_1=13.38$, $\gamma_2=4.24$, and $\gamma_3=5.69$ are the Luttinger parameters for Ge~\cite{terrazosTheoryHolespinQubits2021}.
The strain operators read as
\begin{align}
    P_\varepsilon&= -a_v (\varepsilon_{xx}+\varepsilon_{yy}+\varepsilon_{zz}),\\
    Q_\varepsilon&= -\frac{b_v}{2}(\varepsilon_{xx}+\varepsilon_{yy}-2\varepsilon_{zz}),\\
    R_\varepsilon&= \sqrt{3}\frac{b_v}{2}(\varepsilon_{xx}-\varepsilon_{yy})-id_v \varepsilon_{xy},\\
    S_\varepsilon&= -d_v(\varepsilon_{xz}-i \varepsilon_{yz}),
    \label{eq:strainLKBP}
\end{align}
where $\varepsilon_{ij}$ is the 3D strain tensor, and $a_v=2.0$ eV, $b_v=-2.16$ eV, and $d_v=-6.06$ are the deformation potentials for Ge~\cite{terrazosTheoryHolespinQubits2021}.

The impact of a magnetic field is described by the substitution $\boldsymbol{p}\xrightarrow{} \boldsymbol{p} + e\boldsymbol{A}$, where $\boldsymbol{A}$ is the electromagnetic vector potential and $e$ is the electron charge, and the Zeeman Hamiltonian
\begin{align}
    H_\text{Zeeman} =2\mu_B \kappa\, \boldsymbol{J}\cdot\boldsymbol{B} + 2\mu_B q (J_x^3 B_x + J_y^3 B_y + J_z^3 B_z),
    \label{eq:Zeeman}
\end{align}
where $J_{x,y,z}$ are the spin $\frac{3}{2}$ matrices, $\mu_B$ is Bohr's magneton, $\kappa=3.41$, and $q=0.066$. 

For weak out-of-plane electric fields, harmonic in-plane confinement, and uniaxial strain, the g-tensor of the ground state can be approximated as~\cite{Abadillo2023}
\begin{align}
    \mathcal{G}\approx\left(\begin{array}{ccc}
        3q + \frac{6}{m_0 \Delta_\text{HL}}\left( \lambda \braket{p^2_x} - \lambda^\prime \braket{p^2_y} \right) & 0 & 0  \\ 
        0 & -3q - \frac{6}{m_0 \Delta_\text{HL}}\left( \lambda \braket{p^2_y} - \lambda^\prime \braket{p^2_x} \right) & 0 \\
        0 & 0 & 6 \kappa + \frac{27}{2} q - 2\gamma_h
    \end{array}\right).
\end{align}
Here, $\gamma_h\approx 3.56$, $\lambda=\kappa \gamma_2-2 \eta_h \gamma^2_3\approx 1.51$ and $\lambda^\prime=\kappa \gamma_2-2 \eta_h \gamma_2 \gamma_3\approx 4.81$ with $\eta_h\approx 0.2$ are correction factors from the heavy-hole light-hole coupling~\cite{Abadillo2023}, $\braket{p^2_{\xi}} = -\hbar^2 \int d\boldsymbol{r} \Psi(\boldsymbol{r})^\ast \frac{d^2}{d \xi^2}\Psi(\boldsymbol{r})\approx \frac{\hbar^2}{2 r_{HH}^2} $ is the momentum expectation value, and $r_{HH}\approx 60$ nm is the in-plane Bohr radius of the confined hole. The heavy-hole light-hole splitting is dominated by strain for wide quantum wells and can be approximated by $\Delta_\text{HL}\approx b_v (\epsilon_{xx}+\epsilon_{yy}-2\epsilon_{z,z})$. We can now emulate the variability of the electrostatic environment by varying the in-plane Bohr radius of the confined hole with standard deviation $\sigma_{r_{HH}}$.

Corrections from non-uniaxial strain strongly affect the resulting g-tensor~\cite{Abadillo2023}
\begin{align}
    \Delta\mathcal{G}\approx \frac{\kappa}{\Delta_\text{HL}}\left(\begin{array}{ccc}
        6b_v (\braket{\epsilon_{yy}}-\braket{\epsilon_{xx}}) & 4\sqrt{3}d_v \braket{\epsilon_{xy}} & 0 \\ 
        -4\sqrt{3}d_v \braket{\epsilon_{xy}} & 6b_v (\braket{\epsilon_{yy}}-\braket{\epsilon_{xx}}) & 0 \\
        -4\sqrt{3}d_v \braket{\epsilon_{xz}} & -4\sqrt{3}d_v \braket{\epsilon_{yz}} & 0
    \end{array}\right),
\end{align}
where $\epsilon_{ij}$ is the strain tensor component averaged over the position of the quantum dot. Analogously, we can now emulate the variability of the strain by varying the different components of the stress tensor with standard deviations $\sigma_{\epsilon_{ij}}$.

\begin{figure}
    \centering
    \includegraphics[width=0.99\textwidth]{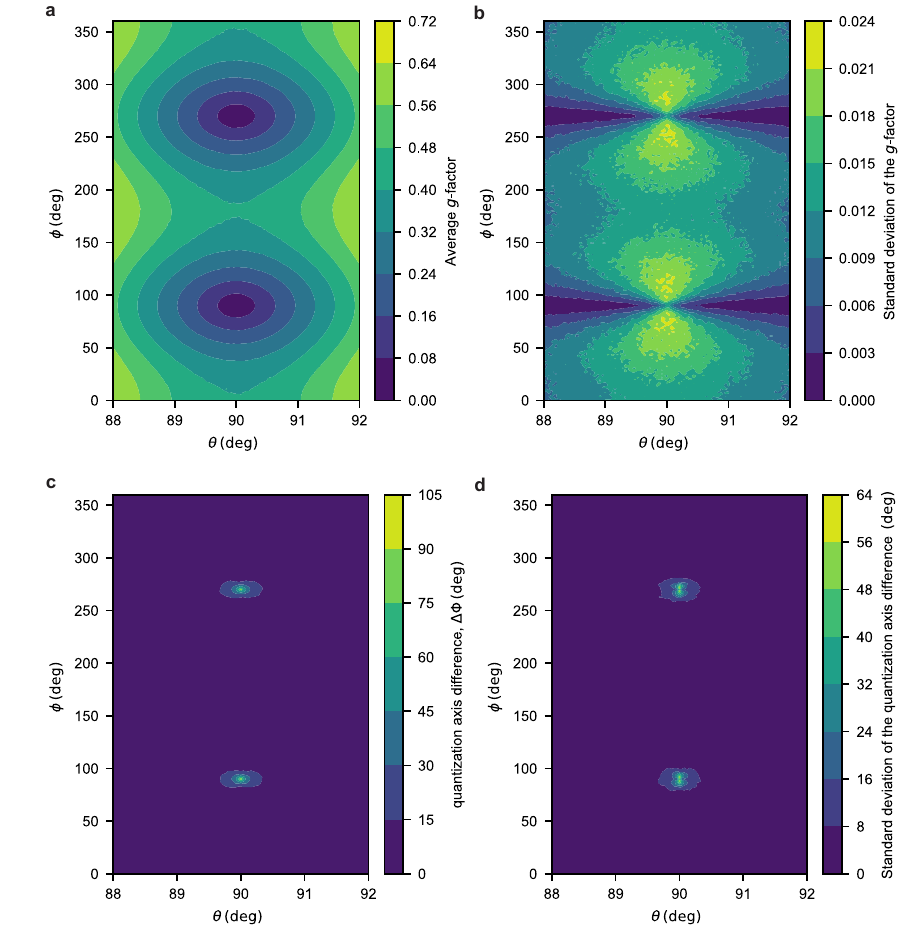}
    \caption{\textbf{Variability of the $g$-factor and spin quantization axis.} 
	\textbf{a, b},  Mean and standard deviation of the simulated $g$-factor as a function of the polar, $\theta$, and azimuth, $\phi$, angles of the magnetic field direction.
	\textbf{c, d}, Same for the difference in quantization axes $\Delta \Phi$.}
    \label{fig:var_gfactor_quant}
\end{figure}

The experimentally observed $g$-factor is given by $g_\text{exp}=|\boldsymbol{B} (\mathcal{G}+\Delta\mathcal{G})|/|\boldsymbol{B}|$ and depends on the magnetic field direction. The mean of the measured devices is $\braket{g}=0.04$ with standard deviation $\sigma_{g}=0.03$. The small $g$-factor can potentially be explained through a very strong electrostatic in-plane confinement with Bohr radius $r_{HH}\ll 45$ nm. We note that a more realistic numerical simulations may alleviate the estimated conditions. Alternatively, the small (large) in-plane $g$-factor can be explained by an asymmetric in-plane strain tensor $|\epsilon_{yy}-\epsilon_{xx}|/|\epsilon_{yy}+\epsilon_{xx}|=1.5-1.9\%$ if the magnetic field is in the direction of the stronger (weaker) strain. We note, that such an asymmetry between the strain components $\epsilon_{yy}$ and $\epsilon_{xx}$ was already measured in a device with a similar heterostructure~\cite{Corley-Wiciak2023}. Since realistic fluctuations in the electrostatic environment have a smaller impact, we now ignore these and only consider fluctuations of the averaged strain tensor. Figs.~\ref{fig:var_gfactor_quant}a, b show the simulation results with a $\sigma_{\epsilon_{ij}}=10^{-5}$, which is on the lower side of measurements and simulations~\cite{Corley-Wiciak2023,Abadillo2023}, as a function of magnetic field direction. Small $g$-factors require $\phi\approx(n+1/2)\pi$ with integer $n$ and $\theta \approx \pi/2$. Here $\phi$ and $\theta$ indicate the azimuthal and polar angles, respectively, of the magnetic field.

We model the misalignment angle of the spin quantization axes $\Delta\Phi$ as
\begin{align}
    \cos(\Delta\Phi) =\frac{\boldsymbol{B} (\mathcal{G}+\braket{\Delta\mathcal{G}})\cdot \boldsymbol{B}\ (\mathcal{G}+\Delta\mathcal{G})}{|\boldsymbol{B} (\mathcal{G}+\braket{\Delta\mathcal{G}})|\,|\boldsymbol{B}(\mathcal{G}+\Delta\mathcal{G})|}
\end{align}
Figs.~\ref{fig:var_gfactor_quant}c,d show the mean and standard deviation of the $\Delta\Phi$ as a function of magnetic field direction using the same parameters as in Figs.~\ref{fig:var_gfactor_quant}a, b. 
We find that large variations of the quantization axis are only possible if the magnetic field orientation is close to in-plane, $\theta \approx \pi/2$, and in the direction of weaker strain, $\phi\approx(n+1/2)\pi$. This opens an avenue to engineer devices with either small or large differences using strain.

\newpage

\bibliographystyle{Science}

\end{document}